\documentclass{jaa}
\usepackage{natbib}
\usepackage{graphicx}

\usepackage{url}
\usepackage{amsmath}

\usepackage{caption}
\usepackage{sidecap}
\sidecaptionvpos{figure}{t}
\usepackage[colorlinks=true,citecolor=blue]{hyperref}
\begin{document}\sloppy

\title{A sample of 25 radio galaxies with highly unusual radio morphologies, selected from the LoTSS-DR2 survey at 144 MHz}


\author{Gopal-Krishna\textsuperscript{1}, Dusmanta\ Patra\textsuperscript{2}, Ravi Joshi\textsuperscript{3,*} }
\affilOne{\textsuperscript{1}UM-DAE Centre for Excellence in Basic Sciences (CEBS), Vidyanagari, Mumbai-400098, India\\}
\affilTwo{\textsuperscript{2}S.~N.~Bose National Centre for Basic Sciences, Kolkata-700106, India.\\}
\affilThree{\textsuperscript{3}Indian Institute of Astrophysics, Sarjapur Rd., Koramangala, Bangalore-560034, India.\\}

\twocolumn[{

\maketitle

\corres{rvjoshirv@gmail.com}

\msinfo{x January xxxx}{x January xxxx}

\begin{abstract}
From a careful visual scrutiny of the radio structures of a well-defined sample of 2428 sources in the LoTSS-DR2 survey made at 144 MHz with a $6''$ beam, we have selected a subset of 25  (i.e.,  1\%) sources showing highly unusual radio structures, mostly not conforming to the prevalent radio morphological classification. Here we present and briefly discuss the basic properties of these rare morphological outliers and attempt to dissect their morphological peculiarities, based on multi-wavelength radio images and radio-optical overlays. Also, we underscore the need to accord due importance to such anomalous radio sources, considering the challenge they pose to the standard theoretical models and simulations of extragalactic double radio sources.

\end{abstract}

\keywords{galaxies: active; galaxies: jets; quasars: supermassive black holes; radio continuum: galaxies; (galaxies:) intergalactic medium; galaxies: nuclei}
}]


\doinum{12.3456/s78910-011-012-3}
\artcitid{\#\#\#\#}
\volnum{000}
\year{0000}
\pgrange{1--}
\setcounter{page}{1}
\lp{1}

\section{Introduction}
\label{sec:int}
From the beginning of extragalactic radio astronomy, arguably with the landmark discovery of a twin-lobed radio structure of the radio galaxy Cygnus A \citep{Jennison1953Natur.172..996J}, morphologies of extended radio structures created by such jetted active galactic nuclei (AGN) have been used as the key to unravelling the physical processes that govern their formation and evolution. A key strand of such studies has concentrated on deviations from the well-aligned twin-lobed structure, as witnessed in `classical' double radio sources like Cygnus A \citep[e.g.,][]{Hargrave1974MNRAS.166..305H, Carilli1991ApJ...383..554C}. Distortions from such a morphology, sometimes appearing as `head-tail' radio sources, were attributed to dynamical pressure acting on the twin-jets due a relative motion between the parent galaxy and external medium \citep[][]{Owen1976ApJ...205L...1O,Rudnick1977AJ.....82....1R, O'Dea1985AJ.....90..954O, Burns1987AJ.....94..587B}.
The rich diversity of such radio sources can be seen, e.g.,  in \citet{O'Donoghue1993ApJ...408..428O}. A second major type of distortion, manifesting as an inversion-symmetry of the twin-lobes, was attributed to a secular change/precession of the jet axis \citep[e.g.][]{Ekers1978Natur.276..588E, Bridle1979ApJ...228L...9B, Hogbom1979A&AS...36..173H}. 
Early reviews of the radio galaxy morphologies include those by \citet{Moffet1966ARA&A...4..145M, Willis1978PhyS...17..243W, Miley1980ARA&A..18..165M} and \citet{Fomalont1981IAUS...94..111F}. In course of time, additional morphological types were recognised, namely: 
(i) X-shaped radio galaxies (XRGs, e.g. \citet{Leahy1984MNRAS.210..929L, Leahy1991AJ....102..537L, Worrall1995ApJ...449...93W}, see, also \citet{Mackay1969MNRAS.145...31M, Hogbom1974A&A....34..341H} and \citet{Fomalont1981IAUS...94..111F}, for the XRG 3C315); 
(ii) Double-Double Radio Galaxies \citep[DDRGs][]{Lara1999A&A...348..699L, Schoenmakers2000MNRAS.315..371S, Jamrozy2008MNRAS.385.1286J} and a variety of radio sources associated with clusters of galaxies, albeit not directly identified with any cluster member \citep[e.g., see][for a recent update]{van-Weeren2019SSRv..215...16V}. Other peculiar morphological types, much rarer but no less puzzling, include the sources designated as `radio rings' \citep[e.g.][]{Buta1996FCPh...17...95B, Proctor2011ApJS..194...31P}, `odd radio circles’ \citep{Norris2021Galax...9...83N} and
`HYMORS' \citep{Gopal-Krishna-Witta2000A&A...363..507G, Gawronski2006A&A...447...63G, Kapinska2017AJ....154..253K, Gopal-Krishna2023JApA...44...44G}.

At the basic level, powerful extragalactic radio sources, i.e., `radio galaxies', 
are the creation of two oppositely-directed `collimated beams' of energy (or, `jets’ of energetic particles), squirted out at relativistic speeds from the vicinity of a spinning supermassive compact body located deep within the stellar core of a massive galaxy \citep[reviewed, e.g., by][]{Begelman1984RvMP...56..255B, Urry-Padvoni1995PASP..107..803U,Blanford2019ARA&A..57..467B}. It has long been suspected \citep{Shklovskii1963SvA.....6..465S, Rees1978Natur.275..516R} that the basic process fuelling this activity is the accretion of gas into the nucleus of a galaxy; probably onto a supermassive black hole situated there \citep{Lynden-Bell1969Natur.223..690L}. Half a century ago, double radio sources were slotted by \citet{Fanaroff1974MNRAS.167P..31F} into two morphological types, distinguished by the location of the brightness peaks within the two radio lobes. Thus, Fanaroff-Riley Class II (FR II), which are statistically more radio powerful, are dominated by edge-brightened lobes, whereas the less powerful ones (FR I sources) have the brightness peaking closer to the parent galaxy \citep[see, also,][]{Bridle1984AJ.....89..979B}. 
It was suggested that this morphological dichotomy may be fundamentally related to the accretion mode of the central engine \citep[][]{Baum1995ApJ...451...88B, Miraghaei2017MNRAS.466.4346M, Hine1979MNRAS.188..111H, Laing1994ASPC...54..227L} (for a recent update on the nuances associated with this proposal, see 
\citet{Hardcastle2020NewAR..8801539H, Mingo2022MNRAS.511.3250M}).
Moreover, an asymmetry in the jets' interaction with the external medium may also play a significant role in the FR dichotomy, a message echoed by HYMORS, as reviewed recently by \citet{Gopal-Krishna2023JApA...44...44G},  but see \citet{Harwood2020MNRAS.491..803H} for an alternative viewpoint. 
In addition, the bulk relativistic speed of the radiating plasma in the jet can dramatically alter the appearance of double radio sources via `relativistic beaming’ of the jet emission on parsec and even kilo-parsec scales \citep{Rees1966Natur.211..468R, Scheuer1979Natur.277..182S, Orr1982MNRAS.200.1067O}, also, 
\citep{Antonucci1986AJ.....92....1A}.
These factors have played a central role in the development of the paradigm unifying several major classes of extragalactic radio sources, taking radio galaxies as the `parent' population \citep[reviewed, e.g., by][]{Antonucci1993ARA&A..31..473A, Barthel1994ASPC...54..175B, Urry-Padvoni1995PASP..107..803U, Gopal-Krishna1995PNAS...9211399G, Tadhunter2016A&ARv..24...10T, Blanford2019ARA&A..57..467B, Hardcastle2020NewAR..8801539H, Antonucci2023Galax..11..102A}. 
In terms of their cosmological role, radio galaxies were proposed to be an important contributor to the process of magnetisation and metallisation of the cosmic-web at high redshifts \citep{Gopal-krishna2001ApJ...560L.115G, Gopal-Krishna2004JKAS...37..517G, Barai2011ApJ...727...54B, Ryu2008Sci...320..909R,Fabian2012ARA&A..50..455F}
(for intergalactic magnetisation by giant radio galaxies, see \citet{koronberg2001ApJ...560..178K}, and by quasar outflows, see \citet{Furlanetto2001ApJ...556..619F}).

As widely documented in the literature, and amply borne out by the classic case of BL Lacs \citep{Blandford1978bllo.conf..328B},
important clues on the `central engine' operating in radio galaxies can be gleaned by paying attention to their rare species/specimen that exhibit some highly uncommon traits, manifested, e.g., in radio morphology, spectrum, polarisation, or time-variability. Such `odd entities' can challenge and repurpose both the theoretical models of double radio sources, as well as their numerical simulation studies, in unanticipated ways. 
Such apparently peculiar radio sources showing rare traits are far from being `fringe players’ and their astrophysical potential is inadequately captured by the casual annotations, like `freaks' \citep{Fomalont1981IAUS...94..111F}, `curiosities' \citep{Proctor2011ApJS..194...31P}, or just `miscellaneous' \citep{Sasmal2022AN....34310083S}, and an upgradation commensurate with their true potential is warranted. With this impetus, we propose a new annotation for extended radio galaxies and quasars whose radio morphologies are clearly identifiable as `anomalies' (i.e., distinctly non-standard); more specifically in the sense of showing a marked departure from the known standard morphological types listed in the recent review article by \citet{Baldi2023A&ARv..31....3B}. 
The general term we propose for such radio morphological outliers is `ANOMERS' (ANOmalous Morphology Extragalactic Radio Sources), and a search for radio
sources under this rubric is the motive behind the present study. Note that in the optical domain, using the SDSS database, an extensive study with a similar motivation has been carried out by \citet{Baron2017MNRAS.465.4530B}, albeit from the perspective of optical spectrum.

We report here a set of 25 ANOMERS, carefully selected from the LoTSS-DR2, a large-area sky survey made at 144 MHz with a $6''$ beam \citep{Shimwell2022A&A...659A...1S}. This work builds on the extensive morphological search carried out by Proctor \citep{Proctor2011ApJS..194...31P} with a similar objective, but using the VLA/FIRST catalogue made at 1.5 GHz with a $5''$ beam \citep{Becker-first1995ApJ...450..559B}. His search involved a visual inspection of 7106 potential multiple-component systems, in order to pick sources of various morphological complexities and abnormalities. Several sources, thus short-listed, were annotated by them as `curiosities’ \footnote{Recently, \citet{Bera2022ApJS..260....7B} have identified 9 radio sources with complex morphologies in the LoTSS-DR1 \citep{Shimwell2019A&A...622A...1S}, although their paper provides no comments on those sources.}
That study is forerunners to the present work, with the basic difference that our search catalogue (LoTSS-DR2) has been made at 10 times lower frequency, yet with a very similar beam and much higher sensitivity to both compact and extended radio structures. It is worth reiterating that LoTSS-DR2 is an exceptionally valuable resource, representing an unprecedented combination of sensitivity (rms 83 $\mu$Jy/beam), resolution (6 arcsec), sky coverage and image fidelity at metre wavelengths (144 MHz), where aged radio features are expected to remain visible for much longer durations, thus offering a distinct advantage for finding sources with rare (non-standard) radio morphologies, among the vast population of extragalactic radio sources.

Throughout this paper, we assume a concordance cosmology with ${{\rm{\Omega }}}_{{\rm{m}}}=0.27$ and ${{\rm{\Omega }}}_{{\rm{\Lambda }}}=0.73$, and , ${H}_{0}\,=70\,\mathrm{km}\,{{\rm{s}}}^{-1}\,{\mathrm{Mpc}}^{-1}$ (see Table \ref{tab:list}). Spectral index $\alpha$ is defined as:  $S \propto \nu^\alpha$.

\section{The sample selection}
\label{sec:The sample}
In our selection process, the first filter applied to the LoTSS-DR2 catalogue \citep{Shimwell2022A&A...659A...1S} was that the catalogue values of major and minor axes of a source should exceed $24^{\prime\prime}$ and $10^{\prime\prime}$, respectively. This was to ensure that, on average, there are at least 2 beamwidths across each primary lobe and a least one beamwidth across each wing. Again, for a reliable detection of the wings, we limited our search to the fields for which the dynamic range given in the catalogue exceeds $\sim$ 40:1 \citep[see,][]{Cheung2007AJ....133.2097C}. Additionally, in order to keep the sample size manageable for the planned visual inspection of individual sources (see below) we considered only the sources whose total flux density given in the catalogue is $>$ 500 mJy at 144 MHz. Application of the above 3 criteria, which is not expected to introduce any bias, led to a basic sample consisting of 2428 sources, whose structures were then inspected by us visually {\footnote {several authors have underscored the importance of visual inspection of the radio image to arrive at a reliable morphological classification, e.g., \citet{Proctor2011ApJS..194...31P,Fanaroff2021MNRAS.505.6003F}}.
As a result, we could short-list 25 sources, deemed highly unusual (ANOMERS) and these are presented in this paper, together with some basic information on each source, mostly taken from the literature (Table \ref{tab:list}). Comments on individual sources are given in sect. 3. The radio maps together with radio-optical overlays for the sample are displayed in Figures 1-25. In doing so, we have also taken into consideration their published radio-optical data, which include the high-resolution radio maps obtained under the radio survey VLASS (FWHM $2.5''$ at $\sim$ 3
GHz, \citep{Lacy2020PASP..132c5001L}), the FIRST survey (FWHM $5''$ at 1.4 GHz, \citep{Becker-first1995ApJ...450..559B}), Pan-STARRS1 survey \citep{Chambers-PanSTARRS2016arXiv161205560C} and the deep large-sky optical survey DECaLS \citep{Dey2019AJ....157..168D}. We wish to clarify that we have used the Pan-STARRS1 survey, instead of DECaLS which goes almost 0.5-mag deeper (R-band), for the sake of uniformity, since DECaLS does not cover 4 sources in our sample of 25 ANOMERS (Table 1). Of the two sources without a counterpart in Pan-STARRS1, J011208+414702 and J093032+311215, the latter source is found to have a very faint counterpart in DECaLS, as mentioned in the text, Fig. \ref{fig:J093032} and Table 1. \\

\begin{SCfigure*}
    \includegraphics[width=14cm, height=11.5cm]{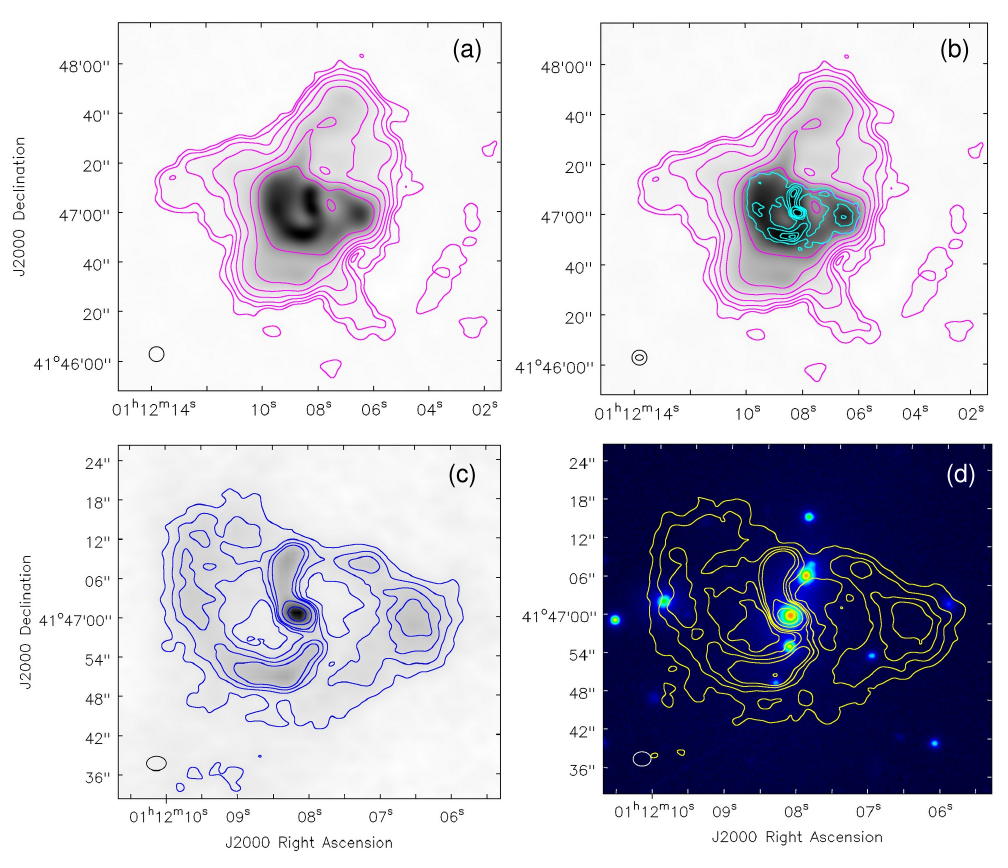}
  \caption{J011208+414702: (a) LoTSS-DR2 contour plot (magenta) \& grey-scale; (b) VLASS contours (cyan) overlaid on LoTSS contours (magenta) \& grey-scale; (c) VLSS contours (blue) \& grey-scale; (d) VLSS contours (yellow) overlaid on the optical (PanSTARRs, I-band) image.}
  \label{fig:J011208}
\end{SCfigure*}
\begin{SCfigure*}
    \includegraphics[width=14cm, height=11.5cm]{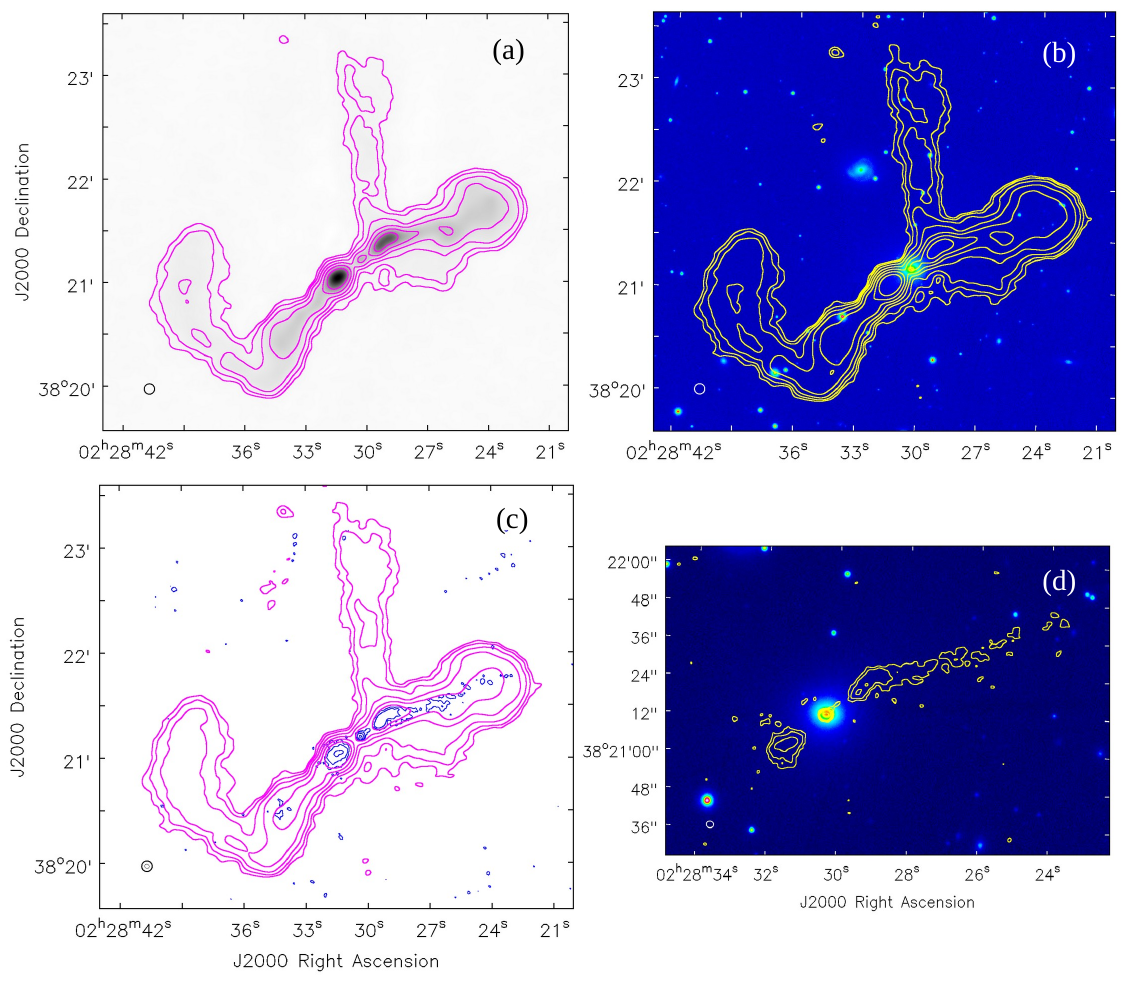}
  \caption{J022830+382108: (a) LoTSS-DR2 contour plot (magenta) \& grey-scale; (b) LoTSS-DR2 contours (yellow) overlaid on the optical (PanSTARRS, I-band) image; (c) VLASS contours (blue) overlaid on LoTSS-DR2 contours (magenta); (d) VLASS contours (yellow) overlaid on the zoomed optical (PanSTARR, I-band) image.}
  \label{fig:J022830}
\end{SCfigure*}

\section{Notes on individual sources}
\label{sec:results}
For our sample of 25 ANOMERS (Table \ref{tab:list}), radio continuum images (both contours and grey-scale), based on various combinations of LoTSS-DR2 (144 MHz), VLASS (3 GHz) and, in a few cases FIRST (1.4 GHz), together with the radio-optical overlays are presented in Figures 1-25. The contour values of these radio maps are given in Table \ref{tab:contour}. Notes on the individual sources are given below.\\

{\bf J011208+414702 (B3~0109+415):}
The source is identified with a galaxy which is the brightest (central) member of a roughly linear chain of galaxies (Fig. \ref{fig:J011208}). From the VLASS map, the structure consists of a radio spiral of diameter $\sim 35''$ embedded within diffuse radio emission of maximum extent $\sim$ $90''$, as measured on the LoTSS-DR2 map (Fig. \ref{fig:J011208}a). The inner spiral could not be discerned in an earlier VLA map at 1.5 GHz \citep{Vigotti1989AJ.....98..419V}, because of its 3 times larger beam. \\

\begin{SCfigure*}
    \includegraphics[width=14cm, height=11.5cm]{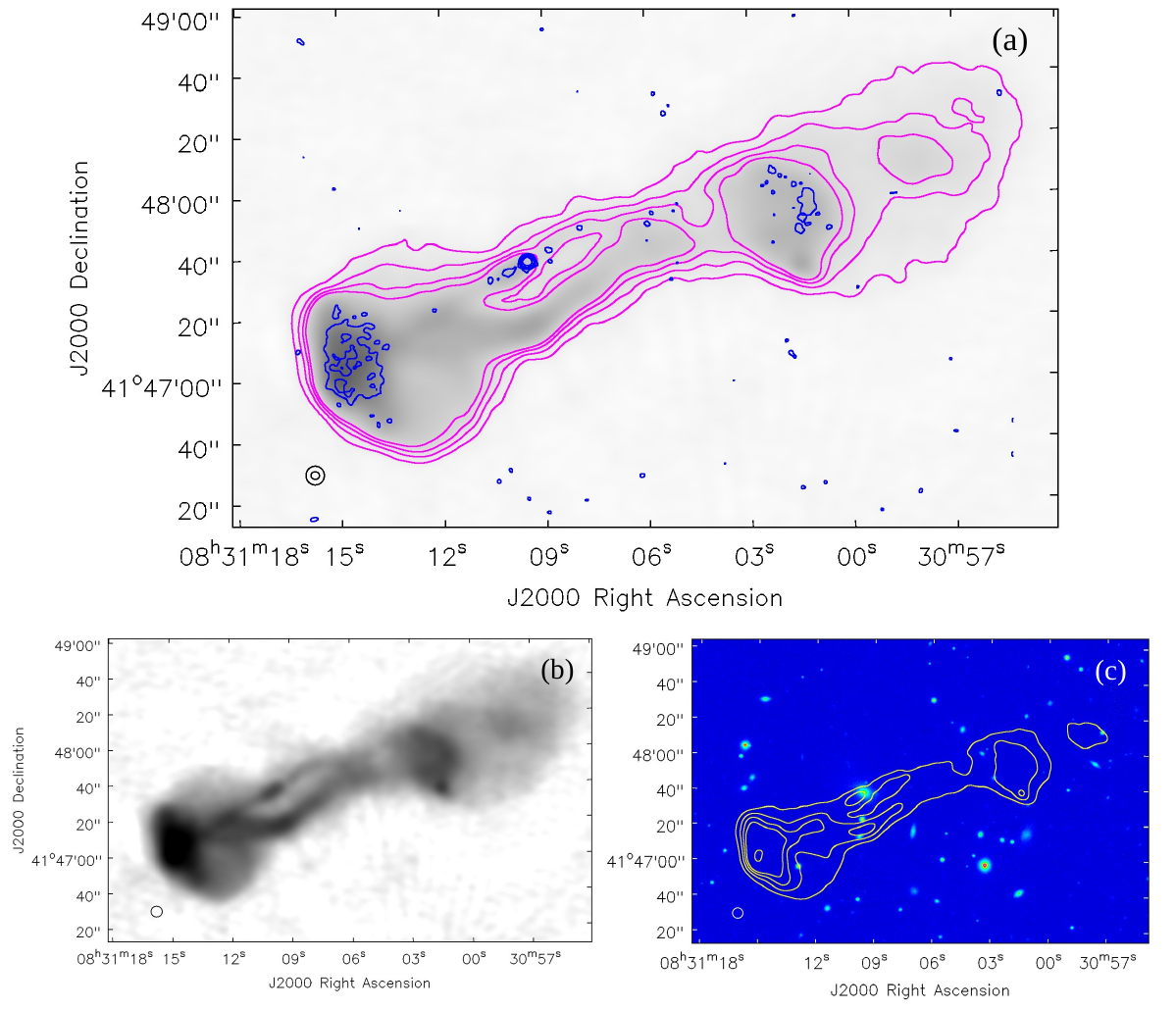}
  \caption{J083106+414741: (a) VLASS contours (blue) overlaid on LoTSS-DR2 contours (magenta) \& grey-scale; (b) LoTSS-DR2 grey-scale only; (c) LoTSS-DR2 contours (yellow) overlaid on the  optical (PanSTARR, I-band) image. }
\label{fig:J083106}
\end{SCfigure*}
\begin{SCfigure*}
    \includegraphics[width=11cm, height=12cm]{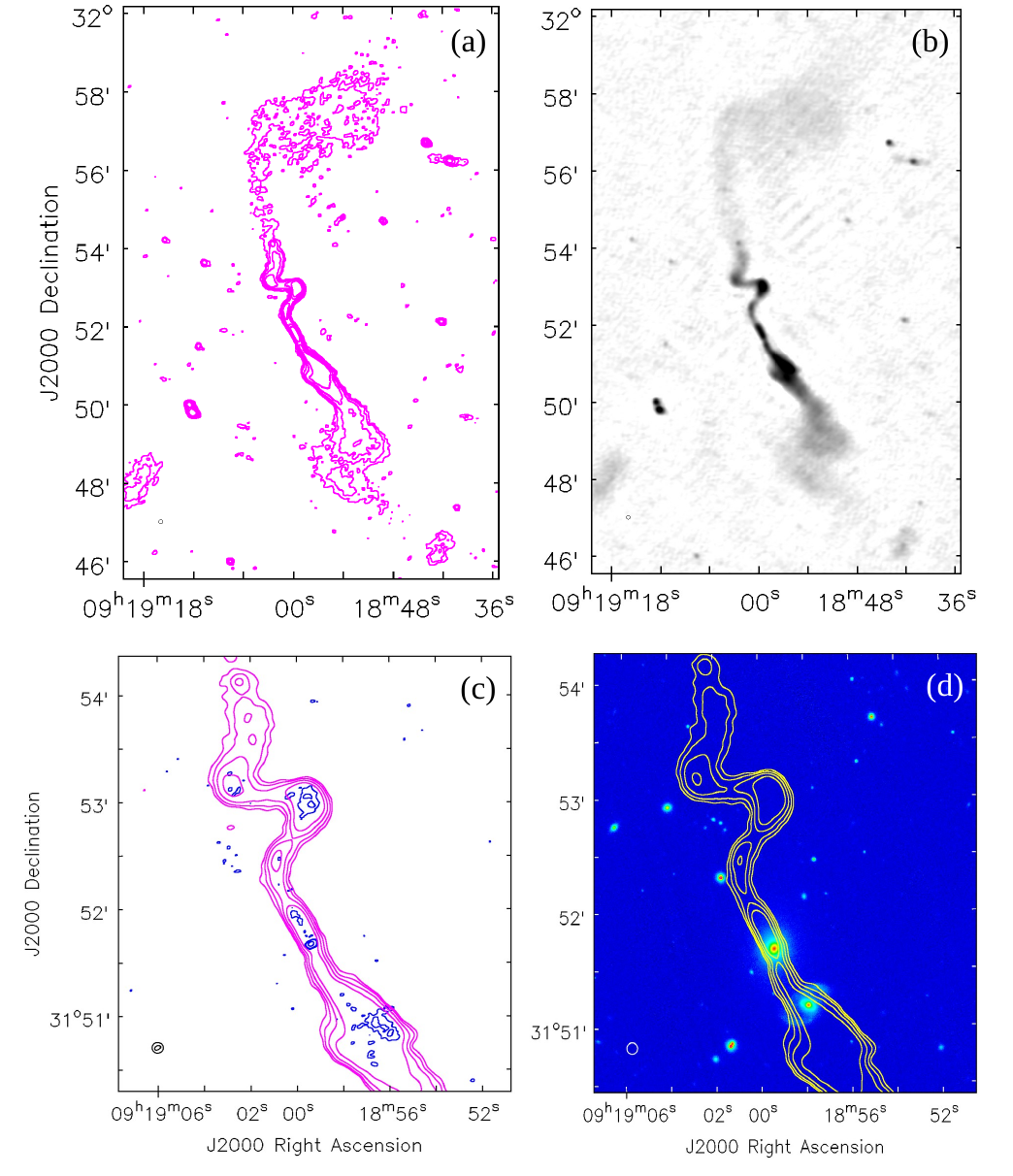}
  \caption{J091900+315254: (a) LoTSS-DR2 contour plot (magenta); (b) LoTSS-DR2 grey-scale only; (c) VLASS contours (blue) overlaid on the zoomed LoTSS-DR2 contours (magenta); (d) VLASS contours (yellow) overlaid on the zoomed optical (PanSTARR, I-band) image.}
\label{fig:J091900}
\end{SCfigure*}
{\bf J022830+382108 (B3 0225+381):}
The source is identified with a bright galaxy ({\it z} = 0.0328) whose radio core is clearly detected in the VLASS map and also visible in the LoTSS-DR2 map. (Fig. \ref{fig:J022830}). This is a FR 1 type radio source, with the brighter eastern jet bending sharply northward by $\sim$ $90^{o}$. Both jets brighten up at $\sim$ $18''$ (12 kpc) from the host galaxy. From the brightened region in the western jet emanates a spectacular `chimney-like’ well-collimated northward radio spur of extent $\sim$ $100''$ (66 kpc).
No optical object is seen near the foot of the chimney. 
A likely explanation for the chimney is blow-out of the jet plasma from a high-pressure region, possibly via a successful formation of a De Laval nozzle, as proposed by \citep{Capetti2002A&A...394...39C} as an explanation for the exceptionally huge radio 'wings’ seen in some X-shaped radio galaxies, e.g., B2 0828+32 \citep{Parma1985A&AS...59..511P,Klein1995A&A...303..427K} and 3C 403 \citep{Black1992MNRAS.256..186B, Condon1995AJ....109.2318C, Dennett-Thorpe1999MNRAS.304..271D}.
Note that the apparent `bifurcation' of the western jet of the present source is reminiscent of the southern jet of the FR II radio galaxy 3C 321 \citep{Evans2008ApJ...675.1057E}, however, that has probably occurred due to a temporary interruption of the jet by a passing galaxy \citep{Gopal2012RAA....12..127G}.\\
\begin{SCfigure*}
    \includegraphics[width=12.5cm, height=11.5cm]{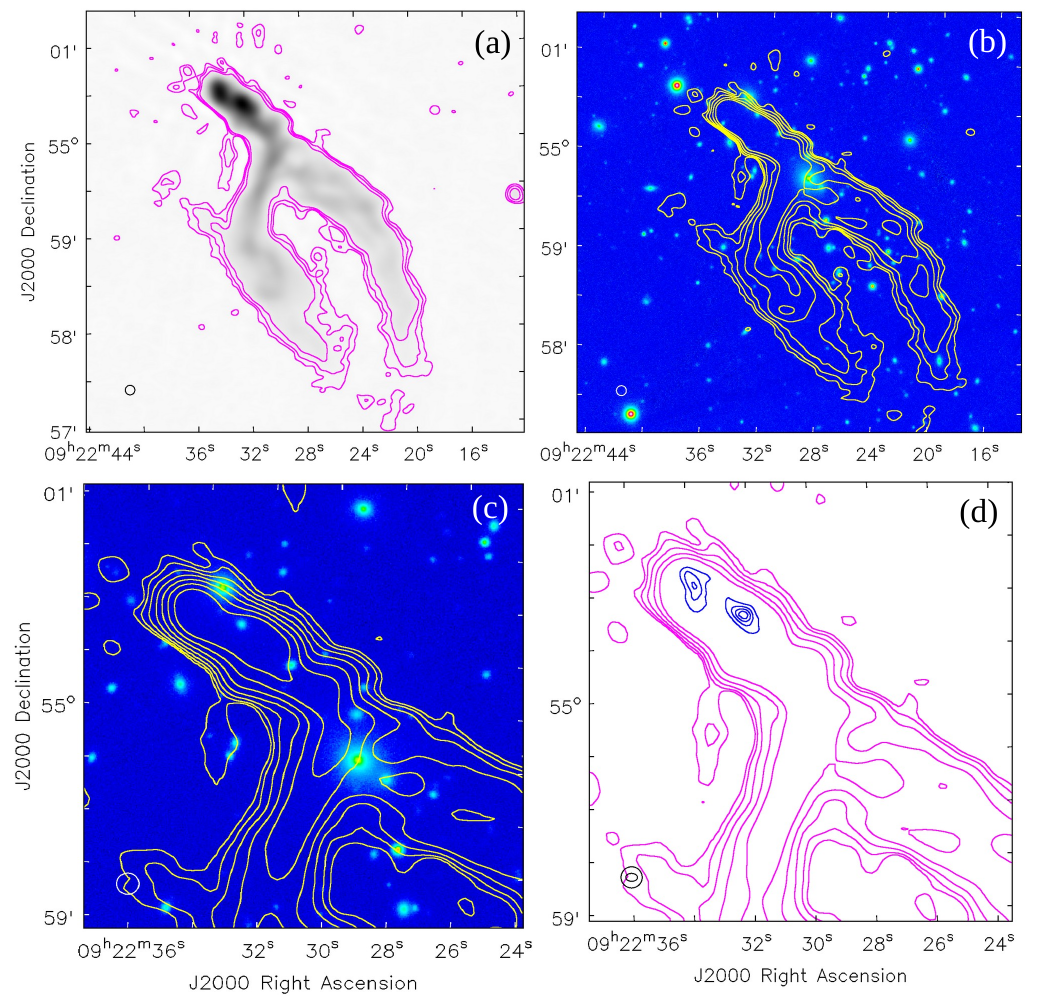}
  \caption{J092229+545943: (a) LoTSS-DR2 contour plot (magenta) \& grey-scale; (b) LoTSS-DR2 contours (yellow) overlaid on the  optical (PanSTARR, I-band) image; (c) LoTSS-DR2 contours (yellow) overlaid on the zoomed optical (PanSTARR, I-band) image; (d) VLASS contours (blue) overlaid on the zoomed LoTSS-DR2 contours (magenta) image.}
\label{fig:J092229}
\end{SCfigure*}
\begin{SCfigure*}
    \includegraphics[width=12.5cm, height=12cm]{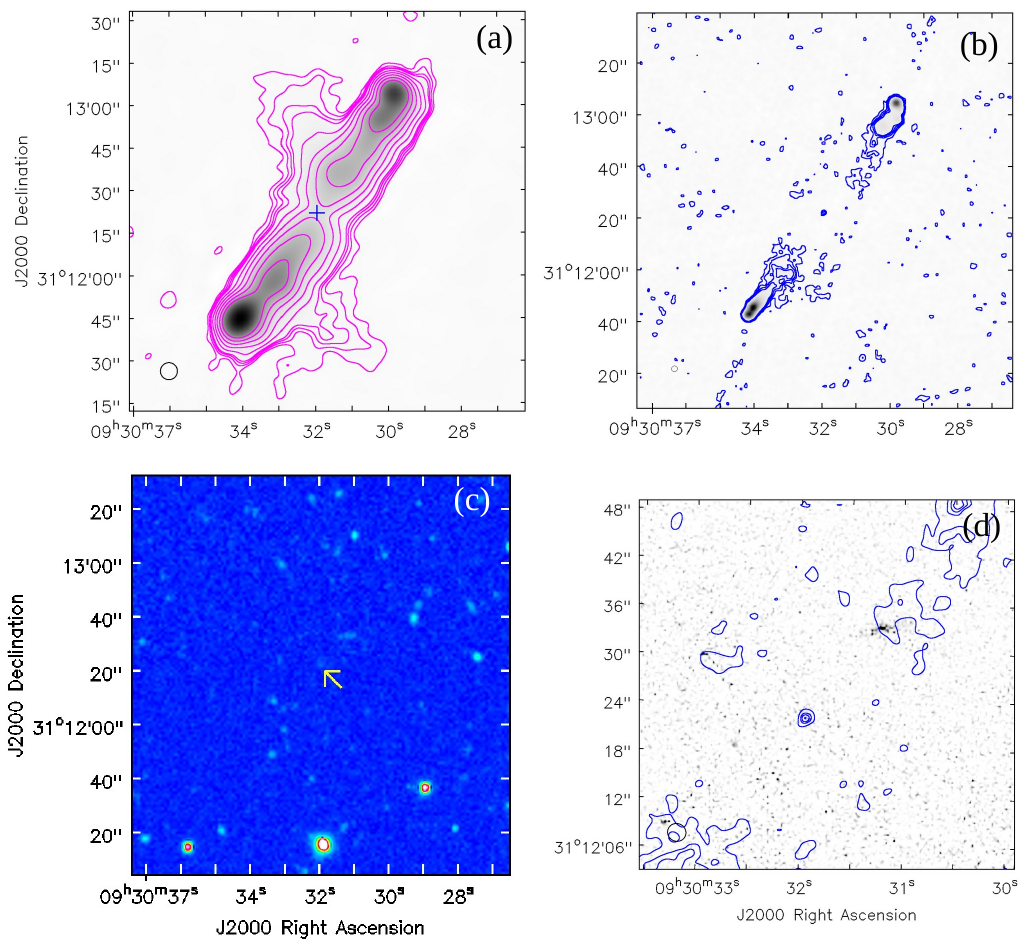}x`
  \caption{J093032+311215: (a) LoTSS-DR2 contour plot (magenta) \& grey-scale; (b) VLASS contour plot (blue) \& grey-scale; (c) DECaLS (r-band) image of the field. The faint object near the detection limit, as indicated by the arrow, lies at the position of the radio core seen in the panel to the right; (d) VLASS contour plot (blue) overlaid on the zoomed grey optical (PanSTARR, I-band) image.}
  \label{fig:J093032}
\end{SCfigure*}

{\bf J083106+414741:}
This large FR 1 radio galaxy is identified with a BCG (brightest cluster galaxy) at {\it z}(spec) = 0.13195 \citep{Ahn2012ApJS..203...21A}. 
Its radio core is clearly detected in the VLASS map and the LoTSS-DR2 map and it has a jet-pair associated with it (e.g., see the grey-scale image in Fig. \ref{fig:J083106}b). Also seen is a bright radio ridge of total length $\sim$ $110''$ (260 kpc), which is offset southward of the jet pair and runs almost parallel to it. This extremely rare morphology is reminiscent of the radio galaxy 3C 338, a wide-angle-tail (WAT) hosted by the BCG of the rich cluster Abell 2199, in which a bright radio ridge of ultra-steep spectrum was detected \citep{Burns1983ApJ...271..575B}. A similar ridge has been identified as a `collimated synchrotron thread’ (CST) connecting the two radio tails of the WAT ESO137-006 \citep{Ramatsoku2020A&A...636L...1R}. Recently, such CSTs have been interpreted as electrical discharges (`cosmic thunderbolts') between the two radio tails \citep{Gopal-Krishna2024MNRAS.529L.135G}. \\

{\bf J091900+315254:}
This FR I radio galaxy with the largest angular size of $\sim$ $10'$  (720 kpc) is hosted by a {\it z} = 0.06194 LINER galaxy whose radio core is clearly detected in the VLASS map at 3 GHz (Fig. \ref{fig:J091900}). The most remarkable feature, seen in its LOFAR map is a huge `kink’ in the northern jet, about 1 arcmin from the core. The kink has an amplitude of $\sim$ $32''$ (38 kpc) and a total length of $\sim 90''$ (110 kpc). The kink has no counterpart in the southern jet. The exceptionality of this `superkink' is that there is no sign of the jet's decollimation/disruption over the kink's huge extent (Fig. \ref{fig:J091900}d), unlike the case with other huge jet kinks reported in recent literature, e.g., the $\sim$ 100 kpc kink in the western jet of the `Barbell’ shaped giant radio galaxy \citep{Dabhade2022A&A...668A..64D}  and the kink in the northern jet of the WAT 3C 40B \citep{Rudnick2022ApJ...935..168R}. 
Possible explanations for the kink formation in radio jets include  Kelvin-Helmholtz instabilities, particularly in low magnetisation jets \citep[e.g.,][]{Mukherjee2020MNRAS.499..681M, Mukherjee2021MNRAS.505.2267M, Loken1995ApJ...445...80L, Lal2021ApJ...915..126L}. Another cause for jet destabilisation is current-driven instabilites \citep[e.g.,][]{Nakamura2007ApJ...656..721N, Mizuno2014ApJ...784..167M}. However, since there is no sign of the jet's decollimation associated with the present superkink, alternative possibilities should be explored, e.g., a gradual eastward dragging of the northern jet by the gaseous envelope of a moving group of galaxies. 
A candidate group of galaxies is in fact visible in the space vacated due jet's diversion at the kink (Fig. \ref{fig:J091900}d). While redshift of this potential galaxy group remains to be determined, we note that likely examples of jet's diversion by a moving galaxy (or a group of galaxies) have been reported, e.g., 3C 321 \citep{Gopal2012RAA....12..127G} and 3C 40B \citep{Rudnick2022ApJ...935..168R}.\\

{\bf J092229+545943 (The `plier'):}
This enigmatic radio source, resembling a plier (or a nutcracker) has a BCG  ({\it z} (spec)= 0.18363) situated at its pivot (tri-junction) (Fig. \ref{fig:J092229}). The overall radio extent seen in the LOFAR map is $\sim$ $200''$ (622 kpc). Much of the peculiar structure showing 3 arms, has remained undetected in the VLASS map which exhibits just two discrete patches coinciding with the northern radio arm. Roughly in the middle of these two radio patches is seen a faint optical object (J092229.12+545947.1) with a redshift of 0.18363. If this identification is established, that would suggest that the two radio patches are an independent double radio source and the remaining two (southerly) radio arms could then be an independent source, probably a narrow-angle-tail (NAT) produced by the BCG. Note that the BCG itself lacks a radio core, even in the VLASS map, which is unexpected for a radio-loud BCG. More extensive observations of this highly peculiar system are needed, in order to verify this proposed scenario of a chance superposition of two independent radio sources. \\

\begin{SCfigure*}
    \includegraphics[width=13cm, height=11.5cm]{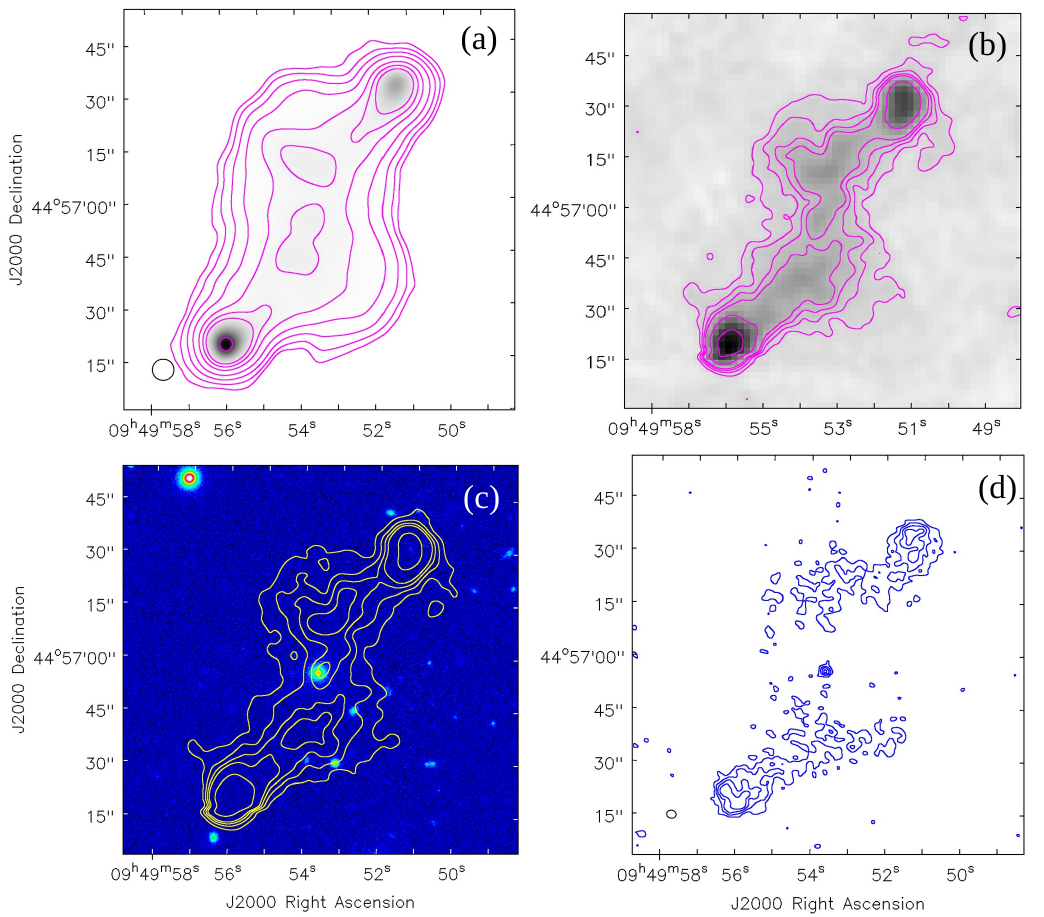}
  \caption{J094953+445658: (a) LoTSS-DR2 contour plot (magenta) \& grey-scale; (b) FIRST contour plot (magenta) \& grey-scale; (c) FIRST contours (yellow) overlaid on the optical (PanSTARR, I-band) image; (d) VLASS contour plot (blue).}
  \label{fig:J094953}
\end{SCfigure*}
\begin{SCfigure*}
    \includegraphics[width=14cm, height=11.5cm]{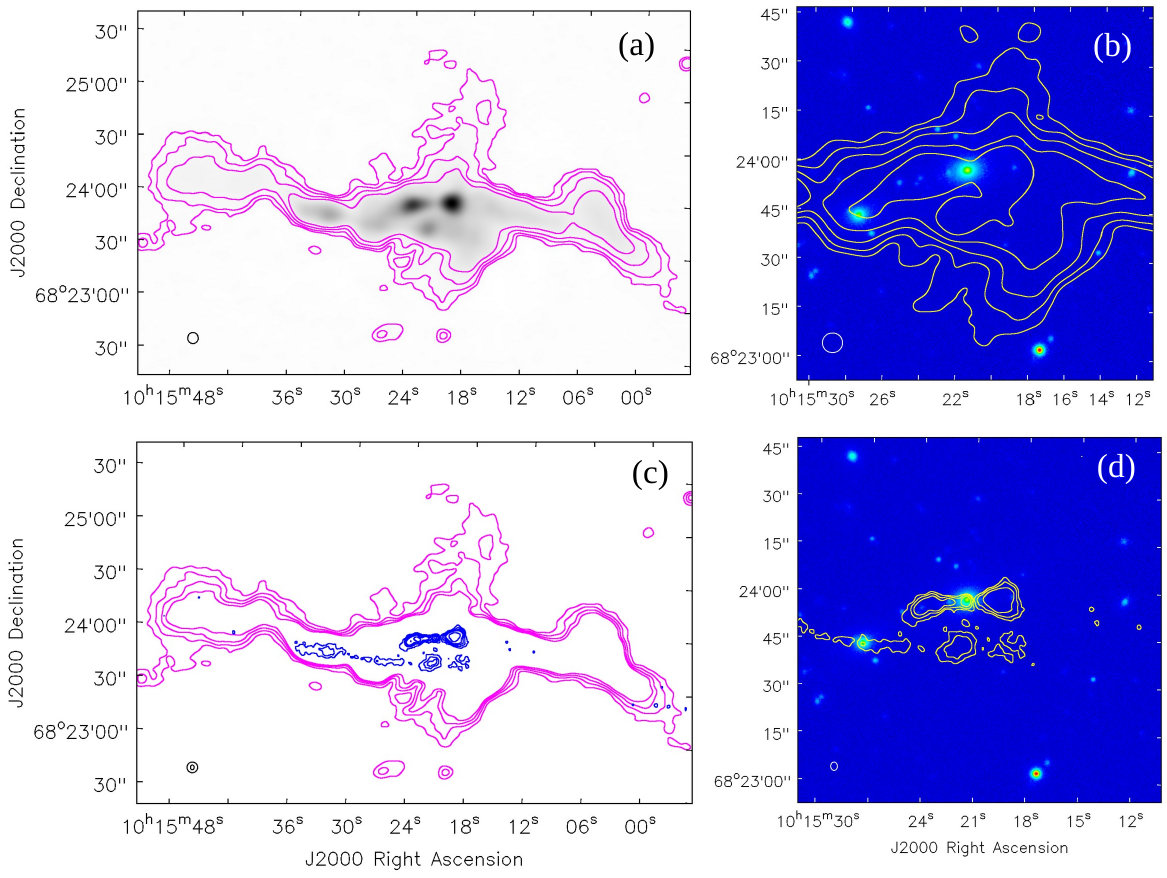}
  \caption{J101522+682351: (a) LoTSS-DR2 contour plot (magenta) \& grey-scale; (b) LoTSS-DR2 contour plot (yellow) overlaid on the zoomed optical (PanSTARR, I-band) image; (c) VLASS contours (blue) overlaid on the LoTSS-DR2 contours (magenta); (d) VLASS contours (yellow) overlaid on the zoomed optical (PanSTARR, I-band) image.  }
  \label{fig:J101522}
\end{SCfigure*}

{\bf J093032+311215 (4C+31.34):}
As seen from Fig. \ref{fig:J093032}, the two lobes of this FR II source are very well aligned with a centrally located compact radio core seen in the VLASS map. At the radio core position a $m{_R}$ $\sim 22.1$ object of unknown redshift is faintly visible in the DECaLS image (Fig. \ref{fig:J093032}c), whose position is given in Table 1.
Remarkably, the LoTSS-DR2 map (Fig. \ref{fig:J093032}a) shows two 
oppositely-directed prominent radio spurs emanating from the two lobes, roughly at their mid-points. As in the case of the similarly large radio
spur in J022830+382108 (Fig. \ref {fig:J022830}), a brightening of the lobe near the foot of each spur is apparent and, once again, no optical object is visible near these two locations. Thus, it is again conceivable that the two radio spurs of the present source have burst out of the high-pressure regions through De Laval nozzle formation. However, since the two radio spurs are {\it oppositely} directed,
an alternative mechanism for their origin is hinted. For instance, they could arise from interaction of the two jets with rotating shells around the host galaxy. Evidence for such jet-shell interaction \citep{Gopal1983Natur.303..217G} has been noted in some radio galaxies, e.g.,  Centaurus A, \citep[see, e.g.,][]{Gopal2010ApJ...720L.155G} and 3C 433 \citep{Gopal2012RAA....12..127G}, which calls for a deep optical/near-infrared imaging of the present system. Note that the overall radio structure of the present source bears some resemblance to the radio galaxy 4C+12.03 imaged with uGMRT and MeerKat \citep{Fanaroff2021MNRAS.505.6003F}, with the difference that at least the radio spur of the southern lobe in the present source does not represent bulk diversion of the entire synchrotron plasma backflowing in that lobe, but of just a fraction of it, leaving the remaining plasma to continue its backflow towards the host galaxy (see the LoTSS map in Fig. \ref{fig:J093032}a), akin to the huge radio spur seen in J022830+382108 (Fig. \ref{fig:J022830}). \\

\begin{SCfigure*}
    \includegraphics[width=14cm, height=11.5cm]{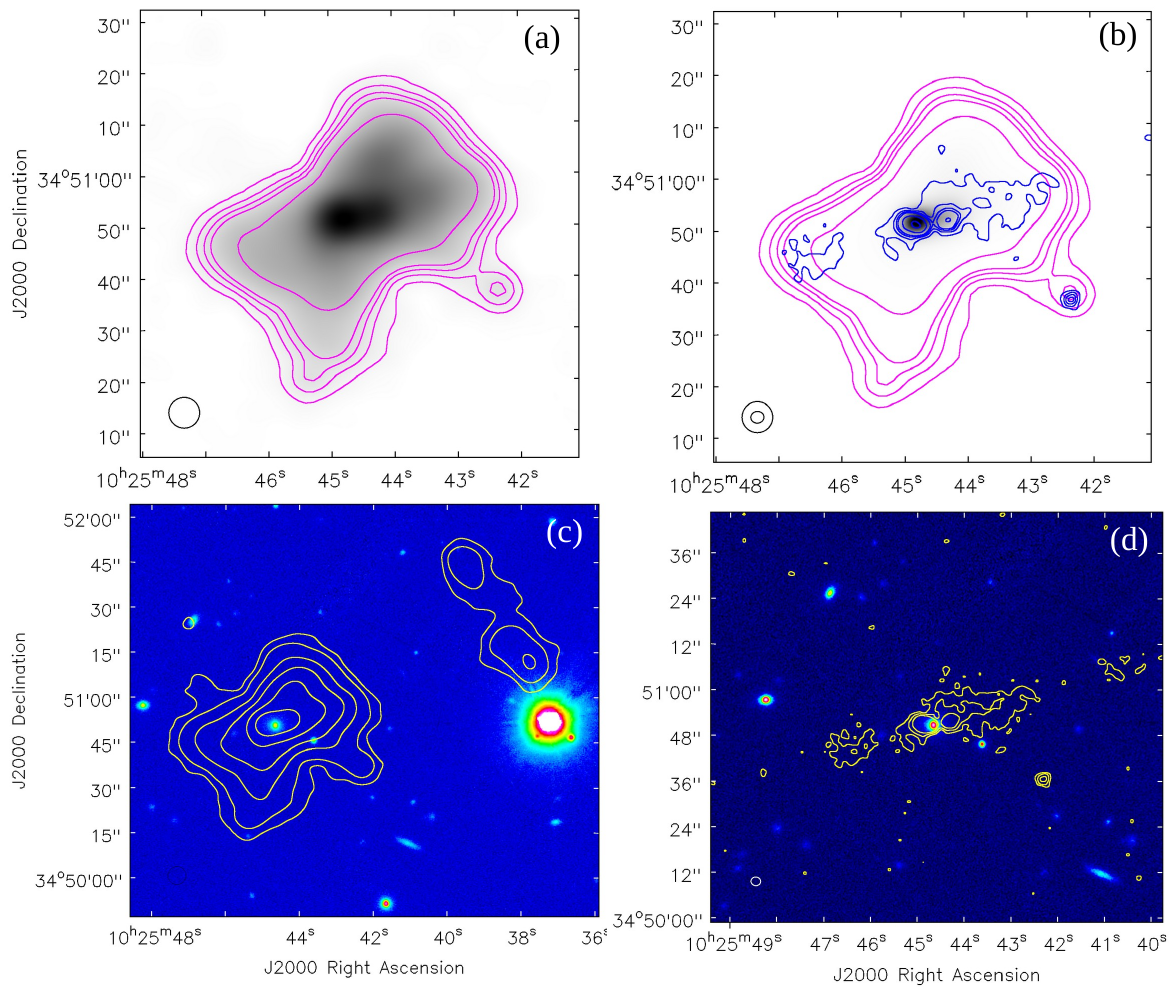}
  \caption{J102544+345051: (a) LoTSS-DR2 contour plot (magenta) \& grey-scale; (b) VLASS contours (blue) overlaid on the LoTSS-DR2 contours (magenta) \& grey; (c) LoTSS-DR2 contour plot (yellow) overlaid on the optical (PanSTARR, I-band) image; (d) VLASS contour plot (yellow) overlaid on the optical (PanSTARR, I-band) image. }
  \label{fig:J102544}
\end{SCfigure*}

\begin{SCfigure*}
    \includegraphics[width=13cm, height=11.5cm]{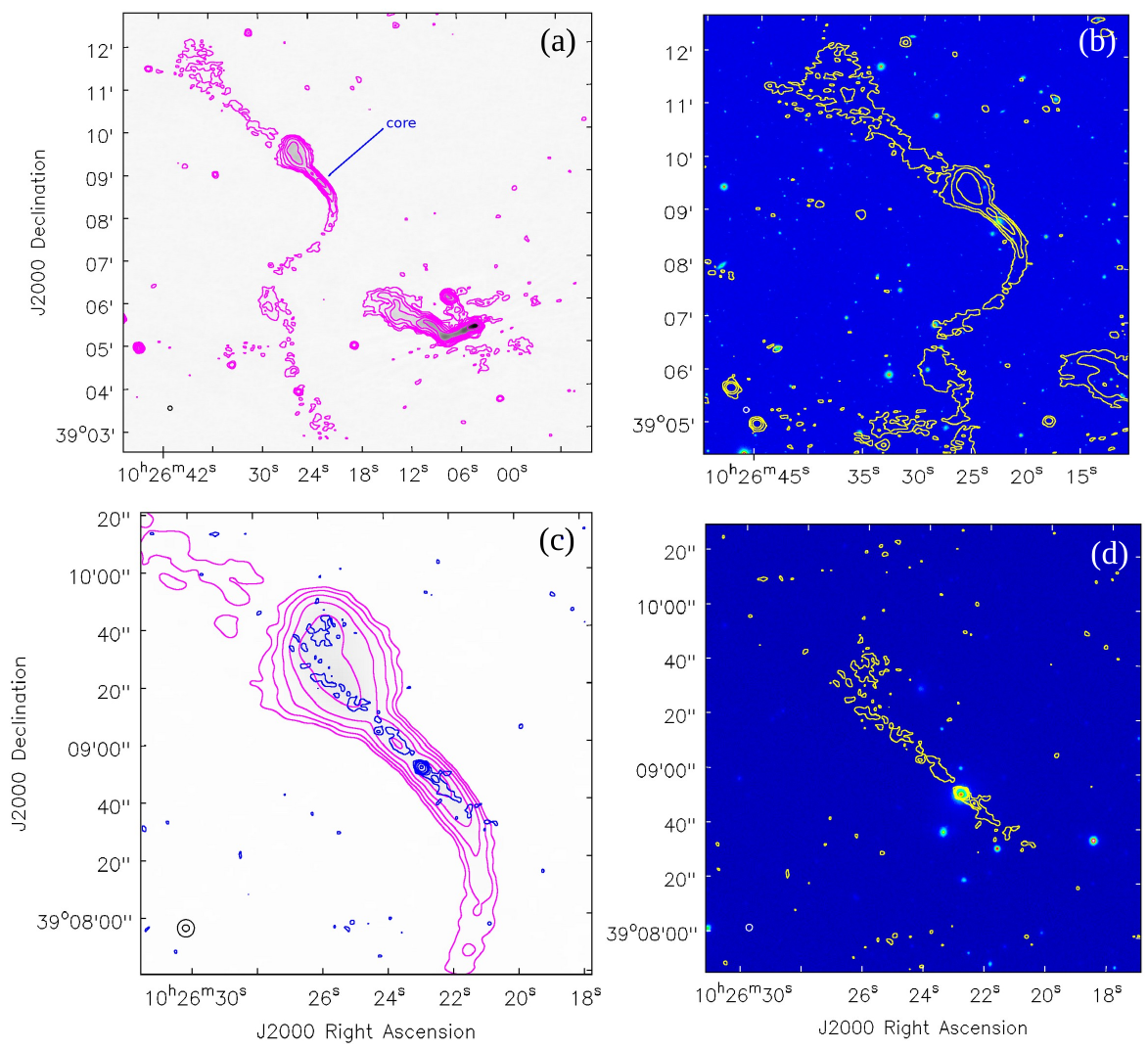}
  \caption{J102625+390929: (a) LoTSS-DR2 contour plot (magenta) \& grey-scale; (b) LoTSS-DR2 contour plot (yellow) overlaid on the optical (PanSTARR, I-band) image; (c) VLASS contours (blue) overlaid on the zoomed LoTSS-DR2 contours (magenta); (d) VLASS contour plot (yellow) overlaid on the optical (PanSTARR, I-band) image.}
  \label{fig:J102625}
\end{SCfigure*}

{\bf J094953+445658:}
This source at {\it z}(spec) = 0.195335 has previously been classified as an XRG \citep{Yang2019ApJS..245...17Y} 
 and a C-shaped radio galaxy \citep{Liu2019RAA....19..127L}. 
 From Fig. \ref{fig:J094953}, it is seen to be a FR II radio galaxy with the core (and associated bright galaxy) located exactly in the middle of the two terminal hotspots. As seen from both LoTSS-DR2 and VLASS maps, the two radio lobes, although parallel to each other, show a lateral offset. An example of such peculiar morphology, XRG J1328+5654, has recently been highlighted by \citet{Gopal-Krishna-3C223.1-2022A&A...663L...8G} who have also pointed out potential difficulty in
explaining the origin of the lateral offset between the two parallel {\it primary} lobes. We have located another such case in the literature, namely $B1249\_092$ 
\citep{Ulvestad1981AJ.....86.1010U}. We shall refer to this radio morphology as `offset parallel twin lobes' (OPTL).\\ 

\begin{table*}
\caption{\bf The  sample of 25 `Anomalous Morphology Extragalactic Radio Sources'}
\vspace{0.5cm}
\begin{centering}
\scriptsize
\addtolength{\tabcolsep}{-0.3em}
\begin{tabular}{cccclrrrrcc}
\hline\hline
Name          & R.A.   & Dec.   & Ref. to  &Redshift & Scale~~  &\multicolumn{2}{c}{LoTSS Flux (144 MHz)}& NVSS Flux & Spectral               &Other Name  \\
              &        &        &Position  & ~~~($z$)&          & (measured) &   (catalog)~~             & (1.4 GHz) & Index                  &   (Simbad)          \\  
              &(J2000) &(J2000) &          &         & $(kpc/')$& (mJy)      &	(mJy)~~~~              & (mJy)~~~~ & ($\alpha_{144}^{1400}$)&     \\
\hline
J011208+414702&01 12 08.16	& +41 47 00.6 &VLASS&           &      & 2296.3& $2366.7\pm 12.8$ & $532.3\pm19.4$ &--0.64 &          \\
J022830+382108&02 28 30.34  & +38 21 12.3 &VLASS&0.03278    & 39.48& 1060.8& $1214.9\pm 9.7$ & $207.7\pm6.9$ &--0.71 & B3 0225+381 \\
J083106+414741&08 31 09.56  & +41 47 37.9 &VLASS&$0.13195^*$&141.78& 1047.8& $\dag$~~~~~~      & $99.8\pm3.7$   &--1.03 & YHW2016  \\
J091900+315254&09 18 59.36  & +31 51 40.7 &OPT  &$0.06194^*$& 72.12& 1144.8& $1087.9\pm9.0$   & $153.7\pm5.0$  &--0.86 & LEDA 1968222  \\
J092229+545943&09 22 29.12  & +54 59 47.1 & OPT &$0.18363^*$&186.48& 1421.7& $\dag$~~~~~~        & $122.4\pm4.3$  &--1.07 &  LEDA 2487098  \\
J093032+311215&09 30 31.95  & +31 12 22.0 &OPT&           &      & 3149.5& $3228.5\pm8.5$   & $405.0\pm8.7$  &--0.91 & 4C 31.34 \\
J094953+445658&09 49 53.61  & +44 56 55.6 &VLASS&$0.19533^*$&195.84&  953.3& $1057.5\pm11.6$  & $207.4\pm7.0$  &--0.71 &       \\
J101522+682351&10 15 21.50  & +68 23 57.7 &OPT  &$0.2098$ &207.12& 1993.3& $2000.9\pm12.3$  & $352.6\pm12.2$ &--0.76 &      \\  
J102544+345051&10 25 44.64  & +34 50 50.4 &OPT&$0.18601^*$&188.40&  588.0& $600.0\pm2.4$    & $109.0\pm4.2$  &--0.74 & B2 1022+35 \\
J102625+390929&10 26 22.88  & +39 08 51.5 &VLASS&$0.14721^*$&155.58&  359.9& $\dag$~~~~~~       & $58.5\pm1.5$   &--0.80 & B3 1023+394  \\
J102704+444547&10 27 04.65  & +44 45 50.1 &VLASS&$0.58605^*$&401.22& 1551.4& $1573.5\pm3.8$   & $200.3\pm7.6$  &--0.90 &   \\
J104234+364040&10 42 33.34  & +36 39 46.6 &OPT&$0.14218^*$&151.08&  686.5& $730.3\pm3.0$    & $56.7\pm2.1$   &--1.12 & LEDA 3097664 \\
J113222+555823&11 32 22.74  & +55 58 18.5   &OPT&$0.05154^*$& 60.72&  618.5& $597.9\pm4.4$    & $52.4\pm2.2$   &--1.08 & MCG+09-19-110  \\
J114429+370915&11 44 27.20  & +37 08 32.1   &OPT&$0.11481^*$&125.76& 8994.1& $\dag$~~~~~~     & 2130.2         &--0.63 & 87GB 114142.2+372406  \\
J114434+672424&11 44 36.67  & +67 24 21.5 &VLASS&$0.11612^*$&127.02& 2008.6& $2020.4\pm8.0$   & $230.8\pm8.1$  &--0.95 & 6C 114151+674026  \\
J115608+441058&11 56 07.78  & +44 10 57.2 &OPT-1&$0.29971^*$ &      &       &                  &                &       &                 \\
              &11 56 08.11  & +44 10 58.2 &OPT-2&$0.29703^*$&267.8 & 737.6& $773.5\pm2.7$     & $31.3\pm1.6$   &--1.41 & 2MASX J11560809+441058  \\
              &11 56 08.25  & +44 11 00.6 &OPT-3&           &      &       &                  &                &       &                  \\
J133116+441851&13 31 16.79  & +44 18 49.5 &VLASS& 0.248     &235.14& 679.1 &$694.0\pm2.9$     & $142.5\pm5.4$  &--0.68 & 6C 132908+443418  \\
J135141+555938&13 51 42.12  & +55 59 43.0 &VLASS&$0.06907^*$&79.74 &1106.3 &$1136.4\pm2.9$    & $220.8\pm7.9$  &--0.71 & SDSS-C4-DR3 3048  \\
J140311+382756&14 03 11.67  & +38 28 03.4 &VLASS& 0.541     &385.74 &4012.1 &$4097.4\pm6.3$    & $662.4\pm13.0$ &--0.79 & 4C 38.38  \\
J143527+550756&14 35 28.48  & +55 07 51.7 &VLASS&$0.13985^*$&148.98&2828.4 &$2908.9\pm8.0$    & $474.5\pm14.9$ &--0.78& 6C 143354+552059 \\
J165732+431941&16 57 31.94  & +43 19 55.2 &VLASS&$0.204^*$    &202.62&491.3  &$512.5\pm2.4$     & $105.6\pm3.9$  &--0.68& LEDA 2218211  \\
J170940+342537&17 09 38.35  & +34 25 52.9 &VLASS&$0.0806^*$ & 91.80&6200.2 &$4401.5\pm29.3$   & $641.8\pm24.0$ &--1.00& 4C 34.45  \\
J221255+220523&22 12 55.06  & +22 05 28.3 &OPT-1&0.466      &      &1200.7 &$1250.9\pm5.7$    & $177.5\pm3.7$  &--0.84&     \\
              &22 12 56.36  & +22 05 08.3 &OPT-2&0.171      &      &       &                  &                &      &      \\
              &22 12 55.99  & +22 05 11.2 &OPT-3&0.868      &      &       &                  &                &      &       \\
J221558+290921&22 15 58.73  & +29 09 08.4 &VLASS & $0.232^*$&223.74&842.3  &$878.5\pm4.1$     &$164.0\pm3.8$   &--0.72&   \\
J235938+400644&23 59 38.63  & +40 06 45.8 &OPT-1 &          &      &       &                  &                &      &    \\
              & 23 59 38.56 & +40 06 32.1 &OPT-2 &0.19670   &196.92&773.6  &$787.9\pm2.7$     &$89.0\pm3.2$    &--0.95&    \\
\hline
\end{tabular}
\label{tab:list}
\end{centering}
\scriptsize
\vspace{0.2cm}
${\ast}$ --- The values marked with an asterisk are spectroscopic redshifts and the remaining values are photometric redshifts.\\
$~~^{\dag}$ --- The missing entry refers to the case where a single source is listed as multiple sources in the LoTSS-DR2 catalog. \\
\end{table*}

{\bf J101522+682351:}
The most remarkable feature of this highly extended FR I radio source of size $\sim$ $4.5'$ (932 kpc) is the conspicuous central radio bulge seen in the LoTSS-DR2 map (Fig. \ref{fig:J101522}). The redshift of the central bright galaxy is 0.2098, and it is detected in the VLASS map (Fig. \ref{fig:J101522}). The galaxy is straddled by a symmetric pair of relatively compact radio lobes separated by $\sim$ $25''$. This appears to be a recently ejected pair of radio lobes, embedded within an older FR I source and sharing the radio axis with it. Note that a thin ridge-line of radio emission visible in the VLASS map (which could be a jet from a previous episode of activity) is significantly displaced southward from the young double source, suggesting that the host galaxy may have moved towards north between the two episodes of jet activity. If so, this system could be an example of `detached double-double radio galaxy' \citep[dDDRG,][]{Gopal-Krishna-dDDRG2022PASA...39...49G}. Alternatively, the thin radio ridge could be associated with the second brightest galaxy which is seen  $\sim$ $35''$ south-east of the brightest galaxy, in the radio-optical overlay (Fig. \ref{fig:J101522}d). 

Another noteworthy feature of this radio source, revealed by the LoTSS-DR2 map is the prominent radio spurs emanating from the central region towards the north and south, i.e., orthogonal to the main radio axis. Both spurs are broad, but do seem to connect to the central galaxy. If confirmed by deeper radio mapping, this source would be a good candidate for a `pure’ XRG, where both radio ejection axes are well aligned with the host active galaxy {\bf \citep{Roberts2015ApJS..220....7R}}, as expected in the basic spin-flip scenario \citep{Rottmann2001PhDT.......173R, Zier2001A&A...377...23Z, Merritt2002Sci...297.1310M},
albeit often not observed in XRGs \citep[e.g.,][]{Gopal-Krishna2003ApJ...594L.103G,Patra2023MNRAS.524.3270P}. \\

{\bf J102544+345051 (B3 1023+394) `The rectangle':}
The LoTSS map of this source exhibits a broadly rectangular shape. However, a closer inspection reveals a quadripolar structure with the two diagonal axes misaligned by $\sim$ 70 degrees (Fig. \ref{fig:J102544}a; c). Hence, it is classifiable as an XRG, with the east-west axis being the primary axis, as evident from the VLASS map (Fig. \ref{fig:J102544}). Note that this map also shows a bright inner double straddling the optical host galaxy ({\it z}(spec) = 0.14721, see Fig. \ref{fig:J102544}d), whose orientation further confirms that the primary axis has continued to be in east-west direction. An early example of `rectangle' shaped appearance is the radio galaxy B0755+379 \citep{Vigotti1989AJ.....98..419V}. \\

\begin{SCfigure*}
    \includegraphics[width=13cm, height=5.5cm]{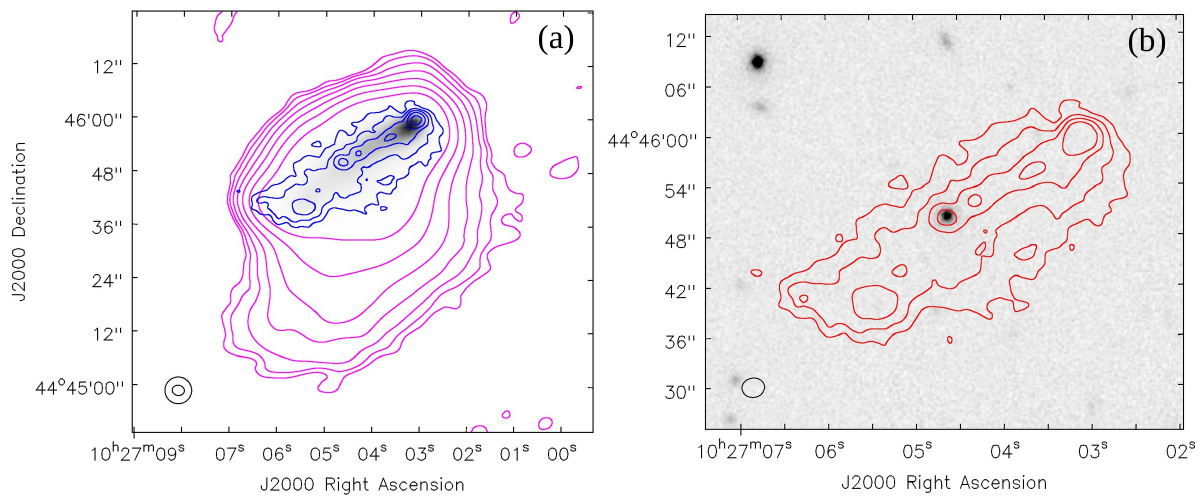}
  \caption{J102704+444547: (a) VLASS contours (blue) overlaid on the LoTSS-DR2 contours (magenta) \& grey-scale; (b) VLASS contour plot (orange) overlaid on the zoomed  optical (PanSTARR, I-band) image}
  \label{fig:J102704}
\end{SCfigure*}

{\bf J102625+390929:}
This WAT associated with a compact group of galaxies at {\it z} = 0.14721 \citep{Best2012MNRAS.421.1569B} shows a smoothly bending kpc-scale jet, tracing a nearly perfect parabolic trajectory. However, the bending is very asymmetric, as it affects only the south-western jet. As a result, the apex of the jet profile is far removed from the host galaxy (Fig. \ref{fig:J102625}b). Such an unusually asymmetric bending of the two jets is possible when the direction of the galaxy’s motion, which is opposite to the direction of the ram-pressure acting on the jet pair, is nearly parallel to the inner jet axis \citep[e.g,][]{Bempong-Manfu2020MNRAS.496..676B}. The straightness of the structure comprising the inner segments of both jets and its alignment with the north-eastern tail, clearly suggests that the ram pressure comes from the south-western direction and it has caused the eventual bending of the south-western jet. This scenario is also broadly consistent with the bending profile of the radio tail of the bright NAT radio galaxy seen to the southwest of the WAT (Fig. \ref{fig:J102625}a), assuming that both sources reside within the same intra-group medium. Curiously, despite the putative much stronger sideway ram-pressure acting on it,  the southwestern radio tail of the WAT is much fainter than the opposite radio tail. Moreover, the same sense of brightness asymmetry is exhibited by the regions where the two jets flare-up (Fig. \ref{fig:J102625}).
\begin{SCfigure*}
    \includegraphics[width=13cm, height=11.5cm]{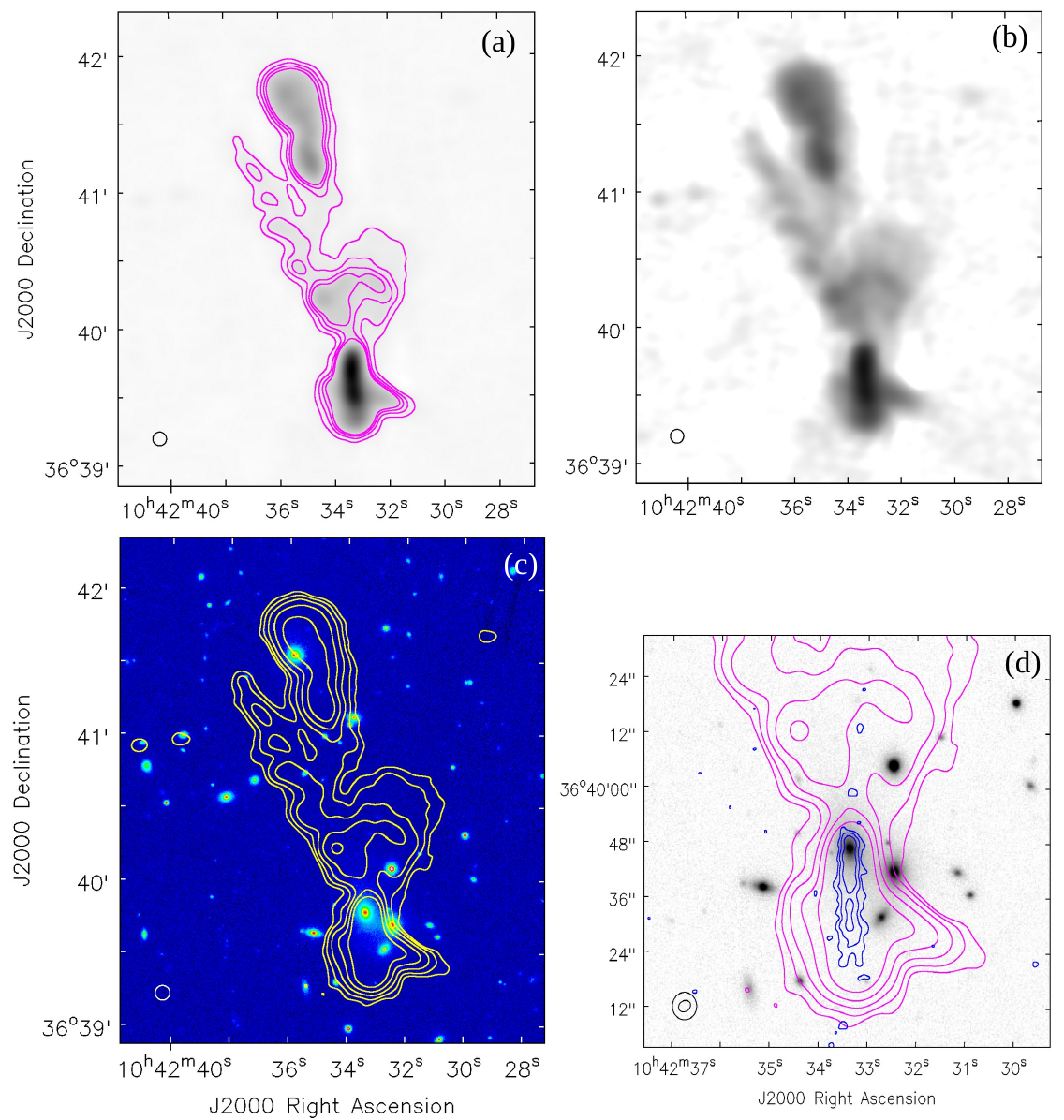}
  \caption{J104234+364040:(a) LoTSS-DR2 contour plot (magenta) \& grey-scale; (b) (a) LoTSS-DR2 grey-scale; (c) LoTSS-DR2 contour plot (yellow) overlaid on the optical (PanSTARR, I-band) image; (d) LoTSS-DR2 contour plot (magenta) overlaid on the zoomed optical (PanSTARR, I-band) image.}
  \label{fig:J104234}
\end{SCfigure*}
\begin{SCfigure*}
    \includegraphics[width=13cm, height=11.5cm]{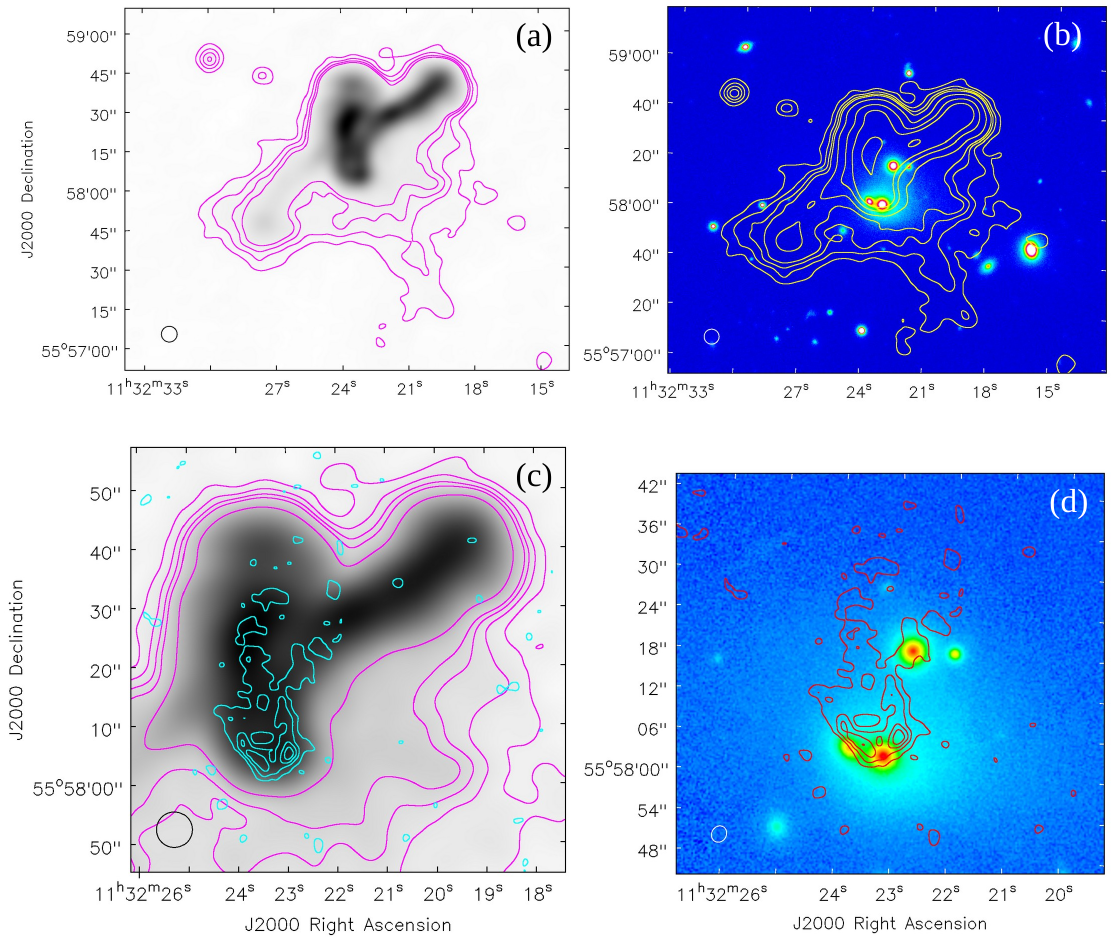}
  \caption{J113222+555823: (a) LoTSS-DR2 contour plot (magenta) \& grey-scale; (b) LoTSS-DR2 contour plot (yellow) overlaid on the optical (PanSTARR, I-band) image; (c) VLASS contours (cyan) overlaid on the zoomed LoTSS-DR2 contours (magenta) \& grey; (d)  VLASS contours (red) overlaid on the zoomed optical (PanSTARR, I-band) image. }
  \label{fig:J113222}
\end{SCfigure*}

We recall that a similar, albeit less spectacular asymmetric jet bending is present in the WAT B3 0247+487 \citep{Vigotti1989AJ.....98..419V}. \\

{\bf J102704+444547:}
This medium-distant radio galaxy [{\it z}(spec) = $0.58605\pm 0.00007$] has been classified as a probable XRG \citep{Yang2019ApJS..245...17Y}.
We find its LoTSS-DR2 map to be fully consistent with FR II classification. It shows a prominent jet to the north-west and a strong radio bridge between the two hot spots. Remarkably, the bridge is highly asymmetric about the radio axis defined by the jets (Fig. \ref{fig:J102704}). Such extreme asymmetry of radio bridge is rare and indicates a large-scale cross-wind. Two previously reported such cases are 3C 430 \citep{Spangler1984AJ.....89.1478S} and B0034+25 \citep{deRuiter1986A&AS...65..111D}. In all three sources, the core is well aligned with the two terminal hot spots, signifying that the twin-jets have remained unbent. \\ 

{\bf J104234+364040:}
This source is a member of the BCG sample of \citet{Furnell2018MNRAS.478.4952F}. Although, at the first glance, it seems to be a regular FR II source comprised of two lobes connected by a bridge of radio emission, a closer scrutiny unveils an exceptional structural complexity (Fig. \ref{fig:J104234}). Firstly, there is no candidate for the host galaxy anywhere near the symmetry-point between the two lobes (this holds even when we use DECaLS image). The only object in the proximity of the radio axis is the bright ${\it z}$ = 0.14133 galaxy located near the northern edge of the southern lobe. As seen from the VLASS map (Fig. \ref{fig:J104234}d), this galaxy has a radio core and a $\sim$ $30''$ (75 kpc) long narrow radio component resembling a one-sided jet (or tail) emanating from it towards the south, i.e., directed opposite to the northern radio lobe! There is no trace of a counter-jet which could possibly be feeding the northern lobe and hence the channel of energy supply to that lobe remains obscure. Moreover, that lobe is itself resolved into two components, but a potential optical host galaxy (situated 10 42 35.86 +36 41 32.6, {\it z} = 0.14133) is undetected in the VLASS map (Fig. \ref{fig:J104234}c). Also, it has a rather large offset from the radio axis defined by the two radio components. 

\begin{figure*}
\begin{center}
\end{center}
    \includegraphics[width=15cm, height=8cm]{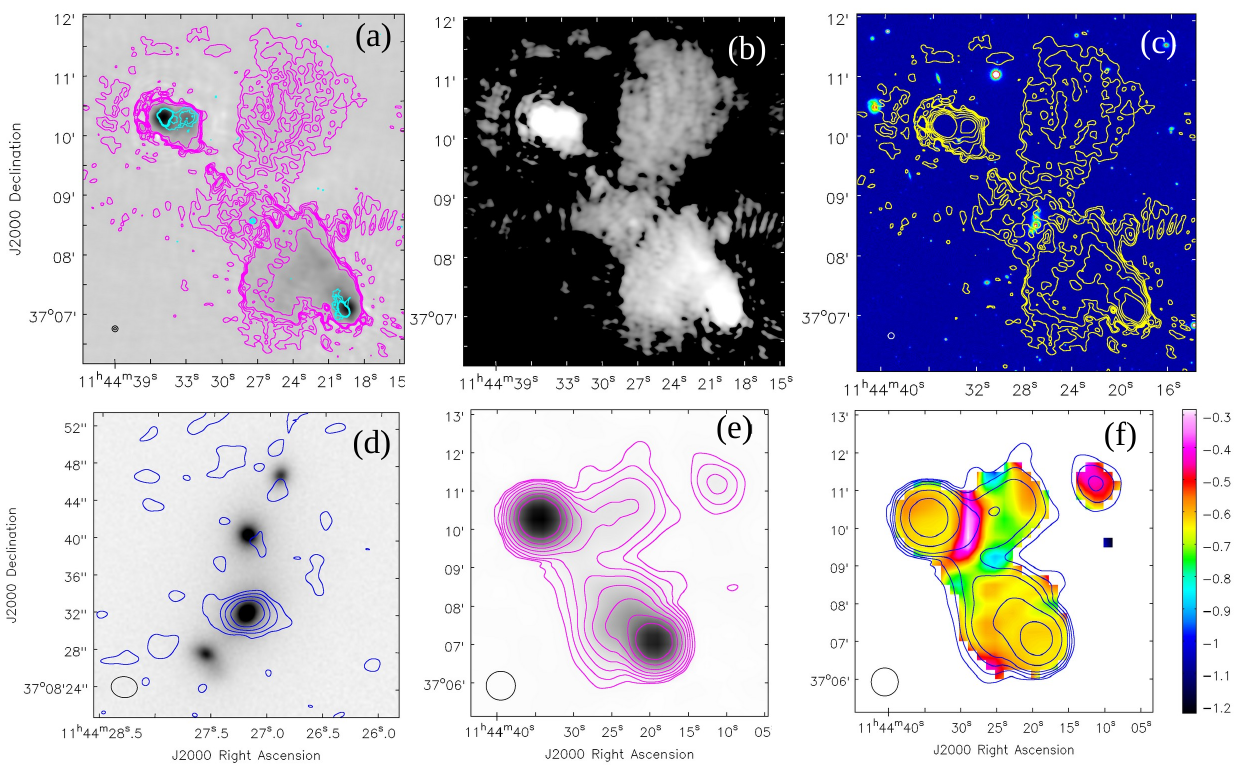}
  \caption{J114429+370915: (a) VLASS contours (cyan) overlaid on the LoTSS-DR2 contours (magenta) \& grey-scale; (b) LoTSS-DR2 image; (c) LoTSS-DR2 contour plot (yellow) overlaid on the optical (PanSTARR, I-band) image; (d) VLASS contours (blue) overlaid on the zoomed optical (PanSTARR, I-band) image; (e) NVSS contours (magenta) \& grey-scale; (f) Spectral index map generated by combining the 1400 GHz NVSS map with the 144 MHz LoTSS-DR2 map, overlaid with NVSS contour (blue).}
  \label{fig:J114429}
\end{figure*}

Even more puzzling is the structure of the lower surface-brightness radio emission observed between the two lobes 
(radio bridge?). This emission is roughly fan-shaped, with the apex near the radio-detected galaxy identified above with the southern lobe. Curiously, the fan is bounded on its eastern side by a highly unusual $\sim$ $100''$ (252 kpc) long radio structure which resembles a helical antenna, or a Christmas candle, emerging out of the radio detected galaxy identified with the southern lobe. All these puzzling features warrant deeper radio imaging of this exceptionally curious radio source. \\
\begin{table*}
\caption{\bf The contour levels of the radio maps shown in Figures 1 - 25.}
\centering
\begin{tabular}{cccl}
\hline\hline
Name          & Panel    &    Map     &  Contour levels (Jy/beam) \\
              &          &            &                           \\  
\hline
J011208+414702&	a        & LoTSS &  0.0023 $\times$ (0.2, 0.4, 0.8, 1.6, 3.2, 6.4, 12.8)   \\
              & b        & LoTSS &  0.0023 $\times$ (0.2, 0.4, 0.8, 1.6, 3.2, 6.4, 12.8)   \\
              & b        & VLASS &  0.0060 $\times$ (0.2, 0.4, 0.8, 1.6, 3.2)   \\
              & c        & VLASS &  0.0037 $\times$ (0.2, 0.4, 0.6, 0.8, 1.6, 3.2, 6.4)   \\
              & d        & VLASS &  0.0030 $\times$ (0.2, 0.4, 0.6, 0.8, 1.6, 3.2)   \\

J022830+382108&a         & LoTSS & 0.0032 $\times$ (0.2, 0.4, 0.8, 1.6, 3.2, 6.4) \\
              &b         & LoTSS & 0.0021 $\times$ (0.2, 0.4, 0.8, 1.2, 3.2) \\
              &c         & LoTSS & 0.0020 $\times$ (0.2, 0.4, 0.8, 1.6, 3.2, 6.4) \\
              &c         & VLASS & 0.0021 $\times$ (0.2, 0.4, 1.2, 3.2) \\
              &d         & LoTSS & 0.0020 $\times$ (0.2, 0.4, 0.8, 3.2) \\
J083106+414741&a         & LoTSS & 0.0037 $\times$ (0.2, 0.4, 0.6,0.8) \\
              &a         & VLASS & 0.0018 $\times$ (0.2, 0.4, 0.8, 1.6) \\
              &c         & LoTSS & 0.0116 $\times$ (0.2, 0.4, 0.6, 0.8, 1.6) \\
J091900+315254&a         & LoTSS & 0.0015 $\times$ (0.2, 0.4, 0.8, 1.6, 3.2) \\
              &c         & LoTSS & 0.0021 $\times$ (0.2, 0.4, 0.8, 1.6)\\
              &c         & VLASS & 0.0026 $\times$ (0.2, 0.4, 1.8) \\
              &d         & LoTSS & 0.0027 $\times$ (0.2, 0.4, 0.8, 1.6) \\
J092229+545943&a         & LoTSS & 0.0014 $\times$ (0.2, 0.4, 0.8) \\
              &b         & LoTSS & 0.0017 $\times$ (0.2, 0.4, 0.8, 1.6, 3.2) \\
              &c         & LoTSS & 0.0020 $\times$ (0.2, 0.4, 0.8, 1.6, 3.2, 6.4,12.8)\\
              &d         & LoTSS & 0.0017 $\times$ (0.2, 0.4, 0.8, 1.6, 3.2) \\
              &d         & VLASS & 0.0070 $\times$ (0.2, 0.4, 0.6, 0.8) \\
J093032+311215&a         & LoTSS & 0.0030 $\times$ (0.2, 0.4, 0.6, 0.8,1.6, 3.2, 6.4, 12.8, 25.6) \\
              &b         & VLASS & 0.0029 $\times$ (0.2, 0.4, 0.8, 1.6) \\
              &c         & VLASS & 0.0018 $\times$ (0.2, 0.4, 0.6, 0.8, 1.0) \\
              &d         & VLASS & 0.0013 $\times$ (0.2, 0.4, 0.5, 0.6, 0.8, 1.0) \\
J094953+445658&a         & LoTSS & 0.0028 $\times$ (0.2, 0.4, 0.6, 0.8, 1.6, 3.2, 4.8, 6.4) \\
              &b         & FIRST & 0.0019 $\times$ (0.2, 0.4, 0.6, 0.8, 1.6, 3.2) \\
              &c         & FIRST & 0.0026 $\times$ (0.2, 0.4, 0.6, 0.8, 1.6, 3.2)  \\
              &d         & VLASS & 0.0014 $\times$ (0.2, 0.4, 0.6,  0.8)\\
J101522+682351&a         & LoTSS & 0.0025 $\times$ (0.2, 0.4, 0.8, 1.6) \\ 
              &b         & LoTSS & 0.0059 $\times$ (0.2, 0.4, 0.8, 1.6, 3.2) \\
              &c         & LoTSS & 0.0025 $\times$ (0.2, 0.4, 0.8) \\
              &c         & VLASS & 0.0026 $\times$ (0.2, 0.4, 0.6, 0.8, 1.6)\\
              &d         & VLASS & 0.0025 $\times$ (0.2, 0.4, 0.8, 1.6, 3.2) \\
J102544+345051&a         & LoTSS & 0.0048 $\times$ (0.2, 0.4, 0.6, 0.8, 1.6) \\
              &b         & LoTSS & 0.0048 $\times$ (0.2, 0.4, 0.6, 0.8, 1.6) \\
              &b         & VLASS & 0.0018 $\times$ (0.2, 0.4, 0.8, 1.6, 3.2) \\
              &c         & LoTSS & 0.0015 $\times$ (0.2, 0.8, 3.2, 6.4, 12.8, 25.6) \\
              &d         & VLASS & 0.0016 $\times$ (0.2, 0.4, 0.8) \\  
J102625+390929&a         & LoTSS & 0.0016 $\times$ (0.2, 0.4, 0.8, 1.6, 3.2, 6.4, 25.6)\\ 
              &b         & LoTSS & 0.0012 $\times$ (0.2, 0.8, 3.2)\\
              &c         & LoTSS & 0.0016 $\times$ (0.2, 0.4, 0.8, 1.6, 3.2, 6.4, 25.6)\\
              &c         & VLASS & 0.0016 $\times$ (0.2, 0.4, 0.6, 0.8, 1.6, 3.2) \\
              &d         & VLASS & 0.0015 $\times$ (0.2, 0.4, 0.6, 0.8) \\
J102704+444547&a         & LoTSS & 0.0015 $\times$ (0.2, 0.4, 0.8, 1.6, 3.2, 6.4, 12.8) \\ 
              &a         & VLASS & 0.0019 $\times$ (0.2, 0.4, 0.8, 1.6, 3.2) \\
              &b         & VLASS & 0.0018 $\times$ (0.2, 0.4, 0.8, 1.6) \\
J104234+364040&a         & LoTSS & 0.0048 $\times$ (0.2, 0.4, 0.6, 0.8) \\ 
              &c         & LoTSS & 0.0023 $\times$ (0.2, 0.4, 0.8, 1.6,  3.2) \\
              &d         & LoTSS & 0.0023 $\times$ (0.2, 0.4, 0.8, 1.6,  3.2) \\
              &d         & VLASS & 0.0020 $\times$ (0.2, 0.4,  0.8) \\
\hline
\end{tabular}
\label{tab:contour}
\vspace{0.2cm}
\end{table*}

\begin{table*}
\caption*{\bf Table 2: continued}
\centering
\begin{tabular}{cccl}
\hline\hline
Name          & Panel    &  Map       &  Contour levels (Jy/beam) \\
              &          &            &                           \\  
\hline
J113222+555823&a         & LoTSS      & 0.0018 $\times$ (0.2, 0.4, 0.6, 0.8) \\ 
              &b         & LoTSS      & 0.0016 $\times$ (0.2, 0.4, 0.8, 1.6,  3.6, 6.4, 12.8) \\
              &c         & LoTSS      & 0.0017 $\times$ (0.2, 0.4, 0.6, 0.8, 1.6) \\
              &c         & VLASS      & 0.0012 $\times$ (0.2, 0.4, 0.6, 0.8, 1.6) \\
              &d         & VLASS      & 0.0012 $\times$ (0.2, 0.4, 0.6, 0.8) \\
J114429+370915&a         & LoTSS      & 0.0030 $\times$ (0.2, 0.4, 0.6, 0.8, 1.6) \\
              &a         & VLASS      & 0.0030 $\times$ (0.2, 0.4, 0.6, 0.8) \\
              &c         & LoTSS      & 0.0030 $\times$ (0.2, 0.4, 0.8, 1.6, 3.2, 6.4, 12.8) \\
              &d         & VLASS      & 0.0013 $\times$ (0.2, 0.4, 0.8, 1.6) \\
              &e         & NVSS       & 0.0100 $\times$ (0.2, 0.4, 0.8, 1.6, 3.2, 6.4, 12.8, 25.6) \\
              &f         & NVSS       & 0.0100 $\times$ (0.2, 0.4, 0.8, 1.6, 3.2, 6.4, 12.8, 25.6) \\
J114434+672424&a         & LoTSS      & 0.0030 $\times$ (0.2, 0.4, 0.8, 1.6) \\
              &b         & LoTSS      & 0.0040 $\times$ (0.2, 0.4, 0.8, 3.2, 12.5, 25.6,  51.2) \\
              &c         & LoTSS      & 0.0030 $\times$ (0.2, 0.4, 0.8, 1.6) \\
              &c         & VLASS      & 0.0027 $\times$ (0.2, 0.4, 0.8, 1.6, 3.2, 6.4, 12.8) \\
              &d         & VLASS      & 0.0022 $\times$ (0.2, 0.8, 3.2, 6.4, 12.8) \\
J115608+441058&a         & LoTSS      & 0.0031 $\times$ (0.2, 0.4, 0.8, 1.6)\\ 
              &a         & VLASS      & 0.0016 $\times$ (0.2, 0.4, 0.8, 1.6) \\
              &c         & LoTSS      & 0.0040 $\times$ (0.2, 0.4, 0.8, 1.6, 3.2, 6.4) \\
              &d         & VLASS      & 0.0010 $\times$ (0.2, 0.4, 0.6, 0.8, 1.6, 2.4) \\
J133116+441851&a         & LoTSS      & 0.0028 $\times$ (0.2, 0.4, 0.6, 0.8, 1.6, 3.2, 6.4,12.8, 25.6)\\ 
              &b         & LoTSS      & 0.0028 $\times$ (0.2, 0.4, 0.6, 0.8, 1.6, 3.2, 6.4,12.8, 25.6)\\
              &c         & FIRST      & 0.0019 $\times$ (0.2, 0.4, 0.6, 0.8, 1.6, 3.2, 6.4)\\
              &d         & LoTSS      & 0.0039 $\times$ (0.2, 0.4, 0.8, 1.6, 3.2, 6.4, 12.8) \\
              &d         & VLASS      & 0.0030 $\times$ (0.2, 0.4, 0.8, 1.6, 3.2)\\
J135141+555938&a         & LoTSS      & 0.0015 $\times$ (0.2, 0.4, 0.8, 1.6, 3.2, 6.4, 12.8, 25.6)\\ 
              &a         & VLASS      & 0.0022 $\times$ (0.2, 0.4, 0.6, 0.8,  1.2, 1.6) \\
              &c         & LoTSS      & 0.0208 $\times$ (0.2, 0.4, 0.6, 0.8, 1.6)\\
              &d         & VLASS      & 0.0024 $\times$ (0.2, 0.4, 0.8, 1.4) \\
J140311+382756&a         & LoTSS      & 0.0390 $\times$ (0.2, 0.4, 0.8, 1.6, 3.2, 6.4, 12.8, 25.6) \\ 
              &a         & VLASS      & 0.0050 $\times$ (0.2, 0.4, 0.6, 0.8) \\
              &b         & VLASS      & 0.0029 $\times$ (0.2, 0.4, 0.6, 0.8, 1.6, 3.2, 6.4) \\
              &c         & LoTSS      & 0.0380 $\times$ (0.2, 0.4, 0.8, 1.6) \\
J143527+550756&a         & LoTSS      & 0.0250 $\times$ (0.2, 0.4, 0.8, 1.6, 3.2, 6.4, 12.8) \\ 
              &a         & VLASS      & 0.0020 $\times$ (0.2, 0.4, 0.8, 1.6, 3.2, 6.4) \\
              &c         & LoTSS      & 0.0140 $\times$ (0.2, 0.4, 0.8, 1.6, 3.2, 6.4, 18.8) \\
              &d         & LoTSS      & 0.0025 $\times$ (0.2, 0.4, 0.8, 1.6, 3.2, 6.4) \\
J165732+431941&a         & LoTSS      & 0.0023 $\times$ (0.2, 0.4, 0.8, 1.6, 3.2, 6.4, 12.8) \\
              &c         & LoTSS      & 0.0030 $\times$ (0.2, 0.4, 0.6, 0.8, 1.6)\\
              &c         & VLASS      & 0.0016 $\times$ (0.2, 0.4, 0.8, 1.6) \\
              &d         & LoTSS      & 0.0021 $\times$ (0.2, 0.4, 0.8) \\
J170940+342537&a         & LoTSS      & 0.0017 $\times$ (0.2, 0.4, 0.8, 1.6, 3.2, 6.4, 12.8) \\ 
              &b         & LoTSS      & 0.0017 $\times$ (0.2, 0.4, 0.8, 1.6, 3.2, 6.4, 12.8, 25.6) \\
              &c         & LoTSS      & 0.0017 $\times$ (0.2, 0.4, 0.8, 1.6, 3.2) \\
              &c         & VLASS      & 0.0040 $\times$ (0.2, 0.4, 0.8, 1.6, 3.2) \\
              &d         & VLASS      & 0.0024 $\times$ (0.2, 0.8, 1.6, 3.2) \\
J221255+220523&a         & LoTSS      & 0.0043 $\times$ (0.2, 0.4, 0.8, 1.6, 3.2, 6.4) \\
              &b         & VLASS      & 0.0025 $\times$ (0.2, 0.4, 0.6, 0.8, 1.6) \\
J221558+290921&a         & LoTSS      & 0.0041 $\times$ (0.2, 0.4, 0.8, 1.6, 3.2, 6.4) \\
              &b         & LoTSS      & 0.0022 $\times$ (0.2, 0.8, 3.2, 6.4, 12.8) \\
              &c         & VLASS      & 0.0020 $\times$ (0.2, 0.4, 0.8, 1.6, 3.2, 6.4) \\
              &d         & VLASS      & 0.0020 $\times$ (0.2, 0.4, 0.8, 1.6, 3.2, 6.4) \\

\hline
\end{tabular}
\vspace{0.2cm}
\end{table*}

\begin{table*}
\caption*{\bf Table 2: continued}
\centering
\begin{tabular}{cccl}
\hline\hline
Name          & Panel    &  Map       &  Contour levels (Jy/beam) \\
              &          &            &                           \\  
\hline
J235938+400644&a         & LoTSS      & 0.0031 $\times$ (0.3, 0.6, 1.2, 2.4) \\
              &a         & VLASS      & 0.0068 $\times$ (0.2, 0.4, 0.6, 0.8)\\
              &b         & LoTSS      & 0.0040 $\times$ (0.2, 0.4, 0.8, 1.6, 3.2, 6.4) \\
              &c         & VLASS      & 0.0028 $\times$ (0.2, 0.4, 0.6, 1.0, 1.6)\\  
\hline
\end{tabular}
\vspace{0.2cm}
\end{table*} 


{\bf J113222+555823:}
Based on its FIRST survey map at 1.4 GHz \citep{Becker-first1995ApJ...450..559B}, \citet{Proctor2011ApJS..194...31P} noted structural peculiarity which he dubbed `Pretzel-with-handle'. More recently, \citet{Bera2022ApJS..260....7B} included this source in their XRG sample, based on its LOFAR map and suggested that the primary lobes of the proposed XRG extend from north to south, whereas the wings lie along the orthogonal axis. If true, this would be the only known XRG with FR II wings. The optical field shows 3 galaxies (all having z $\sim$ 0.051) in the central part of the radio source, the southern two of which appear to blend. The third (northern) galaxy, which is centrally located (at 11h 32m 22.74s, +55d $58'$ $18.4''$), was identified by \citet{Bera2022ApJS..260....7B} as the host galaxy of the XRG proposed by them. 

A closer look at Fig. \ref{fig:J113222} puts a question mark on the above proposal. In particular, the 
high-resolution VLASS map reveals a more complex situation, which indicates a superposition of a NAT on a complex radio structure whose morphology remains unclear. As seen from the radio-optical overlay (Fig. \ref{fig:J113222}d), the proposed NAT can be identified with the southernmost (brightest) optical galaxy, with the twin radio tails sharply bent northward. The remainder of radio emission appears to be associated with the northernmost galaxy, as proposed by \citep{Bera2022ApJS..260....7B}.  Note that none of the 3 galaxies has a radio core detected in the VLASS (or the LoTSS-DR2) maps. \\

\begin{SCfigure*}
    \includegraphics[width=14cm, height=11.5cm]{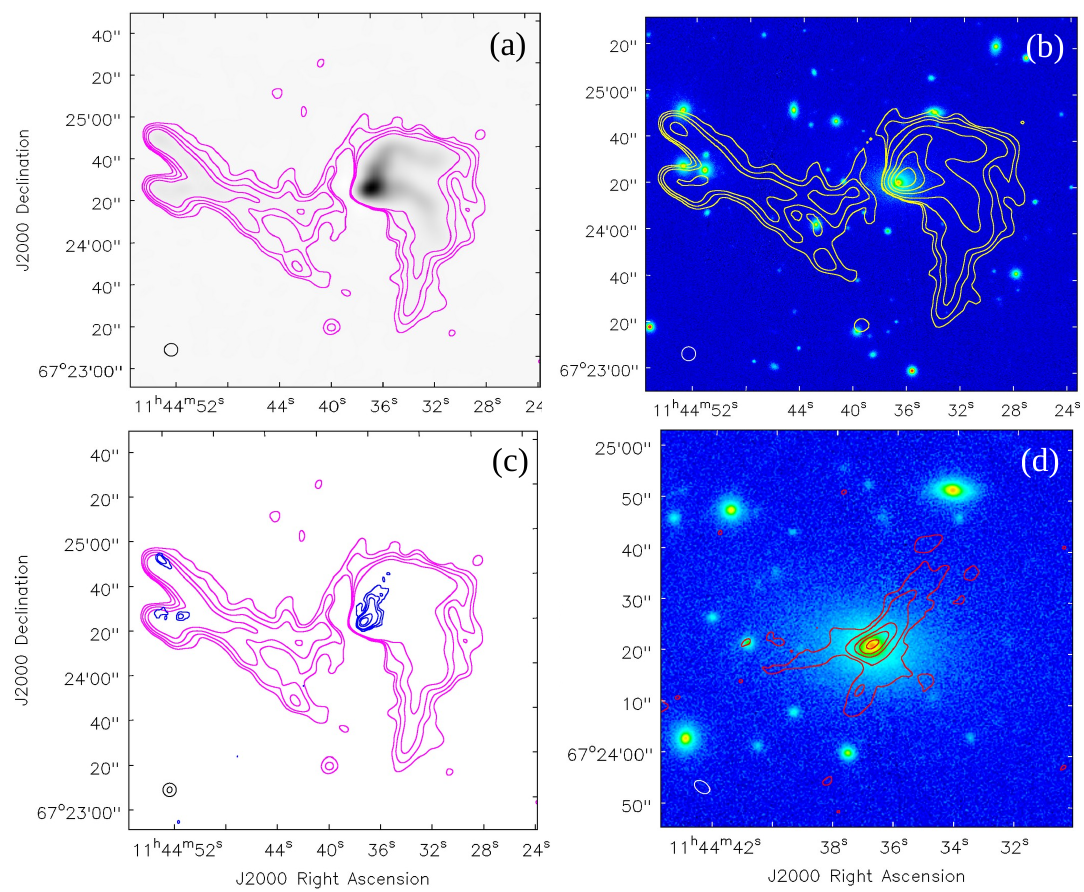}
  \caption{J114434+672424:(a) LoTSS-DR2 contour plot (magenta) \& grey-scale; (c) LoTSS-DR2 contour plot (yellow) overlaid on the optical (PanSTARR, I-band) image; (c) VLASS contours (blue) overlaid on the LoTSS-DR2 contours (magenta); (d) VLASS contours (red) overlaid on the zoomed optical (PanSTARR, I-band) image.}
  \label{fig:J114434}
\end{SCfigure*}
\begin{SCfigure*}
    \includegraphics[width=14cm, height=11.5cm]{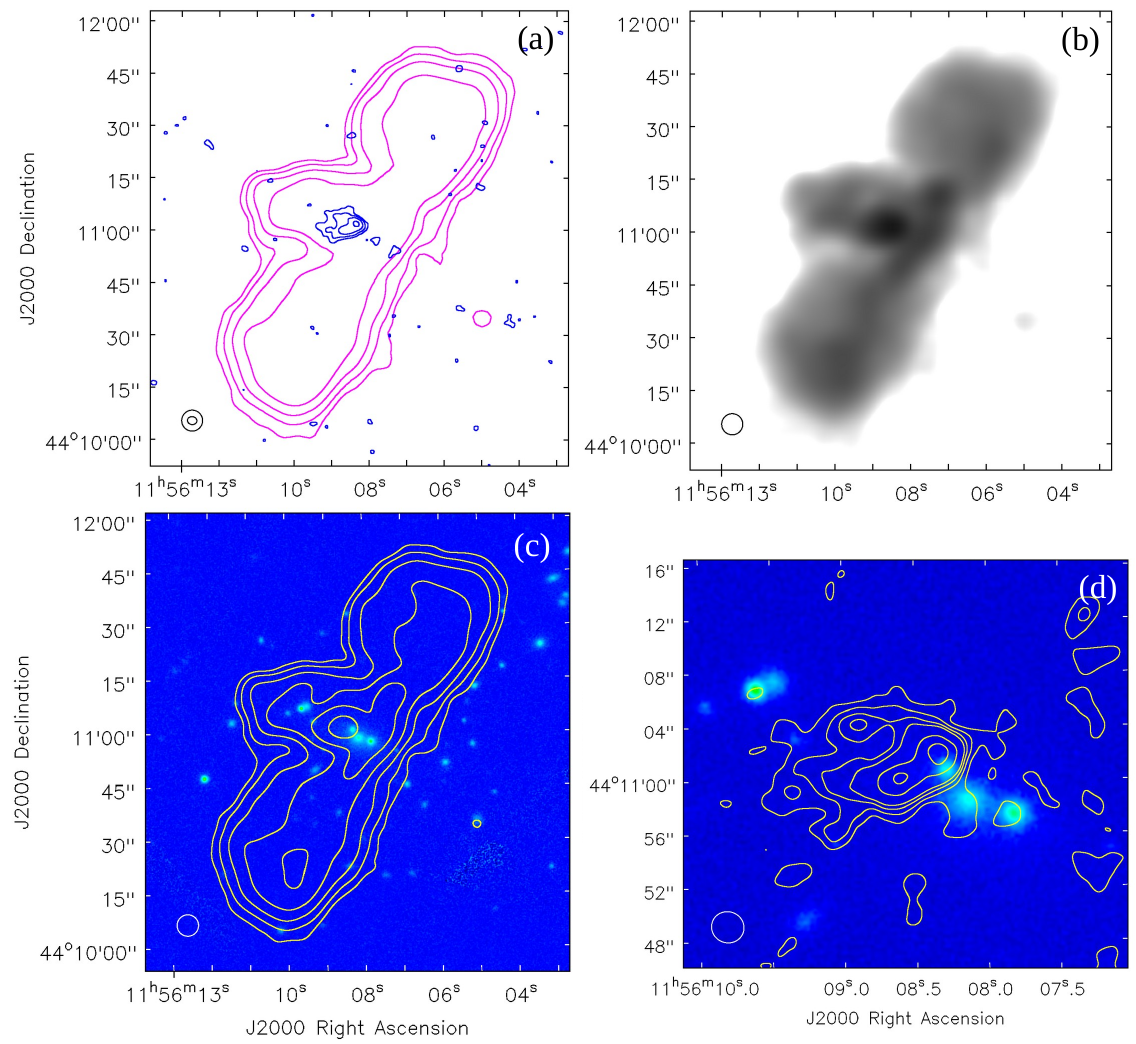}
  \caption{J115608+441058: (a) VLASS contours (blue) overlaid on the LoTSS-DR2 contours (magenta);  (b) LoTSS-DR2 grey-scale image; (c) LoTSS-DR2 contour plot (yellow) overlaid on the optical (PanSTARR, I-band) image; (d) VLASS contour plot (yellow) overlaid on the zoomed optical (PanSTARR, I-band) image.}
  \label{fig:J115608}
\end{SCfigure*}

{\bf J114429+370915 (4C+37.32, B2 1141+37):}
This source has a long history of observations. \citet{Ulrich1980Natur.288..459U} found it to be a GRG with 1 Mpc size, located at {\it z} = 0.115. In their VLA observations, \citet{Machalski1982AJ.....87.1150M} found it to have a simple double-lobed structure, $\sim 267''$  (560 kpc) in size and associated with a 16.5-mag elliptical which is the brightest in a chain of 4 galaxies and is centrally located between the two terminal radio hot spots
(Fig. \ref{fig:J114429}). They also suspected the presence of a very steep spectrum radio component, to be discovered at low frequencies. The source is also a member of the XRG catalogue of \citet{Cheung2007AJ....133.2097C} \citep[see also,][]{Bera2022ApJS..260....7B}. Its host galaxy is found to be a low-excitation object at {\it z} = 0.114804, with position 11h 44m 27.20s, +37d $08'$ $32.4''$ \citep{Koziel-Wierzbowska2020ApJS..247...53K} and has a radio counterpart in the VLASS map (Fig. \ref{fig:J114429}d) 

The LoTSS-DR2 map (Fig. \ref{fig:J114429}) shows the south-western lobe to be much broader than the north-eastern lobe, and rather sharply bounded on the inner side facing the host galaxy.  
A spectacular new feature revealed by the LoTSS-DR2 map in Fig. \ref{fig:J114429}b is a huge cone of radio emission emerging from the galaxy chain along its axis and pointing towards the northwest, roughly orthogonal to the radio axis. This fan-shaped patch of radio emission of size $180''$ (377 kpc) is essentially one-sided, {\bf as} no counterpart is seen on the south-eastern side. Could this {\it radio cone} be the putative very steep spectrum radio emission suspected by \citep{Machalski1982AJ.....87.1150M}(see above)? This is, however, not borne out by the (coarse-resolution) spectral map which we could make by combining the LoTSS-DR2 and NVSS maps (see Fig. \ref{fig:J114429}f). Moreover, this feature is discernible even in the 10.6 GHz pencil-beam map made at Effelsberg \citep{Mack1994A&AS..103..157M}. The origin of this huge one-sided cone of normal-spectrum radio emission remains to be understood.
Very recently, \citet{Kumari2024arXiv240614889K} have reported the case of an XRG, J0011+3217, which exhibits a 0.85 Mpc long one-sided diffuse wing. Since both its primary lobes show a conspicuous bend towards the direction of the wing, these authors have argued that the `missing' wing has been bent and got folded behind the detected wing, due to the same ram-pressure which has bent the primary lobes (the cause of the ram-pressure is the likely infall of the host galaxy towards the cluster Abell 7). However, this scenario can explain neither the conical shape of the wing in the present source J114429+370915, nor the wing's observed one-sidedness because the primary lobes are well-aligned and exhibit essentially no sign of bending and hence little evidence for a side-way ram pressure in action (Fig. \ref{fig:J114429}a).
\\

{\bf J114434+672424 (6C 114151+674026):}
This source is identified with a BCG at {\it z}(spec) = 0.11612 \citep{Ahn2012ApJS..203...21A}. Its LoTSS map is broadly consistent with an S-shaped morphology, with a clear VLASS detection of a radio core in the host galaxy (Fig. \ref{fig:J114434}). A morphological peculiarity revealed by the LoTSS-DR2 map is that the western lobe has a pair of well-resolved ridge lines (jets?) emerging from the host galaxy, both undergoing sharp bending. Likewise, the emission from the eastern lobe, after a sharp bend, is later seen to split into two collimated streams, each containing some compact radio structure visible in the VLASS map (Fig. \ref{fig:J114434}c). \\

{\bf J115608+441058:}
The basic radio axis of this double radio source seems to be defined by a ridge-line (twin-jet?) oriented at position angle (PA) $\sim$150 deg, in the middle of which lies a compact chain of 3 galaxies. These galaxies are aligned roughly orthogonal to the radio ridge and have spectroscopically measured redshifts of $\sim$0.298. Roughly along the galaxy chain lies a radio `spur' emanating from the central region of the double radio source. The optical-radio (VLASS) overlay shown in Fig. \ref{fig:J115608}d suggests that the `spur' may in fact be a NAT associated with the northernmost galaxy of the chain, which is also the least bright/massive of them and could well be moving under the gravitational influence of the other two galaxies. Based on the optical-radio (LoTSS-DR2) overlay in Fig. \ref{fig:J115608}c, the most likely host of the main (double) source is the westernmost member of the triple-galaxy chain. The alternatively possibility is the central galaxy of the chain, which is also the brightest of the three.\\

{\bf J133116+441851 (6C 132908+443418):}
As seen from the LoTSS-DR2 map (Fig.\ref{fig:J133116}a,b), this object is a very rare case of an early-type galaxy having two well developed `giant' spiral arms of synchrotron emission extending over $\sim$ $100''$. Spectroscopic redshift is not available for this `radio spiral'; but its photometric redshift of 0.248 $\pm$ 0.033 (SIMBAD) implies a huge projected size of $\sim$ 392 kpc. The inner radio structure is a FR II double radio source of size $\sim 17''$ (67 kpc), asymmetrically straddling the host elliptical and giving the radio source the appearance of a giant barred radio spiral.

\begin{SCfigure*}
    \includegraphics[width=14cm, height=11.5cm]{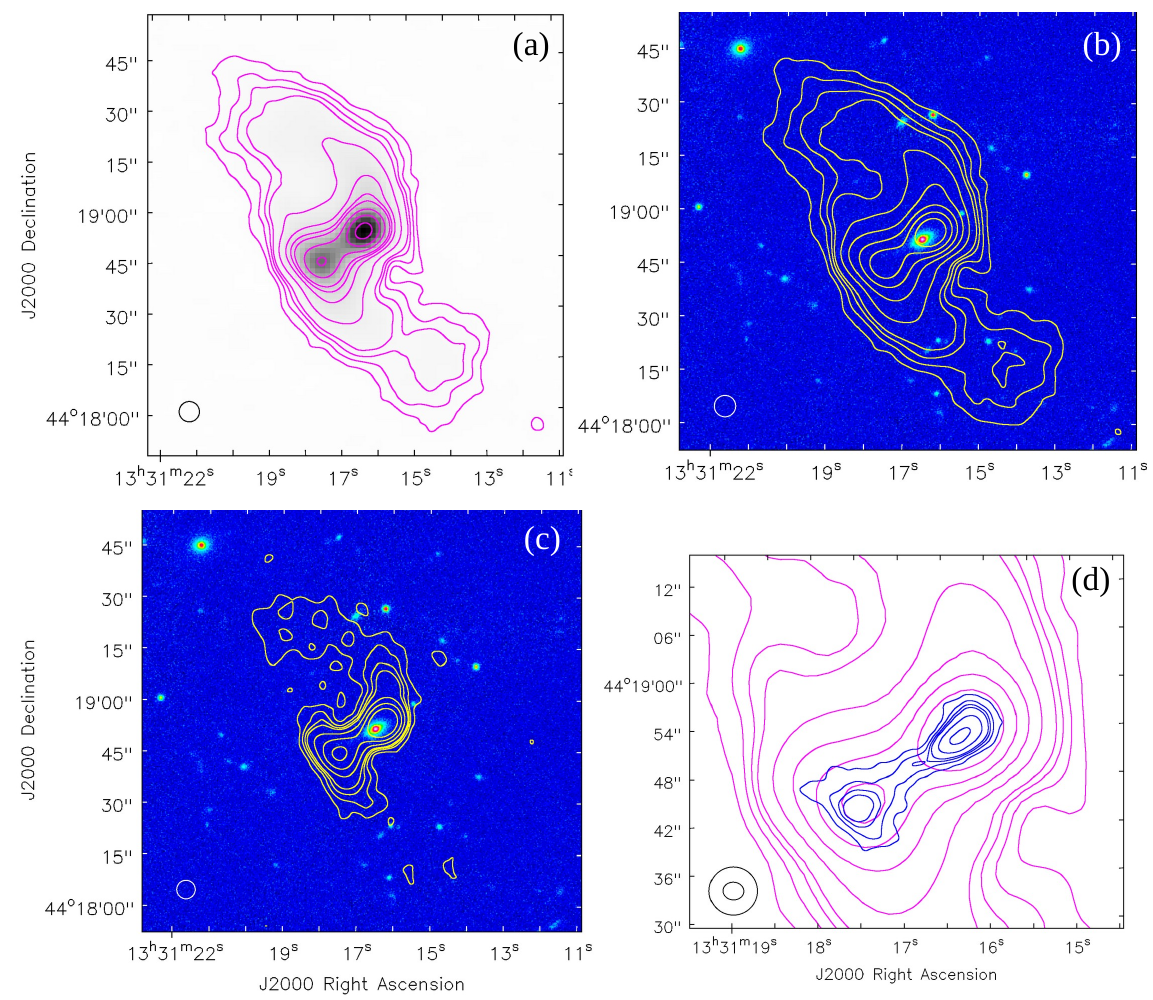}
  \caption{J133116+411851:(a) LoTSS-DR2 contour plot (magenta) \& grey-scale; (b)  LoTSS-DR2 contour plot (yellow) overlaid on the optical (PanSTARR, I-band) image; (c) VLASS contour plot (yellow) overlaid on the zoomed optical (PanSTARR, I-band) image; (d)  VLASS contour plot (blue) overlaid on the zoomed LoTSS-DR2 (magenta).}
  \label{fig:J133116}
\end{SCfigure*}

\begin{SCfigure*}
    \includegraphics[width=14cm, height=11.5cm]{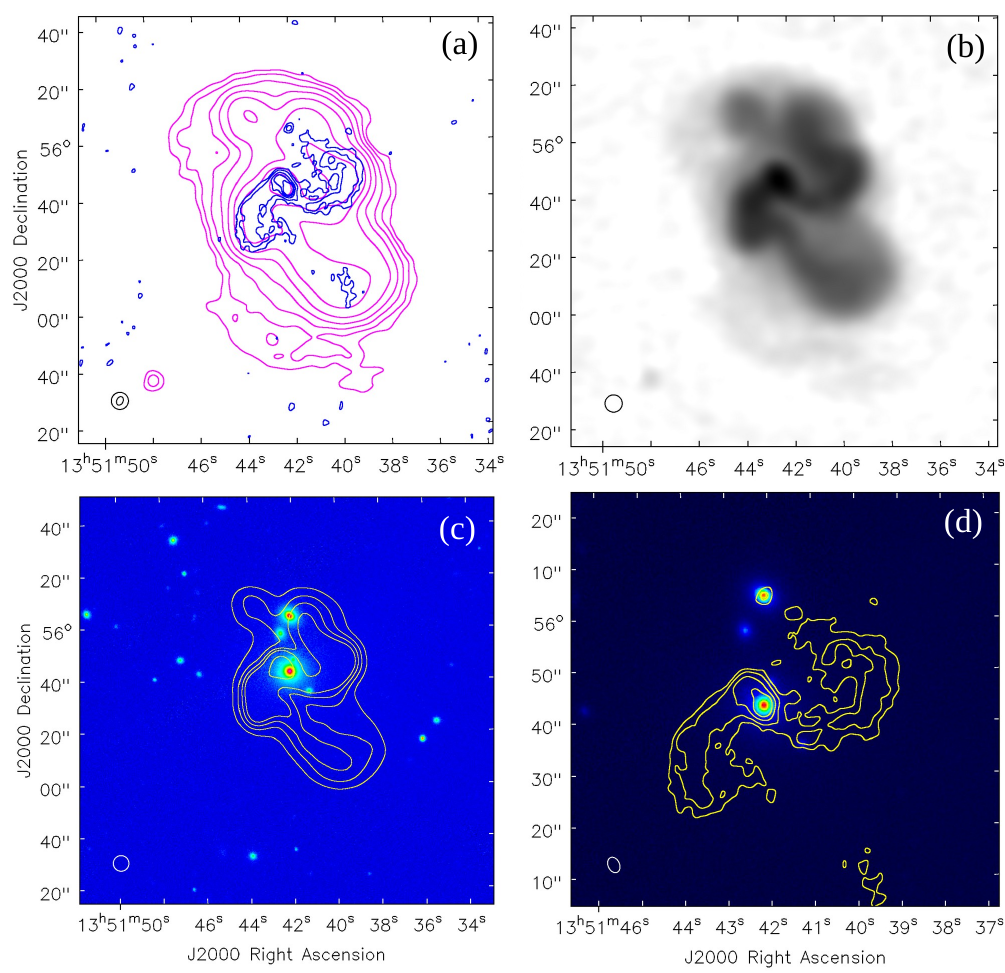}
  \caption{J135141+555938: (a) VLASS contour plot (blue) overlaid on the LoTSS-DR2 (magenta); (b) LoTSS-DR2 grey-scale image; (c) LoTSS-DR2 contour plot (yellow) overlaid on the optical (PanSTARR, I-band) image; (d) VLASS contour plot (yellow) overlaid on the optical (PanSTARR, I-band) image.}
  \label{fig:J135141}
\end{SCfigure*}

\begin{SCfigure*}
    \includegraphics[width=14cm, height=11.5cm]{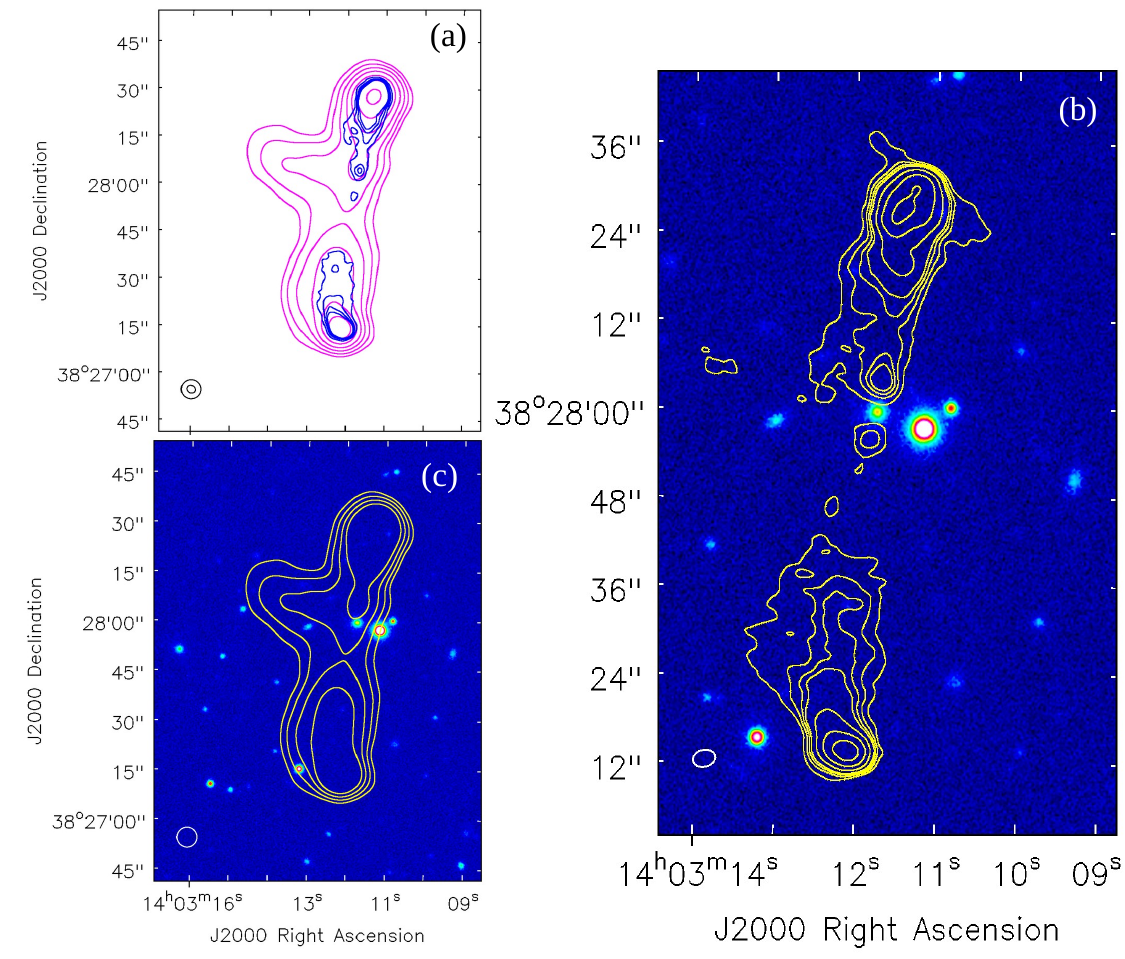}
  \caption{J140311+382756: (a) VLASS contour plot (blue) overlaid on the LoTSS-DR2 (magenta); (b) VLASS contour plot (yellow) overlaid on the optical (PanSTARR, I-band) image; (c)  LoTSS-DR2 contour plot (yellow) overlaid on the optical (PanSTARR, I-band) image.}
  \label{fig:J140311}
\end{SCfigure*}

\begin{SCfigure*}
    \includegraphics[width=14cm, height=11.5cm]{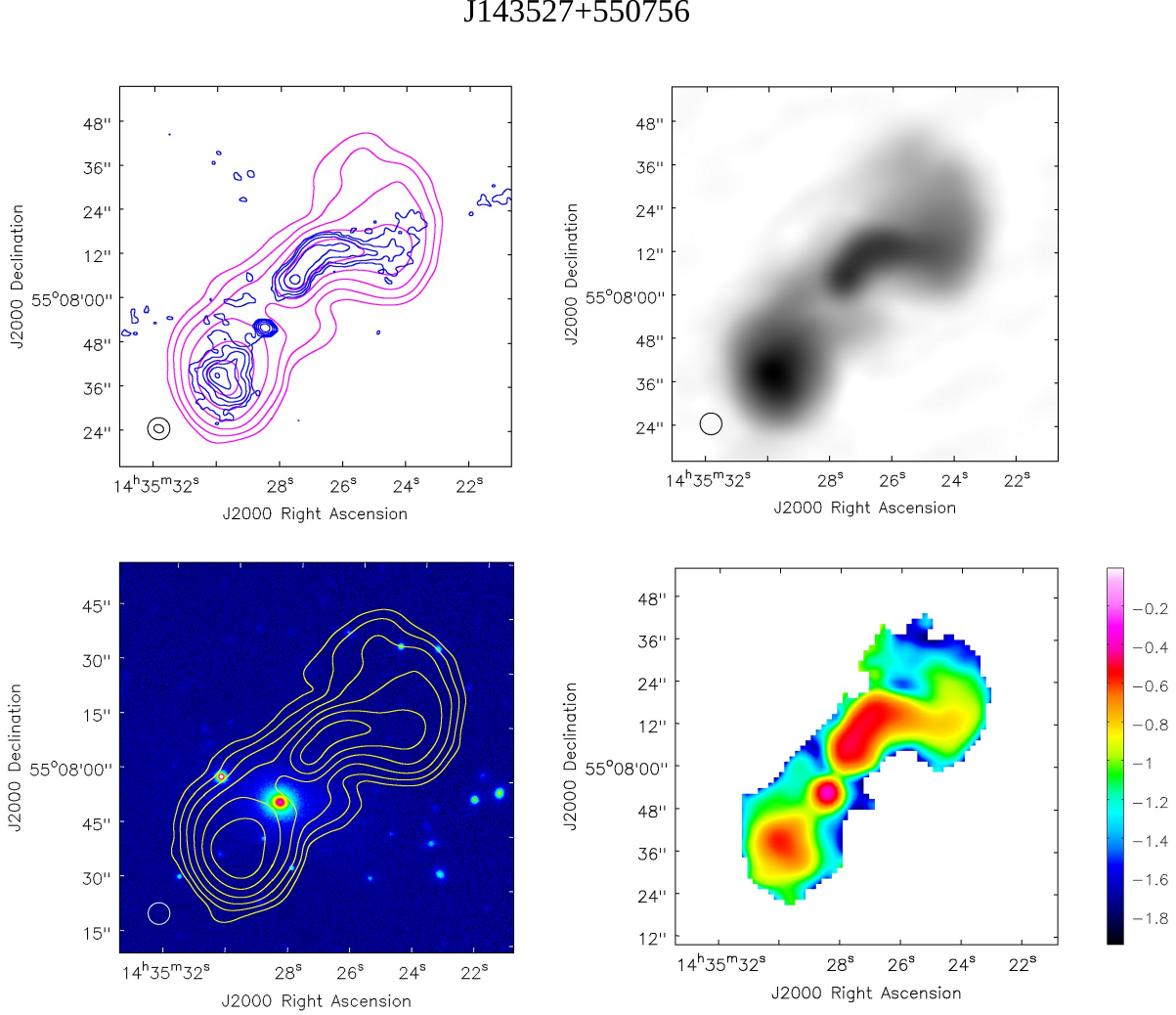}
  \caption{J143527+550756:(a) VLASS contour plot (blue) overlaid on the LoTSS-DR2 (magenta); (b)   LoTSS-DR2 grey-scale; (c) LoTSS-DR2 contour plot (yellow) overlaid on the optical (PanSTARR, I-band) image; (d) $\alpha$ (144 MHz - 1.4 GHz) map (see text).}
  \label{fig:J143527}
\end{SCfigure*}

The inner radio structure, well resolved in the VLASS map, consists of an asymmetric pair of radio lobes extending over $\sim$ $18''$, which straddle the bright elliptical galaxy and bend sharply to join the two giant radio spiral arms (Fig. \ref{fig:J133116} c). This spectacular morphology is probably best explained in terms of a steady clock-wise rotation of the axis of the central engine, a phenomenon previously discussed in the context of, so called `inversion-symmetric', radio galaxies and quasars 
\citep[e.g.,][]{Ekers1976Natur.262..369E, Wills1978A&A....66L...1W, Miley1980ARA&A..18..165M, Wirth1982AJ.....87..602W, Gower1982ApJ...262..478G}.
An early example of radio spiral can be seen in the VLA observations of the Bologna source B1316+29 \citep{deRuiter1986A&AS...65..111D}.\\

\begin{SCfigure*}
    \includegraphics[width=14cm, height=11.5cm]{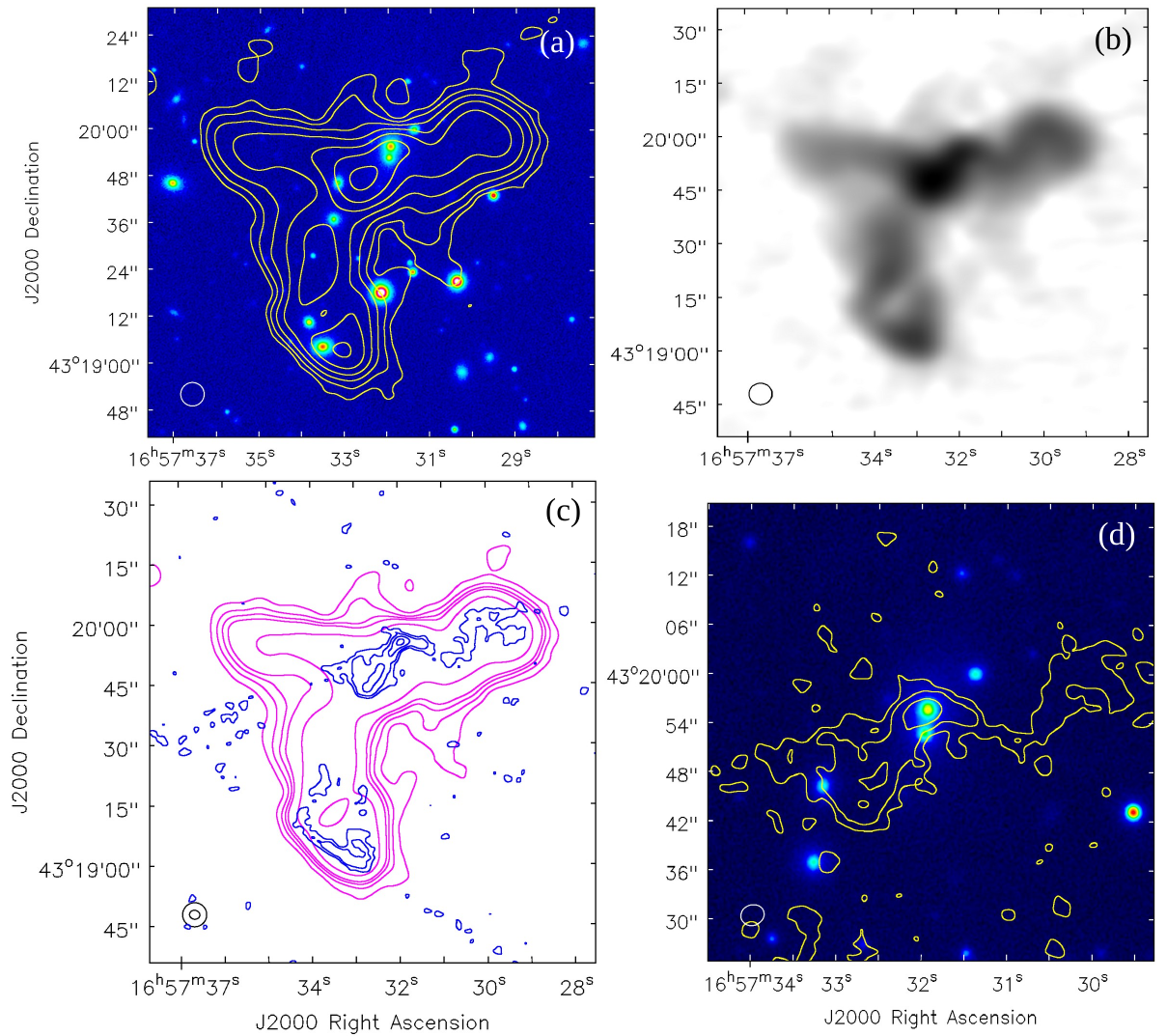}
  \caption{J165732+431941: (a) LoTSS-DR2 contour plot (yellow) overlaid on the optical (PanSTARR, I-band) image; (b) LoTSS-DR2 grey-scale; (c) VLASS contour plot (blue) overlaid on the LoTSS-DR2 (magenta); (d) VLASS contour plot (yellow) overlaid on the zoomed optical (PanSTARR, I-band) image.}
  \label{fig:J165732}
\end{SCfigure*}

\begin{SCfigure*}
    \includegraphics[width=14cm, height=11.5cm]{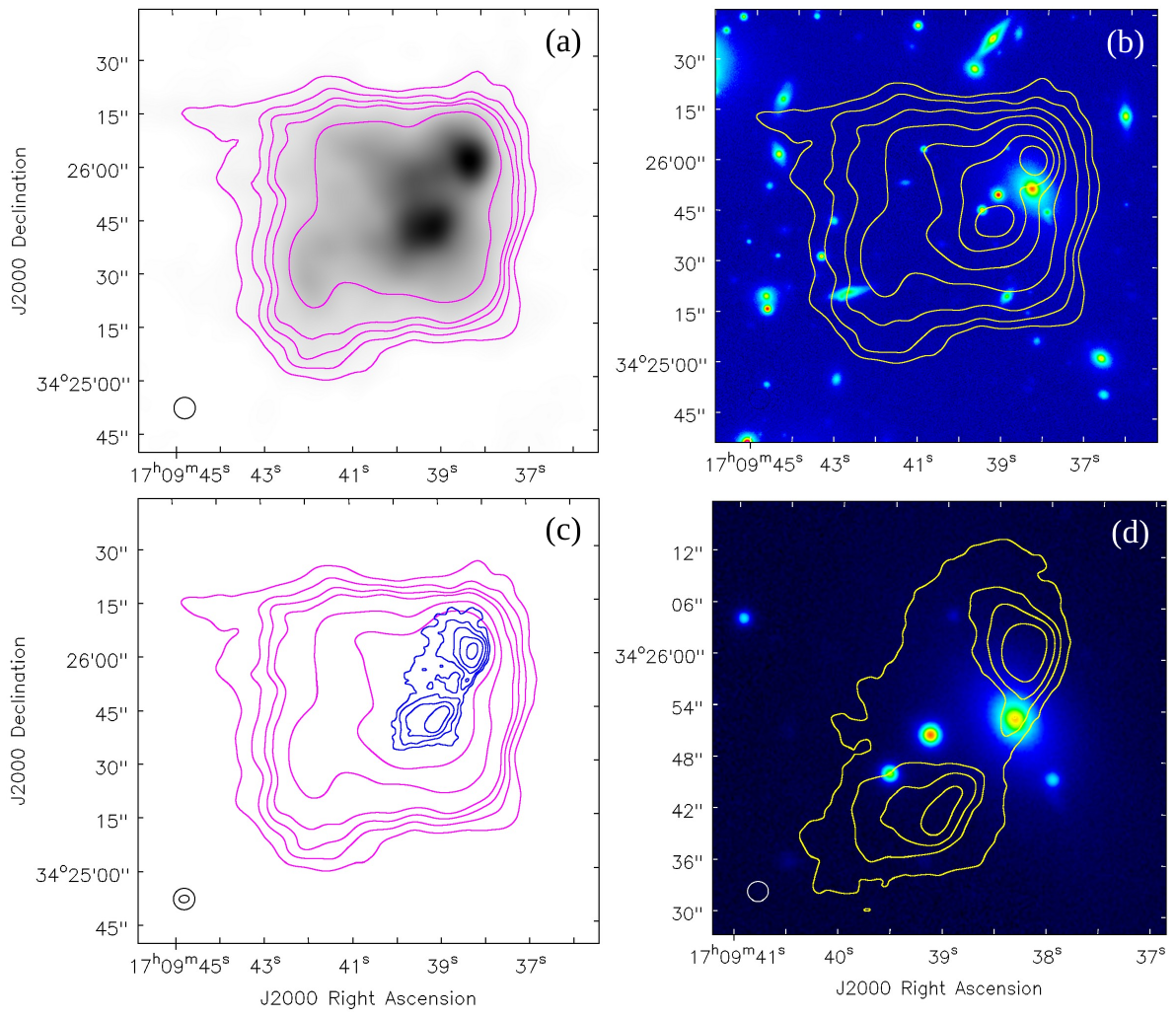}
  \caption{J170940+342537: (a) LoTSS-DR2 contour plot (magenta) \& grey-scale; (b) LoTSS-DR2 contour plot (yellow) overlaid on the optical (PanSTARR, I-band) image; (c) VLASS contour plot (blue) overlaid on the LoTSS-DR2 (magenta); (d) VLASS contour plot (yellow) overlaid on the zoomed optical (PanSTARR, I-band) image.}
  \label{fig:J170940}
\end{SCfigure*}

{\bf J135141+555938:}
This is another rare case of a spiral-shaped radio galaxy, with strongly twisted radio arms. The optical field shows a pair of bright galaxies at {\it z}(spec) $\sim$ 0.07, separated north-south by $\sim$ $21''$ (28 kpc) (Fig. \ref{fig:J135141} c,d). Each galaxy has a radio core detected in the VLASS map where the dominant southern galaxy can be identified as the host of the large spiral structure spanning $\sim$ $75''$ (100 kpc). The northern galaxy is projected on the upper radio spiral arm and is faintly detected in the VLASS map (Fig. \ref{fig:J135141} d). As in the case of the giant radio spiral J133116+441851 (see above), this radio galaxy too appears to manifest a clockwise rotation of the axis of the central engine. Based on VLA observations, 
\citet{Saripalli2018ApJ...852...48S} have classified this source as S-shaped FR I radio galaxy. \\

\begin{SCfigure*}
    \includegraphics[width=14cm, height=6.5cm]{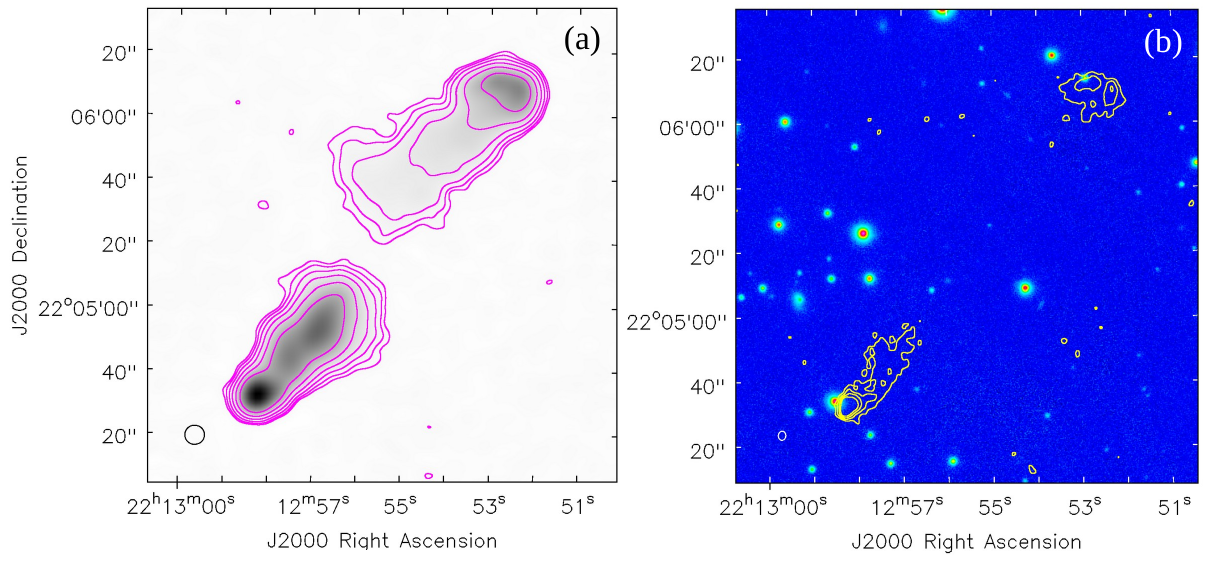}
  \caption{J221255+220523: (a) LoTSS-DR2 contour plot (magenta) \& grey-scale; (b) VLASS contour plot (yellow) overlaid on the optical (PanSTARR, I-band) image. }
  \label{fig:J221255}
\end{SCfigure*}

{\bf J140311+382756 (4C+38.38):}
Basic twin-lobed radio structure of this source became known from the 2.7 GHz VLA map published by \citet{Rudnick1979AJ.....84..437R}. The VLASS map shows it to be actually a well-aligned double-double radio galaxy (DDRG), with a secure optical identification of photometric redshift of 0.541 (Fig. \ref{fig:J140311}b). Its another striking feature unveiled by the LoTSS-DR2 map, is a prominent radio spur jetting out from the host elliptical orthogonally to the DDRG axis (Fig. \ref{fig:J140311} a \& c). This radio spur is not detected in the VLASS map and no optical object is seen to coincide with any part of the spur. The absence of its counterpart on the opposite side rules out an XRG classification for this source. The origin of the one-sided central radio spur in a DDRG poses an interesting situation warranting improved radio imaging. \\

{\bf J143527+550756 (6C 143354+552059):}
This double radio source with LAS $\sim$ $100''$ (246 kpc) is identified with a BCG at {\it z} (spec) = 0.13985
\citep{Ahn2012ApJS..203...21A}. The VLASS map shows a strong radio core associated with the bright galaxy, as well as a prominent radio jet towards the northwest direction (Fig. \ref{fig:J143527}). The jet undergoes a sharp bend at a distance of $\sim$ $27''$ (67 kpc) from the radio nucleus and flares as a plume of decreasing surface brightness, as also evident from the LoTSS-DR2 map, which is highly reminiscent of FR I morphology 
\citep[e.g.,][]{O'Donoghue1993ApJ...408..428O, Laing2008MNRAS.391..521L, Parma1999A&A...344....7P}. 
In contrast, the eastern lobe has a terminal hot spot, hence FR II type. Thus, this source belongs to the rare subclass of double radio sources, called HYMORS in which the two lobes clearly belong to different FR types \citep[see][and references therein, section 4]{Gopal-Krishna2023JApA...44...44G}. Based on its FIRST survey map, this source was already classified as a HYMORS \citep{Kumari2022MNRAS.514.4290K}. Here we show that its hybrid nature is evident not only from the contrasting FR morphologies of its two lobes but also from their spectral gradients, which reinforces the case for its being a genuine HYMORS, rather than a FR II source masquerading as a HYMORS, a possibility suggested by \citet{Harwood2020MNRAS.491..803H}. For making the spectral index map of this source we have combined its LOFAR-DR2 (144 MHz) and the FIRST (1.4 GHz) maps, after ensuring that the FIRST map is essentially free from the missing flux problem (the map's total flux density of 445 mJy, as measured by us, is close to the value of $470\pm15$ mJy given in the 1.4 GHz NVSS catalogue \citep{Condon-nvss-1998AJ....115.1693C}.
As seen from Fig. \ref{fig:J143527}d, the spectral index map of the eastern lobe shows a systematic flattening outward from the parent galaxy until the terminal hot spot, whereas an opposite spectral gradient is observed on the western side. Such spectral gradients are well known to characterise FR II and FR I sources, respectively 
\citep[e.g.,][]{Laing2014MNRAS.437.3405L}. Recall that importance of combining the information on radio brightness and spectral index distributions for identifying genuine HYMORS has been underscored earlier in several studies 
\citep[e.g.,][]{de-Gasperin2017MNRAS.467.2234D, Pirya2011BASI...39..547P, Sebastian2022ApJ...935...59S}.

{\bf J165732+431941:}
This complex radio source is identified with a BCG at a (spectroscopic) redshift of  0.204.
The LoTSS map shows 3 prominent radio arms with a central peak at their node (Fig.\ref{fig:J165732}). The VLASS map shows a radio core at the position of the BCG, both significantly offset from the nodal point by $\sim$ $12''$ (40 kpc) towards northwest. A joint inspection of the VLASS and LoTSS-DR2 maps suggests that the western and southern radio arms together define a WAT associated with the BCG, The nature and role of the $\sim$ $40''$ (135 kpc long eastern radio arm remains unclear). Based on the FIRST map at 1.4 GHz \citep{Becker-first1995ApJ...450..559B, White-first1997ApJ...475..479W}, \citet{Kumari2022MNRAS.514.4290K} have classified it as a HYMORS. However, the above discussion casts doubt on this classification. \\

\begin{SCfigure*}
    \includegraphics[width=13cm, height=11.5cm]{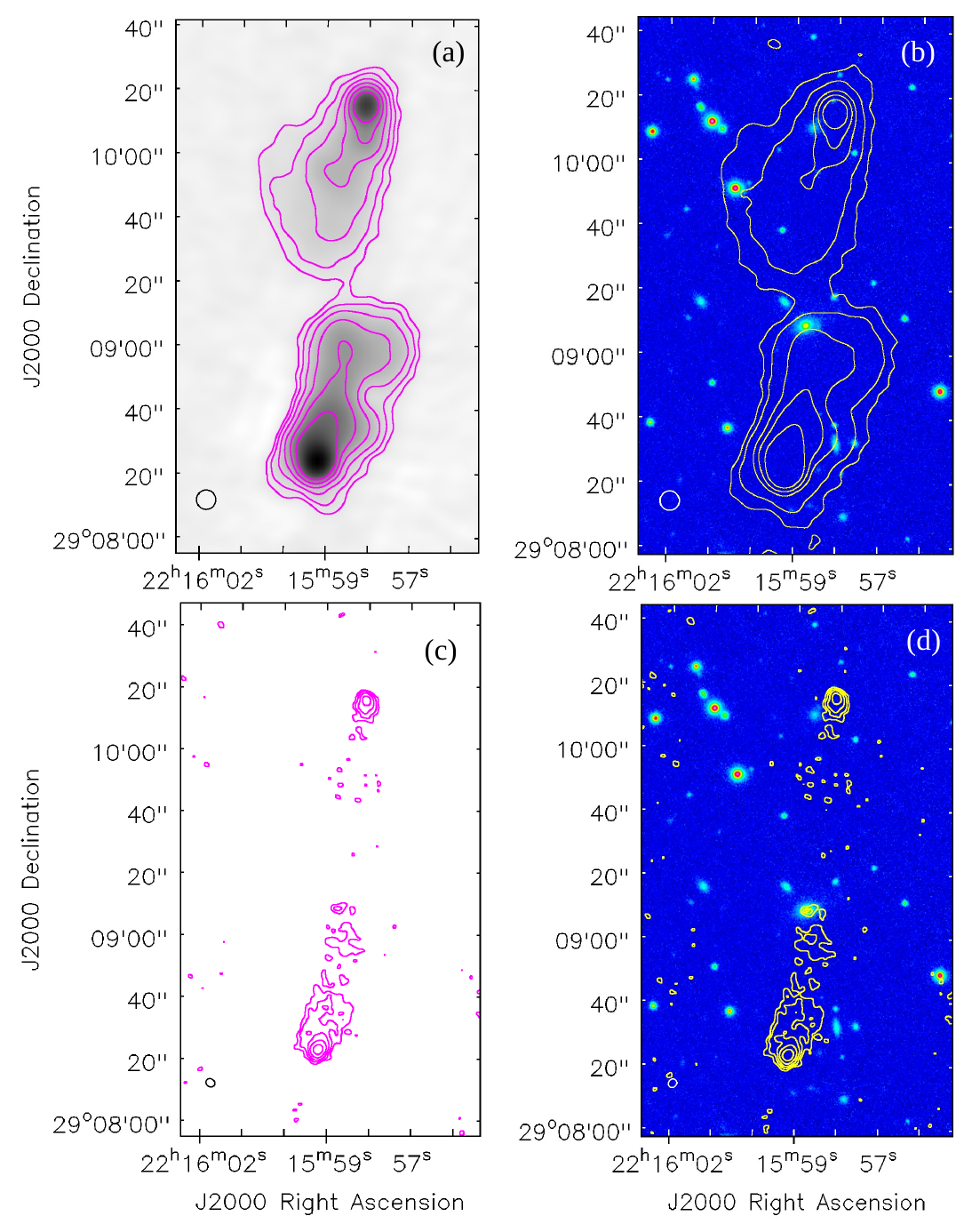}
  \caption{J221558+290921: (a) LoTSS-DR2 contour plot (magenta) \& grey-scale; (b) LoTSS-DR2 contour plot (yellow) overlaid on the optical (PanSTARR, I-band) image; (c) VLASS contour plot (magenta); (d) VLASS contour plot (yellow) overlaid on the zoomed optical (PanSTARR, I-band) image.}
  \label{fig:J221558}
\end{SCfigure*}
\begin{SCfigure*}
    \includegraphics[width=14cm, height=11.5cm]{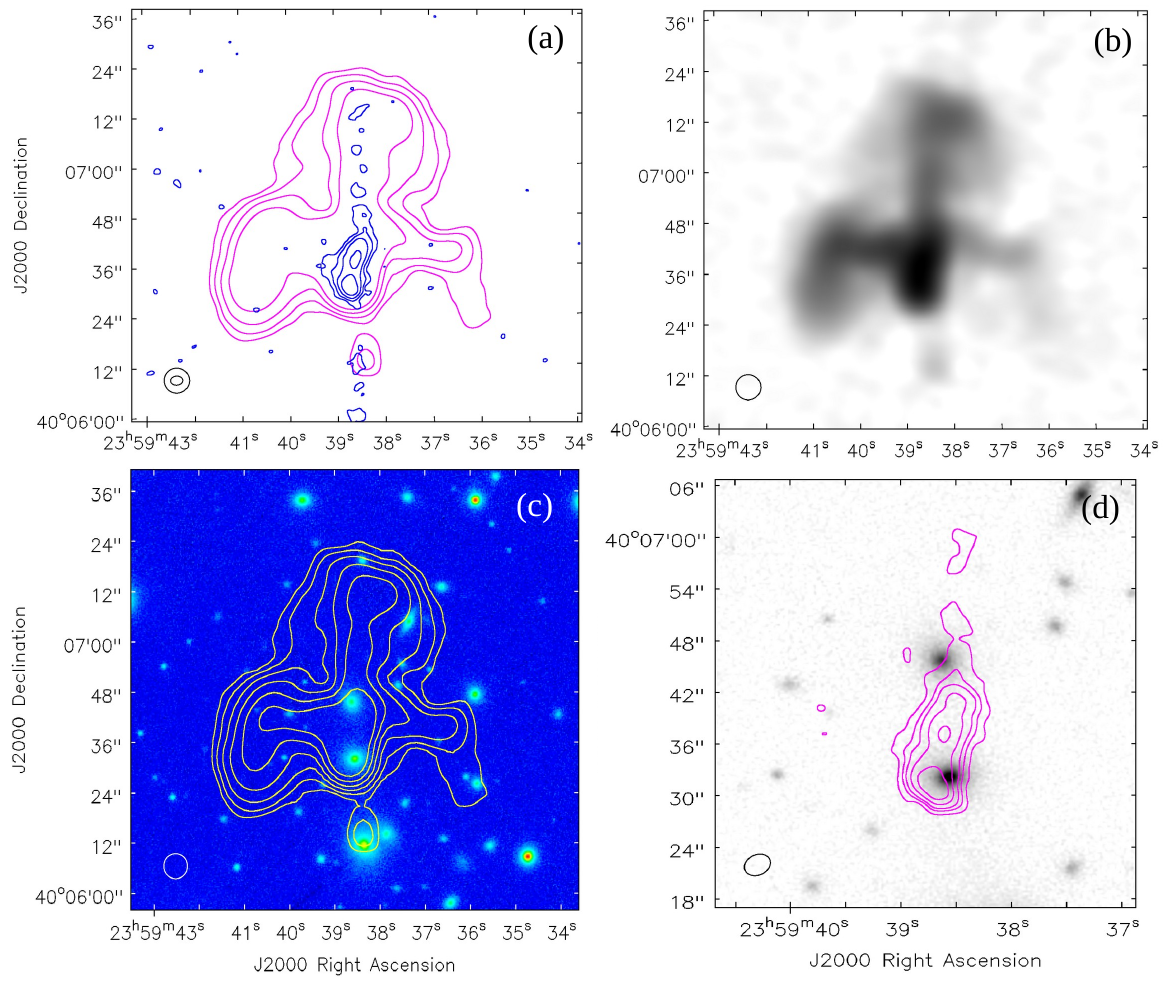}
  \caption{J235938+400644: (a) VLASS contour plot (blue) overlaid on the LoTSS-DR2 (magenta); (b)   LoTSS-DR2 grey-scale; (c) LoTSS-DR2 contour plot (yellow) overlaid on the optical (PanSTARR, I-band) image; (d) VLASS contour (magenta) overlaid on the zoomed optical (PanSTARR, I-band) grey image.}
  \label{fig:J235938}
\end{SCfigure*}

{\bf J170940+342537 (The `Square') (B2 1707+34, 4C+34.45):}
This source is identified with the BCG of the Abell cluster 2249 at {\it z}(photo)=0.0806 \citep{Owen1997ApJS..108...41O}. Using VLA, \citep{Jetha2006MNRAS.368..609J} have mapped this source at 8.5 GHz, with a $3^{\prime\prime}\times2^{\prime\prime}$ beam. While their map is consistent with the VLASS map seen in Fig. \ref{fig:J170940} c\&d, the higher sensitivity of their map make the WAT classification secure, with roughly north-south oriented jets bending sharply towards the east and forming the two tails. On the LoTSS-DR2 map, the tails can be traced farther out to the east and merging together, thus giving this `fat WAT' a square-shaped appearance. \\

{\bf J221255+220523:}
Both VLASS and LoTSS-DR2 maps clearly show this source to be FR II type, albeit without a radio core detection even in the VLASS map (Fig. \ref{fig:J221255}b). The faint $(m\_r \sim 21.93)$ optical object seen near the line joining the two hotspots is a possible host galaxy, with coordinates: RA= 22h 12m 56.37s, DEC= 22d $05'$ $08.2''$. The LoTSS-DR2 map shows an extended lobe behind each hot spot and while these $\sim$ $45''$ long radio lobes run parallel to each other, they exhibit a lateral offset by about $15''$ (Fig. \ref{fig:J221255}a). This FR II source is thus another example of `offset parallel twin-lobes' (OPTL), like the source J094953+445658 mentioned above \citep[see, also][]{Gopal-Krishna-3C223.1-2022A&A...663L...8G} and 
J074144+741440 \citep[Fig. 3 of][]{Cohen2005ApJ...620L...5C}.
Unfortunately, due to radio non-detection, the optical identification of the present source remains unconfirmed. \\

{\bf J221558+290921:}
As seen from the VLASS map, this source has a clear FR II morphology. The bright optical galaxy lying roughly in the middle of the two terminal hotspots is the most probable host galaxy, supported by its radio detection (Fig. \ref{fig:J221558} c\&d). The LoTSS-DR2 map (Fig. \ref{fig:J221558}a) shows both its lobes to extend right up to the parent galaxy, such that they run nearly parallel to each other, but maintaining a clear lateral offset of $\sim$ $30''$ (112 kpc). Hence this source is one more example of the OPTL morphology (see above).

{\bf J235938+400644:} 
The peculiar radio morphology of this source, seen in the LoTSS-DR2 map, shows the emission to be concentrated along two orthogonal axes (Fig. \ref{fig:J235938} a\&b). The north-south radio ridge appears to have a head-tail morphology, as seen from the VLASS map (Fig. \ref{fig:J235938} a,d), with the head coinciding with a bright galaxy at $\it z$ (photo) $\sim$ 0.1967, and the tail pointing to the north. About $13''$ north of this galaxy is seen another fairly bright optical galaxy, which is not clearly detected in the VLASS
(Fig. \ref{fig:J235938}d). 
Consequently, the host galaxy of the east-west radio ridge remains unconfirmed.
\begin{figure*}
\begin{center}
    \includegraphics[width=10cm, height=8cm]{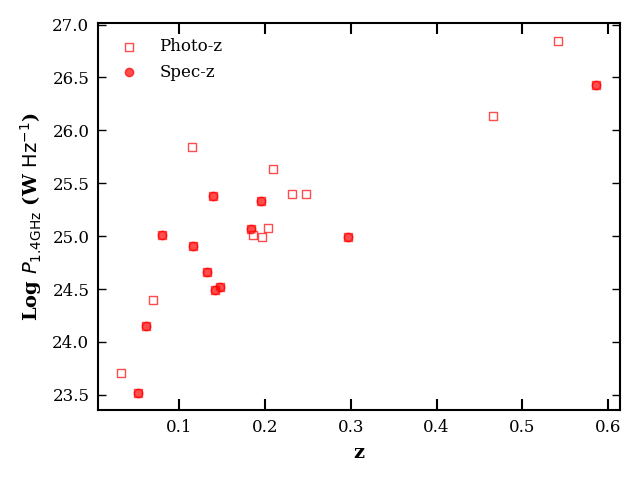}
  \caption{ The redshift-luminosity diagram for the 23 ANOMERS with known redshifts. The radio luminosities have been computed using the NVSS flux densities at 1.4 GHz (see Table 1).}
  \label{fig:lum_red}
\end{center}
\end{figure*}

\section{Discussion and conclusions}
\label{discussion}

Fig. \ref{fig:lum_red} shows the distribution in the redshift-luminosity plane, of our 23 ANOMERS with known redshift (Table 1). A vast majority $(\sim 80\%)$  of them has redshifts below 0.3. Their median radio luminosity is $\sim 10^{25}$ W.Hz$^{-1}$ at 1.4 GHz, which is also the luminosity of the Fanaroff-Riley division 
\citep{Fanaroff1974MNRAS.167P..31F, Ledlow1996AJ....112....9L}. 
Thus, the indication from at least the present sample is that morphologically anomalous radio sources are almost equally likely to occur on both sides of the FRI/II boundary.}

As mentioned in section 1, the impetus for the present work came from the need to highlight the crucial role of the radio galaxies which are morphologically anomalous and, consequently have a high potential to strain/challenge the conventional theoretical models of extragalactic radio sources. In this quest, we have identified 25 such sources by a visual inspection of the morphologies in a well-defined sample of 2428 relatively bright extragalactic radio sources, derived from the LoTSS-DR2 catalogue \citep{Shimwell2022A&A...659A...1S}. We believe that this LOFAR survey is best suited for the present purpose, given its high resolution ($6''$), good sensitivity (83 $\mu$Jy/beam) even to structures extending up to several arc-minutes, and a low observing frequency (144 MHz) at which even highly aged synchrotron plasma is expected to remain visible for a prolonged duration, due to much milder radiative losses. Additionally, we have combined these LOFAR maps with their counterparts in the VLASS survey \citep{Lacy2020PASP..132c5001L} which has not only an even better resolution ($3''$), but also a much higher observing frequency (3 GHz). This greatly enhances its effectiveness in detecting the radio nuclei, hot spots and jets. The complementarity between these two recent radio surveys, augmented with fairly deep optical images available in the  PanSTARRS and DECaLS surveys, has succeeded in clarifying the observed structural complexities for several, albeit not all our sources (see below). In order to underscore the inherent potential of the tiny minority of such morphological outliers, for advancing the understanding of the radio source phenomenon, we have proposed such radio sources to be categorised under a single rubric of `ANOmalous Morphology Extragalactic Radio Sources (ANOMERS)'. Basically, this is meant to consolidate the sources which fall outside the standard taxonomy, as summarised, e.g., in a recent review by \citet{Baldi2023A&ARv..31....3B}, and which also do not belong to the class of diffuse radio sources associated with clusters of galaxies but not directly with any particular cluster member (section 1).

We now briefly highlight 10 species of ANOMERS found in our sample, which mark a stark deviation from the radio structures commonly observed among radio galaxies:

(a) Large collimated radio `chimneys' extending on $\sim$ 100 kpc scale, stemming almost orthogonally from the radio lobe, at a point roughly midway between the nucleus and the hot spot [J022830+382108, Fig. \ref{fig:J022830} \& J093032+311215, Fig. \ref{fig:J093032}]. Such sources may turn out to be good examples of the jet-shell interaction scenario proposed by \citet{Gopal1983Natur.303..217G}  (see, also, \citep{Gopal2010ApJ...720L.155G}). They are also relevant to explaining the XRG phenomenon (section 1).

(b) Potential case of jet bifurcation [J114434+672424, Fig. \ref{fig:J114434}]

(c) Huge radio spurs apparently emanating from the host galaxy, roughly perpendicular to the main radio axis defined by the twin-lobes [J101522+682351, Fig. \ref{fig:J101522}; J140311+382756, Fig. \ref{fig:J140311}]. The latter source is a `double-double radio galaxy'.

(d) A symmetric super-kink in a radio jet, which does not disturb/disrupt the jet's collimation, despite its record length of 110 kpc [J091900+315254, Fig. \ref{fig:J091900}].

(e) A candidate for `detached double-double radio galaxy’ (dDDRG) [J101522+682351, Fig. \ref{fig:J101522}] \citep[see][]{Gopal-Krishna-dDDRG2022PASA...39...49G}.

(f) A Wide-Angle-Tail (WAT) probably having a `collimated synchrotron Thread’ [J083106+414741, Fig. \ref{fig:J083106}] \citep[see,][]{Ramatsoku2020A&A...636L...1R, Gopal-Krishna2024MNRAS.529L.135G}.

(g) Two giant `radio spirals’ of sizes $\sim$ 390 kpc [J133116+441851, Fig. \ref{fig:J133116}] and $\sim 100$ kpc [J135141+555938, Fig. \ref{fig:J135141}].

(h) Three radio galaxies showing `offset parallel twin-lobes’ (OPTL) [J094953+445658, Fig. \ref{fig:J094953}; J221255+220523, Fig. \ref{fig:J221255} \& J221558+290921, Fig. \ref{fig:J221558}].

(i) A FR II radio galaxy with a prominent radio bridge which is swept entirely to one side of the radio axis defined by the bi-polar jets [J102704+444547, Fig. \ref{fig:J102704}].

(j) Three particularly spectacular cases of ANOMERS are: (i) A radio `plier’ [J092229+545943, Fig. \ref{fig:J092229}], (ii) A $\sim$ 250 kpc long radio structure resembling a `helical antenna’ seen between the two radio lobes [J104234+364040, Fig. \ref{fig:J104234}], and a $\sim$ 375 kpc long fan-shaped one-sided ‘radio cone’ extending from the host galaxy perpendicular to the radio axis defined by the twin-lobes [J114429+370915, Fig. \ref{fig:J114429}].
    
(k) The last source in this list, J143527+550756 (Fig. \ref{fig:J143527}), belongs to the very rare class of HYMORS \citep{Gopal-Krishna-Witta2000A&A...363..507G}. As shown in the previous section, this inference is grounded not only in the contrast between the FR types of the two lobes but also in their spectral index gradients. The origin of HYMORS   
has been a subject of debate, including the role of projection effects \citep[][see sect. 3]{Harwood2020MNRAS.491..803H} as summarised in \citet{Gopal-Krishna2023JApA...44...44G}. As argued in the latter paper, the different F-R types of the two lobes can be attributed mainly to an asymmetry in the jet's interaction with the ambient medium. In their preferred model, the FR I lobe is a transformed FR II lobe, probably owing to a stronger interaction on its side, which has decelerated the tip of the jet (hot spot, or `working-surface’) to a subsonic speed. The consequent weakening of the ram-pressure confinement leads to the hot spot's expansion and fading, accompanied with curtailment of the back-flow of the synchrotron plasma out of it. The resulting depletion of the protective cocoon of synchrotron plasma around the relativistic jet, facilitates entrainment of the ambient thermal plasma into the jet, rendering it trans-sonic and therefore increasingly susceptible to instabilities and eventual disruption, instead of terminating in a hot spot \citep{Gopal-Krishna2001A&A...377..827G, Gopal-Krishna2023JApA...44...44G}. The physics of such entrainment-induced instabilities in the jet plasma has been discussed in several studies \citep[e.g.,][]{Bicknell1984ApJ...286...68B, De-young1993ApJ...405L..13D, Laing2002MNRAS.336.1161L, Porth2015MNRAS.452.1089P}. 
Note that, whilst the environmental factor is linked to the FR type \citep[e.g.,][]{Rodman2019MNRAS.482.5625R}, the apparent morphological contrast between the two lobes of a source could also be enhanced by the differential light travel time between the approaching and receding lobes \citep{Gopal1996A&A...316L..13G, Marecki2012A&A...544L...2M, Gopal-Krishna2023JApA...44...44G, Ghosh2023ApJ...958...71G}. Furthermore, it is clear that the above chain of outcomes, initiated due to a stronger interaction of one jet with the ambient gas (and consequent expansion/weakening of that hot spot), would be accentuated by any decline in the jet's kinetic power. If this happens due to a drop in the accretion to the central engine, which is quite plausible, the proposed FR II to FR I transformation of the radio lobe (and associated jet) would also be echoed in a softening of the excitation level of the nuclear optical/UV spectrum. 

In conclusion, the present work should be viewed as a step towards gaining {\it insight from outliers} which should facilitate refinement of the theoretical models of extragalactic radio sources. We have initiated efforts towards radio spectral mapping of these sources, which would enable tracing and modelling of their evolutionary histories. Further progress can also be expected from a systematic and comprehensive scrutiny of the radio images in forthcoming releases of the LOFAR survey.

\section*{Acknowledgements}
We would like to thank the anonymous referees for the constructive comments. GK acknowledges the award of a Senior Scientist position of the Indian National Science Academy. 
This research has made use of the CIRADA cutout service at URL cutouts.cirada.ca, operated by the Canadian Initiative for Radio Astronomy Data Analysis (CIRADA). CIRADA is funded by a grant from the Canada Foundation for Innovation 2017 Innovation Fund (Project 35999), as well as by the Provinces of Ontario, British Columbia, Alberta, Manitoba, and Quebec, in collaboration with the National Research Council of Canada, the US National Radio Astronomy Observatory and Australia's Commonwealth Scientific and Industrial Research Organisation.
The authors would like to acknowledge NASA/IPAC Extragalactic Database (NED) as this work has made use of the NED, which is operated by the Jet Propulsion Laboratory, California Institute of Technology, under contract with the
National Aeronautics and Space Administration.
\vspace{-1em}

\bibliography{references}

\begin{thebibliography}{}
\expandafter\ifx\csname natexlab\endcsname\relax\def\natexlab#1{#1}\fi

\bibitem[{{Ahn} {$et~al$.}(2012){Ahn}, {Alexandroff}, {Allende Prieto}, {Anderson}, {Anderton}, {Andrews}, {Aubourg}, {Bailey}, {Balbinot}, {Barnes}, \& et~al.}]{Ahn2012ApJS..203...21A}
{Ahn}, C.~P., {Alexandroff}, R., {Allende Prieto}, C., {$et~al$.} 2012, \apjs, 203, 21

\bibitem[{{Antonucci}(1993)}]{Antonucci1993ARA&A..31..473A}
{Antonucci}, R. 1993, \araa, 31, 473

\bibitem[{{Antonucci}(2023)}]{Antonucci2023Galax..11..102A}
{Antonucci}, R. R.~J. 2023, Galaxies, 11, 102

\bibitem[{{Antonucci} {$et~al$.}(1986){Antonucci}, {Hickson}, {Olszewski}, \& {Miller}}]{Antonucci1986AJ.....92....1A}
{Antonucci}, R.~R.~J., {Hickson}, P., {Olszewski}, E.~W., \& {Miller}, J.~S. 1986, \aj, 92, 1

\bibitem[{{Baldi}(2023)}]{Baldi2023A&ARv..31....3B}
{Baldi}, R.~D. 2023, \aapr, 31, 3

\bibitem[{{Barai} {$et~al$.}(2011){Barai}, {Martel}, \& {Germain}}]{Barai2011ApJ...727...54B}
{Barai}, P., {Martel}, H., \& {Germain}, J. 2011, \apj, 727, 54

\bibitem[{{Baron} \& {Poznanski}(2017)}]{Baron2017MNRAS.465.4530B}
{Baron}, D., \& {Poznanski}, D. 2017, \mnras, 465, 4530

\bibitem[{{Barthel}(1994)}]{Barthel1994ASPC...54..175B}
{Barthel}, P.~D. 1994, in Astronomical Society of the Pacific Conference Series, Vol.~54, The Physics of Active Galaxies, ed. G.~V. {Bicknell}, M.~A. {Dopita}, \& P.~J. {Quinn}, 175

\bibitem[{{Baum} {$et~al$.}(1995){Baum}, {Zirbel}, \& {O'Dea}}]{Baum1995ApJ...451...88B}
{Baum}, S.~A., {Zirbel}, E.~L., \& {O'Dea}, C.~P. 1995, \apj, 451, 88

\bibitem[{{Becker} {$et~al$.}(1995){Becker}, {White}, \& {Helfand}}]{Becker-first1995ApJ...450..559B}
{Becker}, R.~H., {White}, R.~L., \& {Helfand}, D.~J. 1995, \apj, 450, 559

\bibitem[{{Begelman} {$et~al$.}(1984){Begelman}, {Blandford}, \& {Rees}}]{Begelman1984RvMP...56..255B}
{Begelman}, M.~C., {Blandford}, R.~D., \& {Rees}, M.~J. 1984, Reviews of Modern Physics, 56, 255

\bibitem[{{Bempong-Manful} {$et~al$.}(2020){Bempong-Manful}, {Hardcastle}, {Birkinshaw}, {Laing}, {Leahy}, \& {Worrall}}]{Bempong-Manfu2020MNRAS.496..676B}
{Bempong-Manful}, E., {Hardcastle}, M.~J., {Birkinshaw}, M., {$et~al$.} 2020, \mnras, 496, 676

\bibitem[{{Bera} {$et~al$.}(2022){Bera}, {Sasmal}, {Patra}, \& {Mondal}}]{Bera2022ApJS..260....7B}
{Bera}, S., {Sasmal}, T.~K., {Patra}, D., \& {Mondal}, S. 2022, \apjs, 260, 7

\bibitem[{{Best} \& {Heckman}(2012)}]{Best2012MNRAS.421.1569B}
{Best}, P.~N., \& {Heckman}, T.~M. 2012, \mnras, 421, 1569

\bibitem[{{Bicknell}(1984)}]{Bicknell1984ApJ...286...68B}
{Bicknell}, G.~V. 1984, \apj, 286, 68

\bibitem[{{Black} {$et~al$.}(1992){Black}, {Baum}, {Leahy}, {Perley}, {Riley}, \& {Scheuer}}]{Black1992MNRAS.256..186B}
{Black}, A.~R.~S., {Baum}, S.~A., {Leahy}, J.~P., {$et~al$.} 1992, \mnras, 256, 186

\bibitem[{{Blandford} {$et~al$.}(2019){Blandford}, {Meier}, \& {Readhead}}]{Blanford2019ARA&A..57..467B}
{Blandford}, R., {Meier}, D., \& {Readhead}, A. 2019, \araa, 57, 467

\bibitem[{{Blandford} \& {Rees}(1978)}]{Blandford1978bllo.conf..328B}
{Blandford}, R.~D., \& {Rees}, M.~J. 1978, in BL Lac Objects, ed. A.~M. {Wolfe}, 328--341

\bibitem[{{Bridle}(1984)}]{Bridle1984AJ.....89..979B}
{Bridle}, A.~H. 1984, \aj, 89, 979

\bibitem[{{Bridle} {$et~al$.}(1979){Bridle}, {Davis}, {Fomalont}, {Willis}, \& {Strom}}]{Bridle1979ApJ...228L...9B}
{Bridle}, A.~H., {Davis}, M.~M., {Fomalont}, E.~B., {Willis}, A.~G., \& {Strom}, R.~G. 1979, \apjl, 228, L9

\bibitem[{{Burns} {$et~al$.}(1987){Burns}, {Hanisch}, {White}, {Nelson}, {Morrisette}, \& {Moody}}]{Burns1987AJ.....94..587B}
{Burns}, J.~O., {Hanisch}, R.~J., {White}, R.~A., {$et~al$.} 1987, \aj, 94, 587

\bibitem[{{Burns} {$et~al$.}(1983){Burns}, {Schwendeman}, \& {White}}]{Burns1983ApJ...271..575B}
{Burns}, J.~O., {Schwendeman}, E., \& {White}, R.~A. 1983, \apj, 271, 575

\bibitem[{{Buta} \& {Combes}(1996)}]{Buta1996FCPh...17...95B}
{Buta}, R., \& {Combes}, F. 1996, \fcp, 17, 95

\bibitem[{{Capetti} {$et~al$.}(2002){Capetti}, {Zamfir}, {Rossi}, {Bodo}, {Zanni}, \& {Massaglia}}]{Capetti2002A&A...394...39C}
{Capetti}, A., {Zamfir}, S., {Rossi}, P., {$et~al$.} 2002, \aap, 394, 39

\bibitem[{{Carilli} {$et~al$.}(1991){Carilli}, {Perley}, {Dreher}, \& {Leahy}}]{Carilli1991ApJ...383..554C}
{Carilli}, C.~L., {Perley}, R.~A., {Dreher}, J.~W., \& {Leahy}, J.~P. 1991, \apj, 383, 554

\bibitem[{{Chambers} {$et~al$.}(2016){Chambers}, {Magnier}, {Metcalfe}, {Flewelling}, {Huber}, {Waters}, {Denneau}, {Draper}, {Farrow}, {Finkbeiner}, {Holmberg}, {Koppenhoefer}, {Price}, {Rest}, {Saglia}, {Schlafly}, {Smartt}, {Sweeney}, {Wainscoat}, {Burgett}, {Chastel}, {Grav}, {Heasley}, {Hodapp}, {Jedicke}, {Kaiser}, {Kudritzki}, {Luppino}, {Lupton}, {Monet}, {Morgan}, {Onaka}, {Shiao}, {Stubbs}, {Tonry}, {White}, {Ba{\~n}ados}, {Bell}, {Bender}, {Bernard}, {Boegner}, {Boffi}, {Botticella}, {Calamida}, {Casertano}, {Chen}, {Chen}, {Cole}, {Deacon}, {Frenk}, {Fitzsimmons}, {Gezari}, {Gibbs}, {Goessl}, {Goggia}, {Gourgue}, {Goldman}, {Grant}, {Grebel}, {Hambly}, {Hasinger}, {Heavens}, {Heckman}, {Henderson}, {Henning}, {Holman}, {Hopp}, {Ip}, {Isani}, {Jackson}, {Keyes}, {Koekemoer}, {Kotak}, {Le}, {Liska}, {Long}, {Lucey}, {Liu}, {Martin}, {Masci}, {McLean}, {Mindel}, {Misra}, {Morganson}, {Murphy}, {Obaika}, {Narayan}, {Nieto-Santisteban}, {Norberg}, {Peacock}, {Pier}, {Postman}, {Primak}, {Rae}, {Rai},
  {Riess}, {Riffeser}, {Rix}, {R{\"o}ser}, {Russel}, {Rutz}, {Schilbach}, {Schultz}, {Scolnic}, {Strolger}, {Szalay}, {Seitz}, {Small}, {Smith}, {Soderblom}, {Taylor}, {Thomson}, {Taylor}, {Thakar}, {Thiel}, {Thilker}, {Unger}, {Urata}, {Valenti}, {Wagner}, {Walder}, {Walter}, {Watters}, {Werner}, {Wood-Vasey}, \& {Wyse}}]{Chambers-PanSTARRS2016arXiv161205560C}
{Chambers}, K.~C., {Magnier}, E.~A., {Metcalfe}, N., {$et~al$.} 2016, arXiv e-prints, arXiv:1612.05560

\bibitem[{{Cheung}(2007)}]{Cheung2007AJ....133.2097C}
{Cheung}, C.~C. 2007, \aj, 133, 2097

\bibitem[{{Cohen} {$et~al$.}(2005){Cohen}, {Clarke}, {Feretti}, \& {Kassim}}]{Cohen2005ApJ...620L...5C}
{Cohen}, A.~S., {Clarke}, T.~E., {Feretti}, L., \& {Kassim}, N.~E. 2005, \apjl, 620, L5

\bibitem[{{Condon} {$et~al$.}(1995){Condon}, {Anderson}, \& {Broderick}}]{Condon1995AJ....109.2318C}
{Condon}, J.~J., {Anderson}, E., \& {Broderick}, J.~J. 1995, \aj, 109, 2318

\bibitem[{{Condon} {$et~al$.}(1998){Condon}, {Cotton}, {Greisen}, {Yin}, {Perley}, {Taylor}, \& {Broderick}}]{Condon-nvss-1998AJ....115.1693C}
{Condon}, J.~J., {Cotton}, W.~D., {Greisen}, E.~W., {$et~al$.} 1998, \aj, 115, 1693

\bibitem[{{Dabhade} {$et~al$.}(2022){Dabhade}, {Shimwell}, {Bagchi}, {Saikia}, {Combes}, {Gaikwad}, {R{\"o}ttgering}, {Mohapatra}, {Ishwara-Chandra}, {Intema}, \& {Raychaudhury}}]{Dabhade2022A&A...668A..64D}
{Dabhade}, P., {Shimwell}, T.~W., {Bagchi}, J., {$et~al$.} 2022, \aap, 668, A64

\bibitem[{{de Gasperin}(2017)}]{de-Gasperin2017MNRAS.467.2234D}
{de Gasperin}, F. 2017, \mnras, 467, 2234

\bibitem[{{de Ruiter} {$et~al$.}(1986){de Ruiter}, {Parma}, {Fanti}, \& {Fanti}}]{deRuiter1986A&AS...65..111D}
{de Ruiter}, H.~R., {Parma}, P., {Fanti}, C., \& {Fanti}, R. 1986, \aaps, 65, 111

\bibitem[{{De Young}(1993)}]{De-young1993ApJ...405L..13D}
{De Young}, D.~S. 1993, \apjl, 405, L13

\bibitem[{{Dennett-Thorpe} {$et~al$.}(1999){Dennett-Thorpe}, {Bridle}, {Laing}, \& {Scheuer}}]{Dennett-Thorpe1999MNRAS.304..271D}
{Dennett-Thorpe}, J., {Bridle}, A.~H., {Laing}, R.~A., \& {Scheuer}, P.~A.~G. 1999, \mnras, 304, 271

\bibitem[{{Dey} {$et~al$.}(2019){Dey}, {Schlegel}, {Lang}, {Blum}, {Burleigh}, {Fan}, {Findlay}, {Finkbeiner}, {Herrera}, {Juneau}, {Landriau}, {Levi}, {McGreer}, {Meisner}, {Myers}, {Moustakas}, {Nugent}, {Patej}, {Schlafly}, {Walker}, {Valdes}, {Weaver}, {Y{\`e}che}, {Zou}, {Zhou}, {Abareshi}, {Abbott}, {Abolfathi}, {Aguilera}, {Alam}, {Allen}, {Alvarez}, {Annis}, {Ansarinejad}, {Aubert}, {Beechert}, {Bell}, {BenZvi}, {Beutler}, {Bielby}, {Bolton}, {Brice{\~n}o}, {Buckley-Geer}, {Butler}, {Calamida}, {Carlberg}, {Carter}, {Casas}, {Castander}, {Choi}, {Comparat}, {Cukanovaite}, {Delubac}, {DeVries}, {Dey}, {Dhungana}, {Dickinson}, {Ding}, {Donaldson}, {Duan}, {Duckworth}, {Eftekharzadeh}, {Eisenstein}, {Etourneau}, {Fagrelius}, {Farihi}, {Fitzpatrick}, {Font-Ribera}, {Fulmer}, {G{\"a}nsicke}, {Gaztanaga}, {George}, {Gerdes}, {Gontcho}, {Gorgoni}, {Green}, {Guy}, {Harmer}, {Hernandez}, {Honscheid}, {Huang}, {James}, {Jannuzi}, {Jiang}, {Joyce}, {Karcher}, {Karkar}, {Kehoe}, {Kneib}, {Kueter-Young}, {Lan},
  {Lauer}, {Le Guillou}, {Le Van Suu}, {Lee}, {Lesser}, {Perreault Levasseur}, {Li}, {Mann}, {Marshall}, {Mart{\'\i}nez-V{\'a}zquez}, {Martini}, {du Mas des Bourboux}, {McManus}, {Meier}, {M{\'e}nard}, {Metcalfe}, {Mu{\~n}oz-Guti{\'e}rrez}, {Najita}, {Napier}, {Narayan}, {Newman}, {Nie}, {Nord}, {Norman}, {Olsen}, {Paat}, {Palanque-Delabrouille}, {Peng}, {Poppett}, {Poremba}, {Prakash}, {Rabinowitz}, {Raichoor}, {Rezaie}, {Robertson}, {Roe}, {Ross}, {Ross}, {Rudnick}, {Safonova}, {Saha}, {S{\'a}nchez}, {Savary}, {Schweiker}, {Scott}, {Seo}, {Shan}, {Silva}, {Slepian}, {Soto}, {Sprayberry}, {Staten}, {Stillman}, {Stupak}, {Summers}, {Sien Tie}, {Tirado}, {Vargas-Maga{\~n}a}, {Vivas}, {Wechsler}, {Williams}, {Yang}, {Yang}, {Yapici}, {Zaritsky}, {Zenteno}, {Zhang}, {Zhang}, {Zhou}, \& {Zhou}}]{Dey2019AJ....157..168D}
{Dey}, A., {Schlegel}, D.~J., {Lang}, D., {$et~al$.} 2019, \aj, 157, 168

\bibitem[{{Ekers} {$et~al$.}(1978){Ekers}, {Fanti}, {Lari}, \& {Parma}}]{Ekers1978Natur.276..588E}
{Ekers}, R.~D., {Fanti}, R., {Lari}, C., \& {Parma}, P. 1978, \nat, 276, 588

\bibitem[{{Ekers} {$et~al$.}(1976){Ekers}, {van der Hulst}, \& {Miley}}]{Ekers1976Natur.262..369E}
{Ekers}, R.~D., {van der Hulst}, J.~M., \& {Miley}, G.~K. 1976, \nat, 262, 369

\bibitem[{{Evans} {$et~al$.}(2008){Evans}, {Fong}, {Hardcastle}, {Kraft}, {Lee}, {Worrall}, {Birkinshaw}, {Croston}, \& {Muxlow}}]{Evans2008ApJ...675.1057E}
{Evans}, D.~A., {Fong}, W.-F., {Hardcastle}, M.~J., {$et~al$.} 2008, \apj, 675, 1057

\bibitem[{{Fabian}(2012)}]{Fabian2012ARA&A..50..455F}
{Fabian}, A.~C. 2012, \araa, 50, 455

\bibitem[{{Fanaroff} {$et~al$.}(2021){Fanaroff}, {Lal}, {Venturi}, {Smirnov}, {Bondi}, {Thorat}, {Bester}, {J{\'o}zsa}, {Kleiner}, {Loi}, {Makhathini}, \& {White}}]{Fanaroff2021MNRAS.505.6003F}
{Fanaroff}, B., {Lal}, D.~V., {Venturi}, T., {$et~al$.} 2021, \mnras, 505, 6003

\bibitem[{{Fanaroff} \& {Riley}(1974)}]{Fanaroff1974MNRAS.167P..31F}
{Fanaroff}, B.~L., \& {Riley}, J.~M. 1974, \mnras, 167, 31P

\bibitem[{{Fomalont}(1981)}]{Fomalont1981IAUS...94..111F}
{Fomalont}, E.~B. 1981, in Origin of Cosmic Rays, ed. G.~{Setti}, G.~{Spada}, \& A.~W. {Wolfendale}, Vol.~94, 111--125

\bibitem[{{Furlanetto} \& {Loeb}(2001)}]{Furlanetto2001ApJ...556..619F}
{Furlanetto}, S.~R., \& {Loeb}, A. 2001, \apj, 556, 619

\bibitem[{{Furnell} {$et~al$.}(2018){Furnell}, {Collins}, {Kelvin}, {Clerc}, {Baldry}, {Finoguenov}, {Erfanianfar}, {Comparat}, \& {Schneider}}]{Furnell2018MNRAS.478.4952F}
{Furnell}, K.~E., {Collins}, C.~A., {Kelvin}, L.~S., {$et~al$.} 2018, \mnras, 478, 4952

\bibitem[{{Gawro{\'n}ski} {$et~al$.}(2006){Gawro{\'n}ski}, {Marecki}, {Kunert-Bajraszewska}, \& {Kus}}]{Gawronski2006A&A...447...63G}
{Gawro{\'n}ski}, M.~P., {Marecki}, A., {Kunert-Bajraszewska}, M., \& {Kus}, A.~J. 2006, \aap, 447, 63

\bibitem[{{Ghosh} {$et~al$.}(2023){Ghosh}, {Kharb}, {Baghel}, \& {Silpa}}]{Ghosh2023ApJ...958...71G}
{Ghosh}, S., {Kharb}, P., {Baghel}, J., \& {Silpa}, S. 2023, \apj, 958, 71

\bibitem[{{Gopal-Krishna}(1995)}]{Gopal-Krishna1995PNAS...9211399G}
{Gopal-Krishna}. 1995, Proceedings of the National Academy of Science, 92, 11399

\bibitem[{{Gopal-Krishna} \& {Biermann}(2024)}]{Gopal-Krishna2024MNRAS.529L.135G}
{Gopal-Krishna}, \& {Biermann}, P.~L. 2024, \mnras, 529, L135

\bibitem[{{Gopal-Krishna} {$et~al$.}(2010){Gopal-Krishna}, {Biermann}, {de Souza}, \& {Wiita}}]{Gopal2010ApJ...720L.155G}
{Gopal-Krishna}, {Biermann}, P.~L., {de Souza}, V., \& {Wiita}, P.~J. 2010, \apjl, 720, L155

\bibitem[{{Gopal-Krishna} {$et~al$.}(2012){Gopal-Krishna}, {Biermann}, {Gergely}, \& {Wiita}}]{Gopal2012RAA....12..127G}
{Gopal-Krishna}, {Biermann}, P.~L., {Gergely}, L.~{\'A}., \& {Wiita}, P.~J. 2012, Research in Astronomy and Astrophysics, 12, 127

\bibitem[{{Gopal-Krishna} {$et~al$.}(2003){Gopal-Krishna}, {Biermann}, \& {Wiita}}]{Gopal-Krishna2003ApJ...594L.103G}
{Gopal-Krishna}, {Biermann}, P.~L., \& {Wiita}, P.~J. 2003, \apjl, 594, L103

\bibitem[{{Gopal-Krishna} \& {Chitre}(1983)}]{Gopal1983Natur.303..217G}
{Gopal-Krishna}, \& {Chitre}, S.~M. 1983, \nat, 303, 217

\bibitem[{{Gopal-Krishna} \& {Dabhade}(2022)}]{Gopal-Krishna-3C223.1-2022A&A...663L...8G}
{Gopal-Krishna}, \& {Dabhade}, P. 2022, \aap, 663, L8

\bibitem[{{Gopal-Krishna} {$et~al$.}(2001){Gopal-Krishna}, {Subramanian}, {Wiita}, \& {Becker}}]{Gopal-Krishna2001A&A...377..827G}
{Gopal-Krishna}, {Subramanian}, P., {Wiita}, P.~J., \& {Becker}, P.~A. 2001, \aap, 377, 827

\bibitem[{{Gopal-Krishna} \& {Wiita}(2000)}]{Gopal-Krishna-Witta2000A&A...363..507G}
{Gopal-Krishna}, \& {Wiita}, P.~J. 2000, \aap, 363, 507

\bibitem[{{Gopal-Krishna} \& {Wiita}(2001)}]{Gopal-krishna2001ApJ...560L.115G}
---. 2001, \apjl, 560, L115

\bibitem[{{Gopal-Krishna} {$et~al$.}(2004){Gopal-Krishna}, {Wiita}, \& {Barai}}]{Gopal-Krishna2004JKAS...37..517G}
{Gopal-Krishna}, {Wiita}, P.~J., \& {Barai}, P. 2004, Journal of Korean Astronomical Society, 37, 517

\bibitem[{{Gopal-Krishna} {$et~al$.}(1996){Gopal-Krishna}, {Wiita}, \& {Hooda}}]{Gopal1996A&A...316L..13G}
{Gopal-Krishna}, {Wiita}, P.~J., \& {Hooda}, J.~S. 1996, \aap, 316, L13

\bibitem[{{Gopal-Krishna} {$et~al$.}(2023){Gopal-Krishna}, {Joshi}, \& {Patra}}]{Gopal-Krishna2023JApA...44...44G}
{Gopal-Krishna}, Wiita, P.~J., {Joshi}, R., \& {Patra}, D. 2023, Journal of Astrophysics and Astronomy, 44, 44

\bibitem[{{Gopal-Krishna} {$et~al$.}(2022){Gopal-Krishna}, {Salunkhe}, \& {Sonkamble}}]{Gopal-Krishna-dDDRG2022PASA...39...49G}
{Gopal-Krishna}, Paul, S., {Salunkhe}, S., \& {Sonkamble}, S. 2022, \pasa, 39, e049

\bibitem[{{Gower} {$et~al$.}(1982){Gower}, {Gregory}, {Unruh}, \& {Hutchings}}]{Gower1982ApJ...262..478G}
{Gower}, A.~C., {Gregory}, P.~C., {Unruh}, W.~G., \& {Hutchings}, J.~B. 1982, \apj, 262, 478

\bibitem[{{Hardcastle} \& {Croston}(2020)}]{Hardcastle2020NewAR..8801539H}
{Hardcastle}, M.~J., \& {Croston}, J.~H. 2020, \nar, 88, 101539

\bibitem[{{Hargrave} \& {Ryle}(1974)}]{Hargrave1974MNRAS.166..305H}
{Hargrave}, P.~J., \& {Ryle}, M. 1974, \mnras, 166, 305

\bibitem[{{Harwood} {$et~al$.}(2020){Harwood}, {Vernstrom}, \& {Stroe}}]{Harwood2020MNRAS.491..803H}
{Harwood}, J.~J., {Vernstrom}, T., \& {Stroe}, A. 2020, \mnras, 491, 803

\bibitem[{{Hine} \& {Longair}(1979)}]{Hine1979MNRAS.188..111H}
{Hine}, R.~G., \& {Longair}, M.~S. 1979, \mnras, 188, 111

\bibitem[{{H\"{o}gbom}(1979)}]{Hogbom1979A&AS...36..173H}
{H\"{o}gbom}, J.~A. 1979, \aaps, 36, 173

\bibitem[{{H\"{o}gbom} \& {Carlsson}(1974)}]{Hogbom1974A&A....34..341H}
{H\"{o}gbom}, J.~A., \& {Carlsson}, I. 1974, \aap, 34, 341

\bibitem[{{Jamrozy} {$et~al$.}(2008){Jamrozy}, {Konar}, {Machalski}, \& {Saikia}}]{Jamrozy2008MNRAS.385.1286J}
{Jamrozy}, M., {Konar}, C., {Machalski}, J., \& {Saikia}, D.~J. 2008, \mnras, 385, 1286

\bibitem[{{Jennison} \& {Das Gupta}(1953)}]{Jennison1953Natur.172..996J}
{Jennison}, R.~C., \& {Das Gupta}, M.~K. 1953, \nat, 172, 996

\bibitem[{{Jetha} {$et~al$.}(2006){Jetha}, {Hardcastle}, \& {Sakelliou}}]{Jetha2006MNRAS.368..609J}
{Jetha}, N.~N., {Hardcastle}, M.~J., \& {Sakelliou}, I. 2006, \mnras, 368, 609

\bibitem[{{Kapi{\'n}ska} {$et~al$.}(2017){Kapi{\'n}ska}, {Terentev}, {Wong}, {Shabala}, {Andernach}, {Rudnick}, {Storer}, {Banfield}, {Willett}, {de Gasperin}, {Lintott}, {L{\'o}pez-S{\'a}nchez}, {Middelberg}, {Norris}, {Schawinski}, {Seymour}, \& {Simmons}}]{Kapinska2017AJ....154..253K}
{Kapi{\'n}ska}, A.~D., {Terentev}, I., {Wong}, O.~I., {$et~al$.} 2017, \aj, 154, 253

\bibitem[{{Klein} {$et~al$.}(1995){Klein}, {Mack}, {Gregorini}, \& {Parma}}]{Klein1995A&A...303..427K}
{Klein}, U., {Mack}, K.~H., {Gregorini}, L., \& {Parma}, P. 1995, \aap, 303, 427

\bibitem[{{Kozie{\l}-Wierzbowska} {$et~al$.}(2020){Kozie{\l}-Wierzbowska}, {Goyal}, \& {{\.Z}ywucka}}]{Koziel-Wierzbowska2020ApJS..247...53K}
{Kozie{\l}-Wierzbowska}, D., {Goyal}, A., \& {{\.Z}ywucka}, N. 2020, \apjs, 247, 53

\bibitem[{{Kronberg} {$et~al$.}(2001){Kronberg}, {Dufton}, {Li}, \& {Colgate}}]{koronberg2001ApJ...560..178K}
{Kronberg}, P.~P., {Dufton}, Q.~W., {Li}, H., \& {Colgate}, S.~A. 2001, \apj, 560, 178

\bibitem[{{Kumari} \& {Pal}(2022)}]{Kumari2022MNRAS.514.4290K}
{Kumari}, S., \& {Pal}, S. 2022, \mnras, 514, 4290

\bibitem[{{Kumari} {$et~al$.}(2024){Kumari}, {Pal}, {Hardcastle}, \& {Horton}}]{Kumari2024arXiv240614889K}
{Kumari}, S., {Pal}, S., {Hardcastle}, M.~J., \& {Horton}, M.~A. 2024, arXiv e-prints, arXiv:2406.14889

\bibitem[{{Lacy} {$et~al$.}(2020){Lacy}, {Baum}, {Chandler}, {Chatterjee}, {Clarke}, {Deustua}, {English}, {Farnes}, {Gaensler}, {Gugliucci}, {Hallinan}, {Kent}, {Kimball}, {Law}, {Lazio}, {Marvil}, {Mao}, {Medlin}, {Mooley}, {Murphy}, {Myers}, {Osten}, {Richards}, {Rosolowsky}, {Rudnick}, {Schinzel}, {Sivakoff}, {Sjouwerman}, {Taylor}, {White}, {Wrobel}, {Andernach}, {Beasley}, {Berger}, {Bhatnager}, {Birkinshaw}, {Bower}, {Brandt}, {Brown}, {Burke-Spolaor}, {Butler}, {Comerford}, {Demorest}, {Fu}, {Giacintucci}, {Golap}, {G{\"u}th}, {Hales}, {Hiriart}, {Hodge}, {Horesh}, {Ivezi{\'c}}, {Jarvis}, {Kamble}, {Kassim}, {Liu}, {Loinard}, {Lyons}, {Masters}, {Mezcua}, {Moellenbrock}, {Mroczkowski}, {Nyland}, {O'Dea}, {O'Sullivan}, {Peters}, {Radford}, {Rao}, {Robnett}, {Salcido}, {Shen}, {Sobotka}, {Witz}, {Vaccari}, {van Weeren}, {Vargas}, {Williams}, \& {Yoon}}]{Lacy2020PASP..132c5001L}
{Lacy}, M., {Baum}, S.~A., {Chandler}, C.~J., {$et~al$.} 2020, \pasp, 132, 035001

\bibitem[{{Laing}(1994)}]{Laing1994ASPC...54..227L}
{Laing}, R.~A. 1994, in Astronomical Society of the Pacific Conference Series, Vol.~54, The Physics of Active Galaxies, ed. G.~V. {Bicknell}, M.~A. {Dopita}, \& P.~J. {Quinn}, 227

\bibitem[{{Laing} \& {Bridle}(2002)}]{Laing2002MNRAS.336.1161L}
{Laing}, R.~A., \& {Bridle}, A.~H. 2002, \mnras, 336, 1161

\bibitem[{{Laing} \& {Bridle}(2014)}]{Laing2014MNRAS.437.3405L}
---. 2014, \mnras, 437, 3405

\bibitem[{{Laing} {$et~al$.}(2008){Laing}, {Bridle}, {Parma}, \& {Murgia}}]{Laing2008MNRAS.391..521L}
{Laing}, R.~A., {Bridle}, A.~H., {Parma}, P., \& {Murgia}, M. 2008, \mnras, 391, 521

\bibitem[{{Lal}(2021)}]{Lal2021ApJ...915..126L}
{Lal}, D.~V. 2021, \apj, 915, 126

\bibitem[{{Lara} {$et~al$.}(1999){Lara}, {M{\'a}rquez}, {Cotton}, {Feretti}, {Giovannini}, {Marcaide}, \& {Venturi}}]{Lara1999A&A...348..699L}
{Lara}, L., {M{\'a}rquez}, I., {Cotton}, W.~D., {$et~al$.} 1999, \aap, 348, 699

\bibitem[{{Leahy} \& {Perley}(1991)}]{Leahy1991AJ....102..537L}
{Leahy}, J.~P., \& {Perley}, R.~A. 1991, \aj, 102, 537

\bibitem[{{Leahy} \& {Williams}(1984)}]{Leahy1984MNRAS.210..929L}
{Leahy}, J.~P., \& {Williams}, A.~G. 1984, \mnras, 210, 929

\bibitem[{{Ledlow} \& {Owen}(1996)}]{Ledlow1996AJ....112....9L}
{Ledlow}, M.~J., \& {Owen}, F.~N. 1996, \aj, 112, 9

\bibitem[{{Liu} {$et~al$.}(2019){Liu}, {Xu}, {Zheng}, {Li}, {Zhu}, {Ma}, \& {Lian}}]{Liu2019RAA....19..127L}
{Liu}, Y.-X., {Xu}, H.-G., {Zheng}, D.-C., {$et~al$.} 2019, Research in Astronomy and Astrophysics, 19, 127

\bibitem[{{Loken} {$et~al$.}(1995){Loken}, {Roettiger}, {Burns}, \& {Norman}}]{Loken1995ApJ...445...80L}
{Loken}, C., {Roettiger}, K., {Burns}, J.~O., \& {Norman}, M. 1995, \apj, 445, 80

\bibitem[{{Lynden-Bell}(1969)}]{Lynden-Bell1969Natur.223..690L}
{Lynden-Bell}, D. 1969, \nat, 223, 690

\bibitem[{{Machalski} {$et~al$.}(1982){Machalski}, {Maslowski}, {Condon}, \& {Condon}}]{Machalski1982AJ.....87.1150M}
{Machalski}, J., {Maslowski}, J., {Condon}, J.~J., \& {Condon}, M.~A. 1982, \aj, 87, 1150

\bibitem[{{Mack} {$et~al$.}(1994){Mack}, {Gregorini}, {Parma}, \& {Klein}}]{Mack1994A&AS..103..157M}
{Mack}, K.~H., {Gregorini}, L., {Parma}, P., \& {Klein}, U. 1994, \aaps, 103, 157

\bibitem[{{Mackay}(1969)}]{Mackay1969MNRAS.145...31M}
{Mackay}, C.~D. 1969, \mnras, 145, 31

\bibitem[{{Marecki}(2012)}]{Marecki2012A&A...544L...2M}
{Marecki}, A. 2012, \aap, 544, L2

\bibitem[{{Merritt} \& {Ekers}(2002)}]{Merritt2002Sci...297.1310M}
{Merritt}, D., \& {Ekers}, R.~D. 2002, Science, 297, 1310

\bibitem[{{Miley}(1980)}]{Miley1980ARA&A..18..165M}
{Miley}, G. 1980, \araa, 18, 165

\bibitem[{{Mingo} {$et~al$.}(2022){Mingo}, {Croston}, {Best}, {Duncan}, {Hardcastle}, {Kondapally}, {Prandoni}, {Sabater}, {Shimwell}, {Williams}, {Baldi}, {Bonato}, {Bondi}, {Dabhade}, {G{\"u}rkan}, {Ineson}, {Magliocchetti}, {Miley}, {Pierce}, \& {R{\"o}ttgering}}]{Mingo2022MNRAS.511.3250M}
{Mingo}, B., {Croston}, J.~H., {Best}, P.~N., {$et~al$.} 2022, \mnras, 511, 3250

\bibitem[{{Miraghaei} \& {Best}(2017)}]{Miraghaei2017MNRAS.466.4346M}
{Miraghaei}, H., \& {Best}, P.~N. 2017, \mnras, 466, 4346

\bibitem[{{Mizuno} {$et~al$.}(2014){Mizuno}, {Hardee}, \& {Nishikawa}}]{Mizuno2014ApJ...784..167M}
{Mizuno}, Y., {Hardee}, P.~E., \& {Nishikawa}, K.-I. 2014, \apj, 784, 167

\bibitem[{{Moffet}(1966)}]{Moffet1966ARA&A...4..145M}
{Moffet}, A.~T. 1966, \araa, 4, 145

\bibitem[{{Mukherjee} {$et~al$.}(2020){Mukherjee}, {Bodo}, {Mignone}, {Rossi}, \& {Vaidya}}]{Mukherjee2020MNRAS.499..681M}
{Mukherjee}, D., {Bodo}, G., {Mignone}, A., {Rossi}, P., \& {Vaidya}, B. 2020, \mnras, 499, 681

\bibitem[{{Mukherjee} {$et~al$.}(2021){Mukherjee}, {Bodo}, {Rossi}, {Mignone}, \& {Vaidya}}]{Mukherjee2021MNRAS.505.2267M}
{Mukherjee}, D., {Bodo}, G., {Rossi}, P., {Mignone}, A., \& {Vaidya}, B. 2021, \mnras, 505, 2267

\bibitem[{{Nakamura} {$et~al$.}(2007){Nakamura}, {Li}, \& {Li}}]{Nakamura2007ApJ...656..721N}
{Nakamura}, M., {Li}, H., \& {Li}, S. 2007, \apj, 656, 721

\bibitem[{{Norris} {$et~al$.}(2021){Norris}, {Crawford}, \& {Macgregor}}]{Norris2021Galax...9...83N}
{Norris}, R.~P., {Crawford}, E., \& {Macgregor}, P. 2021, Galaxies, 9, 83

\bibitem[{{O'Dea} \& {Owen}(1985)}]{O'Dea1985AJ.....90..954O}
{O'Dea}, C.~P., \& {Owen}, F.~N. 1985, \aj, 90, 954

\bibitem[{{O'Donoghue} {$et~al$.}(1993){O'Donoghue}, {Eilek}, \& {Owen}}]{O'Donoghue1993ApJ...408..428O}
{O'Donoghue}, A.~A., {Eilek}, J.~A., \& {Owen}, F.~N. 1993, \apj, 408, 428

\bibitem[{{Orr} \& {Browne}(1982)}]{Orr1982MNRAS.200.1067O}
{Orr}, M.~J.~L., \& {Browne}, I.~W.~A. 1982, \mnras, 200, 1067

\bibitem[{{Owen} \& {Ledlow}(1997)}]{Owen1997ApJS..108...41O}
{Owen}, F.~N., \& {Ledlow}, M.~J. 1997, \apjs, 108, 41

\bibitem[{{Owen} \& {Rudnick}(1976)}]{Owen1976ApJ...205L...1O}
{Owen}, F.~N., \& {Rudnick}, L. 1976, \apjl, 205, L1

\bibitem[{{Parma} {$et~al$.}(1985){Parma}, {Ekers}, \& {Fanti}}]{Parma1985A&AS...59..511P}
{Parma}, P., {Ekers}, R.~D., \& {Fanti}, R. 1985, \aaps, 59, 511

\bibitem[{{Parma} {$et~al$.}(1999){Parma}, {Murgia}, {Morganti}, {Capetti}, {de Ruiter}, \& {Fanti}}]{Parma1999A&A...344....7P}
{Parma}, P., {Murgia}, M., {Morganti}, R., {$et~al$.} 1999, \aap, 344, 7

\bibitem[{{Patra} {$et~al$.}(2023){Patra}, {Joshi}, \& {Gopal-Krishna}}]{Patra2023MNRAS.524.3270P}
{Patra}, D., {Joshi}, R., \& {Gopal-Krishna}. 2023, \mnras, 524, 3270

\bibitem[{{Pirya} {$et~al$.}(2011){Pirya}, {Nandi}, {Saikia}, \& {Singh}}]{Pirya2011BASI...39..547P}
{Pirya}, A., {Nandi}, S., {Saikia}, D.~J., \& {Singh}, M. 2011, Bulletin of the Astronomical Society of India, 39, 547

\bibitem[{{Porth} \& {Komissarov}(2015)}]{Porth2015MNRAS.452.1089P}
{Porth}, O., \& {Komissarov}, S.~S. 2015, \mnras, 452, 1089

\bibitem[{{Proctor}(2011)}]{Proctor2011ApJS..194...31P}
{Proctor}, D.~D. 2011, The Astrophysical Journal Supplement Series, 194, 31

\bibitem[{{Ramatsoku} {$et~al$.}(2020){Ramatsoku}, {Murgia}, {Vacca}, {Serra}, {Makhathini}, {Govoni}, {Smirnov}, {Andati}, {de Blok}, {J{\'o}zsa}, {Kamphuis}, {Kleiner}, {Maccagni}, {Moln{\'a}r}, {Ramaila}, {Thorat}, \& {White}}]{Ramatsoku2020A&A...636L...1R}
{Ramatsoku}, M., {Murgia}, M., {Vacca}, V., {$et~al$.} 2020, \aap, 636, L1

\bibitem[{{Rees}(1966)}]{Rees1966Natur.211..468R}
{Rees}, M.~J. 1966, \nat, 211, 468

\bibitem[{{Rees}(1978)}]{Rees1978Natur.275..516R}
---. 1978, \nat, 275, 516

\bibitem[{{Roberts} {$et~al$.}(2015){Roberts}, {Cohen}, {Lu}, {Saripalli}, \& {Subrahmanyan}}]{Roberts2015ApJS..220....7R}
{Roberts}, D.~H., {Cohen}, J.~P., {Lu}, J., {Saripalli}, L., \& {Subrahmanyan}, R. 2015, \apjs, 220, 7

\bibitem[{{Rodman} {$et~al$.}(2019){Rodman}, {Turner}, {Shabala}, {Banfield}, {Wong}, {Andernach}, {Garon}, {Kapi{\'n}ska}, {Norris}, \& {Rudnick}}]{Rodman2019MNRAS.482.5625R}
{Rodman}, P.~E., {Turner}, R.~J., {Shabala}, S.~S., {$et~al$.} 2019, \mnras, 482, 5625

\bibitem[{{Rottmann}(2001)}]{Rottmann2001PhDT.......173R}
{Rottmann}, H. 2001, PhD thesis, -

\bibitem[{{Rudnick} \& {Adams}(1979)}]{Rudnick1979AJ.....84..437R}
{Rudnick}, L., \& {Adams}, M.~T. 1979, \aj, 84, 437

\bibitem[{{Rudnick} \& {Owen}(1977)}]{Rudnick1977AJ.....82....1R}
{Rudnick}, L., \& {Owen}, F.~N. 1977, \aj, 82, 1

\bibitem[{{Rudnick} {$et~al$.}(2022){Rudnick}, {Br{\"u}ggen}, {Brunetti}, {Cotton}, {Forman}, {Jones}, {Nolting}, {Schellenberger}, \& {van Weeren}}]{Rudnick2022ApJ...935..168R}
{Rudnick}, L., {Br{\"u}ggen}, M., {Brunetti}, G., {$et~al$.} 2022, \apj, 935, 168

\bibitem[{{Ryu} {$et~al$.}(2008){Ryu}, {Kang}, {Cho}, \& {Das}}]{Ryu2008Sci...320..909R}
{Ryu}, D., {Kang}, H., {Cho}, J., \& {Das}, S. 2008, Science, 320, 909

\bibitem[{{Saripalli} \& {Roberts}(2018)}]{Saripalli2018ApJ...852...48S}
{Saripalli}, L., \& {Roberts}, D.~H. 2018, \apj, 852, 48

\bibitem[{{Sasmal} {$et~al$.}(2022){Sasmal}, {Bera}, \& {Mondal}}]{Sasmal2022AN....34310083S}
{Sasmal}, T.~K., {Bera}, S., \& {Mondal}, S. 2022, Astronomische Nachrichten, 343, e20210083

\bibitem[{{Scheuer} \& {Readhead}(1979)}]{Scheuer1979Natur.277..182S}
{Scheuer}, P.~A.~G., \& {Readhead}, A.~C.~S. 1979, \nat, 277, 182

\bibitem[{{Schoenmakers} {$et~al$.}(2000){Schoenmakers}, {de Bruyn}, {R{\"o}ttgering}, {van der Laan}, \& {Kaiser}}]{Schoenmakers2000MNRAS.315..371S}
{Schoenmakers}, A.~P., {de Bruyn}, A.~G., {R{\"o}ttgering}, H.~J.~A., {van der Laan}, H., \& {Kaiser}, C.~R. 2000, \mnras, 315, 371

\bibitem[{{Sebastian} {$et~al$.}(2022){Sebastian}, {Kharb}, {Lister}, {Marshall}, {O'Dea}, \& {Baum}}]{Sebastian2022ApJ...935...59S}
{Sebastian}, B., {Kharb}, P., {Lister}, M.~L., {$et~al$.} 2022, \apj, 935, 59

\bibitem[{{Shimwell} {$et~al$.}(2019){Shimwell}, {Tasse}, {Hardcastle}, {Mechev}, {Williams}, {Best}, {R{\"o}ttgering}, {Callingham}, {Dijkema}, {de Gasperin}, {Hoang}, {Hugo}, {Mirmont}, {Oonk}, {Prandoni}, {Rafferty}, {Sabater}, {Smirnov}, {van Weeren}, {White}, {Atemkeng}, {Bester}, {Bonnassieux}, {Br{\"u}ggen}, {Brunetti}, {Chy{\.z}y}, {Cochrane}, {Conway}, {Croston}, {Danezi}, {Duncan}, {Haverkorn}, {Heald}, {Iacobelli}, {Intema}, {Jackson}, {Jamrozy}, {Jarvis}, {Lakhoo}, {Mevius}, {Miley}, {Morabito}, {Morganti}, {Nisbet}, {Orr{\'u}}, {Perkins}, {Pizzo}, {Schrijvers}, {Smith}, {Vermeulen}, {Wise}, {Alegre}, {Bacon}, {van Bemmel}, {Beswick}, {Bonafede}, {Botteon}, {Bourke}, {Brienza}, {Calistro Rivera}, {Cassano}, {Clarke}, {Conselice}, {Dettmar}, {Drabent}, {Dumba}, {Emig}, {En{\ss}lin}, {Ferrari}, {Garrett}, {G{\'e}nova-Santos}, {Goyal}, {G{\"u}rkan}, {Hale}, {Harwood}, {Heesen}, {Hoeft}, {Horellou}, {Jackson}, {Kokotanekov}, {Kondapally}, {Kunert-Bajraszewska}, {Mahatma}, {Mahony}, {Mandal}, {McKean},
  {Merloni}, {Mingo}, {Miskolczi}, {Mooney}, {Nikiel-Wroczy{\'n}ski}, {O'Sullivan}, {Quinn}, {Reich}, {Roskowi{\'n}ski}, {Rowlinson}, {Savini}, {Saxena}, {Schwarz}, {Shulevski}, {Sridhar}, {Stacey}, {Urquhart}, {van der Wiel}, {Varenius}, {Webster}, \& {Wilber}}]{Shimwell2019A&A...622A...1S}
{Shimwell}, T.~W., {Tasse}, C., {Hardcastle}, M.~J., {$et~al$.} 2019, \aap, 622, A1

\bibitem[{{Shimwell} {$et~al$.}(2022){Shimwell}, {Hardcastle}, {Tasse}, {Best}, {R{\"o}ttgering}, {Williams}, {Botteon}, {Drabent}, {Mechev}, {Shulevski}, {van Weeren}, {Bester}, {Br{\"u}ggen}, {Brunetti}, {Callingham}, {Chy{\.z}y}, {Conway}, {Dijkema}, {Duncan}, {de Gasperin}, {Hale}, {Haverkorn}, {Hugo}, {Jackson}, {Mevius}, {Miley}, {Morabito}, {Morganti}, {Offringa}, {Oonk}, {Rafferty}, {Sabater}, {Smith}, {Schwarz}, {Smirnov}, {O'Sullivan}, {Vedantham}, {White}, {Albert}, {Alegre}, {Asabere}, {Bacon}, {Bonafede}, {Bonnassieux}, {Brienza}, {Bilicki}, {Bonato}, {Calistro Rivera}, {Cassano}, {Cochrane}, {Croston}, {Cuciti}, {Dallacasa}, {Danezi}, {Dettmar}, {Di Gennaro}, {Edler}, {En{\ss}lin}, {Emig}, {Franzen}, {Garc{\'\i}a-Vergara}, {Grange}, {G{\"u}rkan}, {Hajduk}, {Heald}, {Heesen}, {Hoang}, {Hoeft}, {Horellou}, {Iacobelli}, {Jamrozy}, {Jeli{\'c}}, {Kondapally}, {Kukreti}, {Kunert-Bajraszewska}, {Magliocchetti}, {Mahatma}, {Ma{\l}ek}, {Mandal}, {Massaro}, {Meyer-Zhao}, {Mingo}, {Mostert}, {Nair},
  {Nakoneczny}, {Nikiel-Wroczy{\'n}ski}, {Orr{\'u}}, {Pajdosz-{\'S}mierciak}, {Pasini}, {Prandoni}, {van Piggelen}, {Rajpurohit}, {Retana-Montenegro}, {Riseley}, {Rowlinson}, {Saxena}, {Schrijvers}, {Sweijen}, {Siewert}, {Timmerman}, {Vaccari}, {Vink}, {West}, {Wo{\l}owska}, {Zhang}, \& {Zheng}}]{Shimwell2022A&A...659A...1S}
{Shimwell}, T.~W., {Hardcastle}, M.~J., {Tasse}, C., {$et~al$.} 2022, \aap, 659, A1

\bibitem[{{Shklovskii}(1963)}]{Shklovskii1963SvA.....6..465S}
{Shklovskii}, I.~S. 1963, \sovast, 6, 465

\bibitem[{{Spangler} {$et~al$.}(1984){Spangler}, {Myers}, \& {Pogge}}]{Spangler1984AJ.....89.1478S}
{Spangler}, S.~R., {Myers}, S.~T., \& {Pogge}, J.~J. 1984, \aj, 89, 1478

\bibitem[{{Tadhunter}(2016)}]{Tadhunter2016A&ARv..24...10T}
{Tadhunter}, C. 2016, \aapr, 24, 10

\bibitem[{{Ulrich} {$et~al$.}(1980){Ulrich}, {Butcher}, \& {Meier}}]{Ulrich1980Natur.288..459U}
{Ulrich}, M.~H., {Butcher}, H., \& {Meier}, D.~L. 1980, \nat, 288, 459

\bibitem[{{Ulvestad} {$et~al$.}(1981){Ulvestad}, {Johnston}, {Perley}, \& {Fomalont}}]{Ulvestad1981AJ.....86.1010U}
{Ulvestad}, J., {Johnston}, K., {Perley}, R., \& {Fomalont}, E. 1981, \aj, 86, 1010

\bibitem[{{Urry} \& {Padovani}(1995)}]{Urry-Padvoni1995PASP..107..803U}
{Urry}, C.~M., \& {Padovani}, P. 1995, Publications of the Astronomical Society of the Pacific, 107, 803

\bibitem[{{van Weeren} {$et~al$.}(2019){van Weeren}, {de Gasperin}, {Akamatsu}, {Br{\"u}ggen}, {Feretti}, {Kang}, {Stroe}, \& {Zandanel}}]{van-Weeren2019SSRv..215...16V}
{van Weeren}, R.~J., {de Gasperin}, F., {Akamatsu}, H., {$et~al$.} 2019, \ssr, 215, 16

\bibitem[{{Vigotti} {$et~al$.}(1989){Vigotti}, {Grueff}, {Perley}, {Clark}, \& {Bridle}}]{Vigotti1989AJ.....98..419V}
{Vigotti}, M., {Grueff}, G., {Perley}, R., {Clark}, B.~G., \& {Bridle}, A.~H. 1989, \aj, 98, 419

\bibitem[{{White} {$et~al$.}(1997){White}, {Becker}, {Helfand}, \& {Gregg}}]{White-first1997ApJ...475..479W}
{White}, R.~L., {Becker}, R.~H., {Helfand}, D.~J., \& {Gregg}, M.~D. 1997, \apj, 475, 479

\bibitem[{{Willis}(1978)}]{Willis1978PhyS...17..243W}
{Willis}, A.~G. 1978, \physscr, 17, 243

\bibitem[{{Willis} {$et~al$.}(1978){Willis}, {Wilson}, \& {Strom}}]{Wills1978A&A....66L...1W}
{Willis}, A.~G., {Wilson}, A.~S., \& {Strom}, R.~G. 1978, \aap, 66, L1

\bibitem[{{Wirth} {$et~al$.}(1982){Wirth}, {Smarr}, \& {Gallagher}}]{Wirth1982AJ.....87..602W}
{Wirth}, A., {Smarr}, L., \& {Gallagher}, J.~S. 1982, \aj, 87, 401

\bibitem[{{Worrall} {$et~al$.}(1995){Worrall}, {Birkinshaw}, \& {Cameron}}]{Worrall1995ApJ...449...93W}
{Worrall}, D.~M., {Birkinshaw}, M., \& {Cameron}, R.~A. 1995, \apj, 449, 93

\bibitem[{{Yang} {$et~al$.}(2019){Yang}, {Joshi}, {Gopal-Krishna}, {An}, {Ho}, {Wiita}, {Liu}, {Yang}, {Wang}, {Wu}, \& {Yang}}]{Yang2019ApJS..245...17Y}
{Yang}, X., {Joshi}, R., {Gopal-Krishna}, {$et~al$.} 2019, \apjs, 245, 17

\bibitem[{{Zier} \& {Biermann}(2001)}]{Zier2001A&A...377...23Z}
{Zier}, C., \& {Biermann}, P.~L. 2001, \aap, 377, 23

\end{thebibliography}

\end{document}